\documentclass[preprint]{aastex}
\usepackage{multirow}
\usepackage{subfigure}
\usepackage{natbib}

\newcommand{\kms}{km s$^{-1}$ }
\newcommand{\kkms}{K km s$^{-1}$ }
\newcommand{\kkmspc}{K km s$^{-1}$ pc$^{2}$ }

\newcommand{\Msun}{M$_{\odot}$ }

\def\ts   {\thinspace}
\def\kms  {\ifmmode{{\rm \ts km\ts s}^{-1}}\else{\ts km\ts s$^{-1}$}\fi}
\def\kkms  {\ifmmode{{\rm \ts K\ts km\ts s}^{-1}}\else{\ts K\ts km\ts s$^{-1}$}\fi}
\def\lcou  {\ifmmode{{\rm \ts K\ts km\ts s}^{-1}\ts {\rm pc}^{2}}\else{\ts K\ts km\ts s$^{-1}$\ts
pc$^{2}$}\fi}
\def\msol   {\ifmmode{{\rm M}_{\odot}}\else{M$_{\odot}$}\fi}
\def\punit  {\ifmmode{{\rm \ts cm}^{-3} {\rm \ts K}}\else{\ts cm$^{-3}$ \ts K}\fi}

\slugcomment{ApJ. in press}

\shorttitle{Environmental Dependence of Giant Molecular
Cloud Properties in M51}
\shortauthors{Colombo et al.}

\begin{document}

\title{The PdBI Arcsecond Whirlpool Survey (PAWS): Environmental Dependence of Giant Molecular
Cloud Properties in M51\footnote{Based on
observations carried out with the IRAM Plateau de Bure Interferometer and 30m telescope. IRAM is
operated by INSY/CNRS (France), MPG (Germany) and IGN (Spain).}}

\author{Dario Colombo\altaffilmark{1}, Annie
Hughes\altaffilmark{1}, Eva Schinnerer\altaffilmark{1}, Sharon E. Meidt\altaffilmark{1}, Adam K. Leroy\altaffilmark{2}, J\'er\^ome
Pety\altaffilmark{3,}\altaffilmark{4}, Clare L. Dobbs\altaffilmark{5}, Santiago
Garc\'{i}a-Burillo\altaffilmark{6},
Ga\"{e}lle Dumas\altaffilmark{3}, Todd A. Thompson\altaffilmark{7,}\altaffilmark{8}, Karl
F. Schuster\altaffilmark{3} and Carsten Kramer\altaffilmark{9}.}

\altaffiltext{1}{Max Planck Institute for Astronomy, K\"onigstuhl 17, 69117 Heidelberg, Germany}
\altaffiltext{2}{National Radio Astronomy Observatory, 520 Edgemont Road, Charlottesville, VA
22903, USA}
\altaffiltext{3}{Institut de Radioastronomie Millim\'etrique, 300 Rue de la Piscine, F-38406 Saint
Martin d'H\`eres, France}
\altaffiltext{4}{Observatoire de Paris, 61 Avenue de l'Observatoire, F-75014 Paris, France.}
\altaffiltext{5}{School of Physics and Astronomy, University of Exeter, Stocker Road, Exeter EX4
4QL, UK}
\altaffiltext{6}{Observatorio Astron\'{o}mico Nacional - OAN, Observatorio de Madrid Alfonso XII, 3,
28014 - Madrid, Spain}
\altaffiltext{7}{Department of Astronomy, The Ohio State University, 140 W. 18th Ave., Columbus,
OH 43210, USA} 
\altaffiltext{8}{Center for Cosmology and AstroParticle Physics, The Ohio State University, 191 W.
Woodruff Ave., Columbus, OH 43210, USA}
\altaffiltext{9}{Instituto Radioastronom\'{i}a Milim\'{e}trica, Av. Divina Pastora 7, Nucleo
Central, 18012 Granada, Spain}

\begin{abstract}
\noindent Using data from the PdBI Arcsecond Whirlpool Survey (PAWS),
we have generated the largest extragalactic Giant Molecular Cloud
(GMC) catalog to date, containing 1,507 individual objects. GMCs in
the inner M51 disk account for only 54\% of the total $^{12}$CO(1-0)
luminosity of the survey, but on average they exhibit physical
properties similar to Galactic GMCs. We do not find a strong
correlation between the GMC size and velocity dispersion, and a simple
virial analysis suggests that $\sim30\%$ of GMCs in M51 are
unbound. We have analyzed the GMC properties within seven
dynamically-motivated galactic environments, finding that GMCs in the
spiral arms and in the central region are brighter and have higher
velocity dispersions than inter-arm clouds. Globally, the GMC mass
distribution does not follow a simple power-law shape. Instead, we
find that the shape of the mass distribution varies with galactic
environment: the distribution is steeper in inter-arm region than in
the spiral arms, and exhibits a sharp truncation at high masses for
the nuclear bar region. We propose that the observed environmental
variations in the GMC properties and mass distributions are a
consequence of the combined action of large-scale dynamical processes
and feedback from high mass star formation. We describe some
challenges of using existing GMC identification techniques for
decomposing the $^{12}$CO(1-0) emission in molecule-rich environments,
such as M51's inner disk.
\end{abstract}

\section{Introduction}\label{sec:intro}
\noindent The interstellar medium (ISM) is a dynamic and complex
system that is subject to numerous physical processes acting across a
wide range of spatial and temporal scales. Of these, understanding how
stars form out of the ISM is especially important since star formation
determines the appearance and evolution of galaxies. In enriched
systems (with metallicity Z$\geq$Z$_{\odot}$), stars preferentially
form in molecular gas (e.g \citealt{young91},
\citealt{glover12}). Milky Way surveys using CO emission lines as a
tracer for molecular gas (e.g \citealt{solomon87}, hereafter S87;
\citealt{dame01}), have shown that most of the Galactic molecular gas
is organized in large, discrete structures called giant molecular
clouds (GMCs). These clouds host virtually all star formation in the
Galaxy, but their formation, evolution and the processes that regulate
the conversion of molecular gas into stars remain poorly understood (for
a recent review, see \citealt{mckee07}). \\

\noindent GMCs in the Galaxy have typical sizes of $\sim50$ pc, masses
of $\sim1-2\times10^{5}$ $M_{\odot}$, temperatures of $\sim10$ K and
and number densities of $\sim50$ cm$^{-3}$ (e.g. \citealt{blitz93}).
As first described by \citealt{larson81}, Galactic GMCs show
correlations between their size, line width and luminosity. S87
determined these empirical relations using a catalog of 273 inner
Milky Way GMCs, establishing that GMCs are virialized objects with a
velocity dispersion proportional to the square root of their radius,
and a roughly constant surface density of $\sim100$ M$_{\odot}$
pc$^{-2}$ (\citealt{heyer09}). GMCs in the Galaxy show a power-law
mass spectrum with index $\gamma\sim-1.5$, which indicates that most
of the molecular gas is located in high mass clouds. \\

\noindent High resolution surveys of the CO emission in nearby
galaxies provide the opportunity to address the universality of GMC
properties and the relationship between GMCs and star formation across
a wide range of environments. To date, several CO surveys of Local
Group galaxies have achieved sufficient resolution to identify
individual GMCs (e.g. \citealt{fukui01}, \citealt{{engargiola03}},
\citealt{fukui05}, \citealt{leroy06}, \citealt{mizuno06},
\citealt{rosolowsky07}, \citealt{hughes10}, \citealt{wong11},
\citealt{hirota11}, \citealt{gratier12}, \citealt{rebolledo12},
Donovan-Meyer et al. 2013). Some studies (e.g. \citealt{sheth08},
\citealt{fukui10}) have concluded that GMCs are insensitive to the
physical conditions in their surroundings, while others have reported
environment-dependent variations in GMC properties. Several authors
have observed that quiescent GMCs are typically less luminous than
clouds that are actively forming stars (e.g. \citealt{hughes10},
\citealt{hirota11}, \citealt{gratier12}, \citealt{rebolledo12}),
 Bolatto et al. (2008; hereafter B08) found that GMC
  populations in Local Group galaxies followed similar Larson-type
  scaling relations as Milky Way GMCs, concluding that GMCs have
  similar physical properties (as traced through their CO emission)
  throughout the Local Group. Yet the universality of Larson's Laws
  has also been questioned: \cite{heyer09} showed that Milky Way
  clouds with higher mass surface densities typically have a larger
  velocity dispersion at a fixed size scale. In the LMC, \cite{wong11}
  found no obvious relation between cloud size and velocity
  dispersion, while \cite{gratier12} also obtained a poor
  size-linewidth correlation for GMCs in M33. A comparative study of
  Local Group galaxies using a consistent methodology to identify and
  parametrize GMCs suggested that the GMC mass distribution is
  steeper in the low-mass galaxies than in the inner Milky Way
  (\citealt{blitz07}). The more recent surveys of CO emission in the LMC
  and M33 by \cite{wong11} and \cite{gratier12} -- which identify a
  greater number of GMCs across a wider mass range than the datasets
  analyzed by \cite{blitz07} -- also find mass distributions steeper
  than in the Milky Way, with power-law slopes of $\gamma \sim -2$.
  \cite{wong11} demonstrate that the value of $\gamma$ in the LMC
  depends on the decomposition method, while \cite{gratier12} find
  that the GMC mass spectrum steepens with increasing galactocentric
  radius in M33.\\

\noindent To date, studies of extragalactic GMC populations have
mostly probed low-mass galaxies where atomic gas dominates the neutral
ISM. This is because it is difficult to achieve the angular resolution
required to identify individual GMCs in any galaxy outside the Local
Group with current telescopes. As a result, there are almost no maps of
the CO emission in massive star-forming spiral galaxies where
individual GMCs can be distinguished (the recent CARMA-Nobeyama Nearby
galaxies CO(1-0) survey (CANON) described by \citealt{donovan13} is a
notable exception). This is a major lack, because massive star-forming
spirals dominate the mass and light budget of blue galaxies and host
most of the star formation in the present-day universe
(e.g. \citealt{schiminovich07}). Understanding the formation and
evolution of GMCs in such systems will help us to understand the
physical processes that regulate the bulk of present-day massive star
formation, something that studies of HI-dominated, low-mass Local
Group galaxies with weak or absent spiral structure cannot do.\\

\noindent M51 represents one of the best targets to study the
properties of GMCs in a molecular gas dominated environment, since it
is a face-on (inclination $\sim22^{\circ}$, e.g. \cite{miyamoto2013},
Colombo et al., submitted), nearby (distance=7.6\,Mpc;
\citealt{ciardullo02}), interacting galaxy with prominent spiral arms,
a weak starburst, a LINER core and with a wealth of multi-wavelength
ancillary data. For these reasons, the molecular gas in M51 has
already been extensively studied (\citealt{vogel88},
\citealt{garcia93a}a, \citealt{garcia93b}b, \citealt{kuno95},
\citealt{kuno97}, \citealt{aalto99}, \citealt{helfer03},
\citealt{schuster07}, \citealt{hitschfeld09}, \citealt{koda09},
\citealt{schinnerer10}, \citealt{egusa11}). Among the more recent
works, the CARMA-NRO survey by \citealt{koda09} with a resolution of
$\sim150$\,pc allowed them to distinguish -- but not resolve -- GMCs
in M51. The authors identified a number of high mass objects
($M_{H_{2}}\approx10^{6}-10^{7}$ $M_{\odot}$) in the spiral arms, and
smaller clouds of $M_{H_{2}}\approx4\times10^{5}$ $M_{\odot}$
constituting $\sim30\%$ of the molecular mass in the
inter-arm. However previous studies of M51 have not had sufficient
resolution to analyze individual GMCs. One of the major goals of the
Plateau de Bure Interferometer Arcsecond Whirlpool Survey (PAWS,
\citealt{schinnerer13}) is to identify and describe the GMC population
in this prototypical massive star-forming spiral galaxy.\\

\noindent This paper is structured as follows. In
Section~\ref{sec:data} we briefly describe the PAWS dataset. In
Section~\ref{sec:constr} we summarize the method used to identify M51
GMCs and derive their physical properties. The GMC catalog is
presented in Section~\ref{sec:cat}. Our analysis of how cloud
properties, scaling relations and mass spectra vary between the
different dynamical environments is presented in
Sections~\ref{sec:props} to~\ref{sec:masspect}. In Section~\ref{sec:disc_mass} we
discuss a possible origin for the environmental differences in the GMC
properties and mass distributions, and summarize the evidence against
the universality of the GMC properties and Larson's laws (Section~\ref{sec:disc_larson}). Our conclusions are presented
in Section~\ref{summary}. The tests that we conducted to determine the
optimal parameters for our cloud decomposition and identification
algorithm are presented in the Appendix.

\section{Data}\label{sec:data}

\noindent The PdBI Arcsecond Whirlpool Survey (PAWS, Schinnerer et
al. 2013) is a large IRAM program involving 210 hours of
observations with the Plateau de Bure Interferometer (PdBI) and IRAM
30\,m telescope to conduct a sensitive, high angular resolution
($1''.16\times0''.97$), $^{12}$CO\,(1-0) survey of the inner disk of
M51a (field-of-view, FoV $\sim270''\times170''$). The spatial resolution at
our assumed distance to M51 of $7.6$\,Mpc (\citealt{ciardullo02})
is $\rm \sim40\,pc$. The inclusion of the 30\,m single dish
data during joint deconvolution ensures that flux information on all
spatial scales is conserved. The RMS of the noise fluctuations in the
cube is $\sim 0.4$ K per 5 km s$^{-1}$ channel. This sensitivity is
sufficient to detect an object with a gas mass of $\rm
1.2\times10^{5}\,M_{\odot}$ at the $5\sigma_{RMS}$ level. The PAWS
data cube covers the LSR velocity range between 173 to 769 km
s$^{-1}$. A detailed description of the observing strategy,
calibration and data reduction is presented by \cite{pety13}.

\section{Construction of the GMC catalog}\label{sec:constr}

\subsection{Identification of Significant Emission and Decomposition into GMCs}\label{sec:constr_meth_cprops}

\noindent We used the CPROPS package (\citealt{rl06}; herafter RL06)
to identify GMCs and measure their physical properties. CPROPS has
been fully described in RL06. In this Section, we provide a brief
summary of CPROPS in order to explain the construction of the PAWS GMC
catalog. \\

\noindent CPROPS begins by identifying a ``working area'',
i.e. regions of significant emission within the data cube. This is
done by masking pixels in two consecutive velocity channels in which
the signal is above $t\sigma_{RMS}$ (the \verb"THRESHOLD" parameter in
CPROPS). These regions are then extended to include all adjacent
pixels in which the signal is above $e\sigma_{RMS}$ (the \verb"EDGE"
parameter in CPROPS) in at least two consecutive channels. The RMS
noise $\sigma_{RMS}$ is estimated from the median absolute deviation
(MAD) of each spectrum. To be consistent with previous GMC studies
(e.g. B08) we adopted $t=4$ and $e=1.5$. After defining the working
area, CPROPS proceeds to generate a catalog of \emph{islands},
emission structures within the working area with a projected area of
at least one telescope beam and spanning one or more velocity
channels. This kind of approach can be sufficient to catalog discrete
molecular structures in irregular and flocculent galaxies, where the
emission is typically sparsely distributed within the observed field
(e.g. the LMC, \citealt{wong11}). For the PAWS data cube, by contrast,
bright CO emission is present throughout the inner spiral arms and across
the central region, and is hence identified as a single island. We
present a catalog of islands within the PAWS FoV in
Appendix~\ref{app:island}. \\

\noindent To identify structures that resemble Galactic GMCs, we used
a ``data-based'' decomposition to further segment the islands. These
objects are defined using a modified watershed algorithm: local maxima
(called ``kernels'' in CPROPS) within a box of 120 pc $\times$ 120 pc
and 15 km\,s$^{-1}$ are recognized as independent objects if they lie at
least $2\sigma_{RMS}$ above the shared contour (called the ``merge
level'' in CPROPS) with any other maximum. By default, CPROPS requires
that the moments associated with other maxima differ by 100\%,
otherwise the two maxima are merged into a single cloud. We found that
this condition does not work well for the PAWS data, causing CPROPS to
reject a large number of objects that visual inspection would suggest
are GMCs. In brief, this is because CPROPS attempts to compare all the
local maxima within the bright region of contiguous emission that
encompasses the spiral arms, even when the local maxima are spatially
well-separated. We disable this step of the decomposition algorithm by
setting the parameter \verb"SIGDISCONT"=0. We explain our tests of the
CPROPS decomposition algorithm in more detail in
Appendix~\ref{app:test_sd}.\\

\subsection{Definition of GMC properties}\label{sec:constr_props}

\noindent CPROPS uses an extrapolated moment method to measure the
physical properties of the clouds that it identifies. To reduce
observational bias, CPROPS extrapolates the cloud property
measurements to values that would be expected in the case of perfect
sensitivity by performing a growth-type analysis on the observed
emission. CPROPS also corrects for finite resolution in the spatial
and spectral domain by deconvolving the telescope beam and the width
of a spectral channel from the measured cloud size and line
width. CPROPS estimates the uncertainty in measured cloud properties
via bootstrapping of the assigned pixels. We tested that 50
bootstrapping measurements provide a reliable estimate of the
uncertainty. This bootstrapping approach captures the dominant
uncertainty for bright clouds, but neglects the statistical
uncertainty due to noise fluctuations that can be significant for low
$S/N$ data. To check that the bootstrapping
  uncertainties provide a reliable estimate of the uncertainty in our
  cloud properties, we generated 100 synthetic datacubes each
  containing a barely resolved, round model cloud, to which we added
  different realizations of noise at the beam scale. We ran CPROPS on
  these cubes, and compared the standard deviation of the cloud
  property measurements to the uncertainties estimated by the
  bootstrapping procedure. We found that the bootstrapping
  uncertainties were approximately equal to the standard deviation of
  the cloud property measurements for clouds with low $S/N$ ratios
  ($S/N \in [3,5]$), while for brighter clouds ($S/N \in [10,20]$), the
  bootstrapping uncertainties were larger than the standard deviation
  of the cloud property measurements by a factor of $\sim2$ or
  more. In what follows, we refer to all objects whose properties
have been calculated by these procedures as GMCs, and we quote the
bootstrapping uncertainties only. We distinguish them from the
entities that are initially identified by CPROPS (i.e. prior to the
application of sensitivity and resolution corrections), which we call
``identified objects''. In the rest of this Section, we summarize the
cloud property definitions that are used by CPROPS. \\

\subsubsection{Basic GMC properties}\label{sec:basic_def}

\noindent \emph{Peak brightness temperature}. The peak brightness
temperature of a GMC is the CO brightness at the local maximum within
the cloud. It is measured directly from the data, i.e. without
extrapolation or deconvolution.\\

\noindent \emph{Effective radius}. CPROPS calculates the major and
minor axes of the identified objects using a moment method that takes
into account the intensity profile of the emission. In this technique,
the cloud root-mean-square (RMS) size, $\sigma_{r}$, is calculated as the
geometric mean of the second spatial moment of the intensity distribution
along the major ($\sigma_{a}$(0 K)) and minor ($\sigma_{b}$(0 K)) axes
extrapolated for perfect sensitivity:
\begin{equation}\label{sigmar}
 \sigma_{r}=\sqrt{\sigma_{a}(0\textrm{ K})\sigma_{b}(0\textrm{ K})},
\end{equation}
\noindent Assuming that the cloud is a sphere, its \emph{effective
radius}, $R$, is related to $\sigma_{r}$ through
the sphere's density profile, $\rho\propto r^{-\beta}$. CPROPS uses a truncated
density profile with $\beta=1$, in which case the object's effective
radius is $R = 1.91\sigma_{r}$. The effective
radius is then deconvolved by the beam size $\theta_{FWHM}$:
\begin{equation}\label{beamdec}
    R = 1.91\sqrt{\left(\sigma_{a}^{2}(0\textrm{ K})-\left(\frac{\theta_{FWHM}}{\sqrt{8\ln(2)}}\right)^{2}\right)^{1/2}\left(\sigma_{b}^{2}(0\textrm{ K})-\left(\frac{\theta_{FWHM}}{\sqrt{8\ln(2)}}\right)^{2}\right)^{1/2}},
\end{equation}

\noindent If one or both axes of the cloud are
  smaller than the beam ($\theta_{FWHM}/\sqrt{8\ln(2)}$), then the
  deconvolution correction results in an undefined radius. The cloud
  is not rejected by CPROPS since it consists of more pixels than a
  cylinder with dimensions of one beam area $\times$ one channel
  width.  For these objects we define an upper limit to the effective
  radius:
\begin{equation}\label{rupplim}
 R=1.91\frac{\theta_{FWHM}}{\sqrt{8\ln(2)}}.
\end{equation}
\noindent Approximately $\sim35\%$ of the GMCs in the PAWS catalog have
  only an upper limit to their radius. We exclude these clouds from the
  analysis in this paper.\\

\noindent \emph{Velocity dispersion}. To estimate the FWHM line width of a GMC,
$\Delta V$, CPROPS assumes a Gaussian velocity profile. In this case,
$\Delta V$ is related to the velocity dispersion $\sigma_{v}$ as:
\begin{equation}\label{deltaV}
    \Delta V=\sqrt{8\textrm{ln}(2)}\sigma_{v}.
\end{equation}
\noindent The velocity dispersion $\sigma_{v}$ is obtained from its extrapolated
value for perfect sensitivity, $\sigma_{v}$(0 K), deconvolved by the
channel width $\Delta V_{chan}$:
\begin{equation}\label{sigmav}
 \sigma_{v}=\sqrt{\sigma_{v}^{2}(0\textrm{ K})-\frac{\Delta V_{chan}^{2}}{2\pi}}.
\end{equation}

\noindent As for the GMC radius, the deconvolution can
  result in clouds with line widths narrower than a single
  channel. However, we note that if the initially identified object
  spans less than two channels, then it is automatically discarded
  from the catalog.\\

\noindent \emph{Axis ratio}. The ratio between the major and minor
axis is obtained directly from the spatial moments $\sigma_{b}$(0 K)
and $\sigma_{a}$(0 K) without conversion into their physical
quantities. The axis ratio, $b/a$, parametrizes the shape of the
cloud: for a round cloud $b/a=1$, while $b/a<1$ corresponds to an
elongated cloud.\\

\noindent \emph{Position angle and orientation}. The position angle
$PA$ of each cloud's major axis is measured clockwise, i.e. from North
through West, with North set to $PA=0^{\circ}$. In a spiral galaxy, it
is often more instructive to study the position angle of the clouds
with respect to the spiral arm frame. Thus we define the cloud
orientation $\phi$ as the angle between the cloud major axis and a
double logarithmic spiral with a pitch angle $i_{p}=21^{\circ}$. This
pitch angle is conventionally adopted to define M51's spiral arms
(e.g. \citealt{kuno97}). A GMC population with major axes perfectly
aligned with the spiral arms would yield a delta function distribution
of $\phi$ values, centered at $\phi=0^{\circ}$.

\subsubsection{Derived GMC properties}

\noindent \emph{Cloud mass}. CPROPS estimates the cloud mass in two
ways: from the CO luminosity and from the virial theorem.  The CO
luminosity of the cloud, $L_{CO}$, is the integrated flux scaled by
the square of the distance $D$ in parsec:
\begin{equation}\label{Lco}
    L_{CO}[\textrm{K km s}^{-1}\textrm{ pc}^{2}]=\sum_{i}T_{i}\delta_{v}\delta_{x}\delta_{y}\times
D^{2}\times\left(\frac{\pi}{180\cdot3600}\right)^{2},
\end{equation}
\noindent where $\delta x$ and $\delta y$ are the pixel scale in
arcsec, and $\delta v$ is the channel width in \kms. We use the same
formula to calculate the total CO luminosity within the cube (or part
thereof). The CO luminosity of each GMC is corrected for finite
sensitivity using the standard CPROPS procedure to extrapolate
$L_{CO}$. \\

\noindent Assuming that the CO integrated intensity $I_{CO}$ is
related to the underlying molecular hydrogen column density
$N_{H_{2}}$ by a constant conversion factor, $X_{CO}=I_{CO}/N_{H_{2}}$
(e.g. \citealt{dickman78}), the cloud's CO luminosity $L_{CO}$ can be
  used to estimate its total mass $M_{lum}$. That is,
\begin{equation}\label{mlum}
    M_{lum}[\textrm{M}_{\odot}]=\frac{X_{CO}}{2\times10^{20}\textrm{cm$^{-2}$ (K km
s$^{-1}$)$^{-1}$}}\times4.4L_{CO}[\textrm{K km s}^{-1}\textrm{ pc}^{2}].
\end{equation}
\noindent An appropriate value of $X_{CO}$ is often chosen to
  bring a cloud population close to virial equilibrium
  (\citealt{hughes10}, \citealt{fukui08}). By contrast, we calculate
  $M_{lum}$ using the fiducial CPROPS conversion factor
  $X_{CO}=2\times10^{20}$ cm$^{-2}$ (K km s$^{-1}$)$^{-1}$, consistent with the 
  recent estimations of M51 $X_{CO}$ obtained by \cite{schinnerer10} and \cite{tan11}.\\

\noindent The virial mass, $M_{vir}$, depends on the density profile of the
cloud. For a cloud with a density profile of $\rho\propto r^{-1}$ the
virial mass is:
\begin{equation}\label{Mvir}
    M_{vir}[\textrm{M}_{\odot}]=1040\sigma_{v}^{2}R,
\end{equation}
\noindent where $R$ is the cloud radius in parsec, and $\sigma_{v}$ is
the velocity dispersion in km s$^{-1}$.\\

\noindent \emph{H$_{2}$ mass surface density}. The effective radius of
the cloud $R$ is defined as the radius of a circle that encompasses an
area equivalent to the projected area of the cloud. The molecular gas
surface density $\Sigma_{H_{2}}$ is then:
\begin{equation}\label{surdens}
 \Sigma_{H_{2}} = \frac{M_{lum}}{\pi R^{2}}.
\end{equation}\\

\noindent \emph{Scaling coefficient}. The scaling coefficient, $c$, parametrizes
the scaling between size and velocity dispersion of a cloud. It is defined as:
\begin{equation}\label{scalecoeff}
 c\equiv\frac{\sigma_{v}}{R^{1/2}}.
\end{equation}
\noindent For a cloud in virial equilibrium ($M_{lum}\approx M_{vir}$), the
scaling coefficient is related to the cloud surface density as:
\begin{equation}\label{scalecoeffsigma}
 c=\sqrt{\frac{\pi\Sigma_{H_{2}}}{1040}}.
\end{equation}\\

\noindent \emph{Virial parameter}. The dimensionless virial parameter
$\alpha$ has a value of order unity and characterizes deviations from
the virial theorem applied to a non-magnetized cloud with no external
pressure and constant density (see \citealt{bertoldi92}). This
parameter quantifies the ratio of the cloud's kinetic to gravitational
energy, i.e.:

\begin{equation}\label{virpar}
    \alpha = \frac{5\sigma_{v}^{2}R}{GM_{lum}} = \frac{1161\sigma_{v}^{2}R}{M_{lum}}.
\end{equation}

\noindent In the literature, clouds with $\alpha\sim1$ are considered
as gravitationally bound and stabilized by internal thermal and
turbulent pressure against collapse. Clouds with $\alpha>>1$ are
either externally bound or transient features of the ISM. In general
$\alpha=2$ is regarded as the threshold between gravitationally bound
and unbound objects. If long-lived, clouds with $\alpha<<1$ must be
supported against collapse by something more than their internal
turbulent motions, such as the magnetic field.

\section{PAWS GMC catalog}\label{sec:cat}

\noindent The final GMC catalog of the PAWS project contains 1,507
objects.  Table~\ref{gmcs_catalog} presents the first 10 entries of
the PAWS GMC catalog. The complete version is available in electronic
format to the dedicated web-page \verb"http://www.mpia-hd.mpg.de/home/PAWS/PAWS/Data.html".
Here we provide a brief description of the information
contained in the catalog.

\begin{itemize}
  \item \footnotesize \verb"Column 1": \emph{ID}, cloud identification number;
  \item \footnotesize \verb"Column 2": \emph{RA (J2000)}, cloud's Right
Ascension in sexagesimal format;
  \item \footnotesize \verb"Column 3": \emph{Dec (J2000)}, cloud's Declination in sexagesimal
format;
  \item \footnotesize \verb"Column 4": $V_{LSR}$, cloud's radial velocity with respect to M51
systemic velocity in the Local Standard of Rest in km\,s$^{-1}$;
  \item \footnotesize \verb"Column 5": $T_{max}$, cloud's peak temperature in K; 
  \item \footnotesize \verb"Column 6": $S/N$, cloud's peak signal-to-noise ratio; 
  \item \footnotesize \verb"Column 7": \emph{R}, cloud's deconvolved, extrapolated
effective radius in pc including uncertainty;
  \item \footnotesize \verb"Column 8": $\sigma_{v}$, cloud's deconvolved, extrapolated velocity
dispersion in km\,s$^{-1}$ including uncertainty;
  \item \footnotesize \verb"Column 9": $L_{CO}$, cloud's integrated and extrapolated CO
luminosity in K\,km\,s$^{-1}$\,pc$^{2}$ including uncertainty;
  \item \footnotesize \verb"Column 10": $M_{vir}$, cloud's mass inferred from the virial theorem in
M$_{\odot}$ including uncertainty;
  \item \footnotesize \verb"Column 11": $\alpha$, cloud's virial parameter;
  \item \footnotesize \verb"Column 12": \emph{PA}, cloud's position angle in degrees;
  \item \footnotesize \verb"Column 13": \emph{b/a}, the cloud's minor-to-major axis ratio;
  \item \footnotesize \verb"Column 14": Region where a given GMC has been identified, i.e. center
(CR), spiral arms (SA), inter-arm (IA);
  \item \footnotesize \verb"Column 15": Flag for radius measurement: 0 = measurement of radius, 1
= upper limit (see
Section~\ref{sec:constr_props} for details).
\end{itemize}

\noindent The values tabulated for the cloud's location in space and
velocity (Column 2 to 4) refer to the weighted mean position within
the cloud, which is not necessarily coincident with the location of
the brightness temperature peak within the cloud. We consider the
catalog to be complete down to a mass equivalent to $3\times$ the
survey's $5\sigma_{RMS}$ sensitivity limit. Our adopted mass
\emph{completeness limit} is therefore $3.6\times10^{5}$ $M_{\odot}$.\\

\noindent The initial list of clouds identified by
  CPROPS includes some objects in regions of the data cube where no CO
  emission associated with M51 is expected. These detections are
  likely to be noise peaks that are falsely identified as GMCs.
  To eliminate obvious false positives from the catalog, we inspected
  the line profiles from each cloud candidate visually, and rejected 99
  objects that lie outside the CLEAN mask that was used in the joint
  deconvolution of the PAWS cube (\citealt{pety13}). The CLEAN mask
  includes $\sim50\%$ of the total number of (x,y,v) pixels in the
  cube, which is large compared to the number of pixels corresponding
  to identified islands ($\sim3\%$). Objects that fall on the edge of
  the mask are retained in the catalog if their centers are inside the
  mask. Fig.~1 presented histograms of the $S/N$ ratio of false
  positives and the objects identified inside the deconvolution
  mask. The $S/N$ of the false positives ranges between $4$ and
  $6.5$. Since the number of pixels inside and outside the CLEAN mask
  is roughly equal, we expect $\sim100$ of the cataloged GMCs to be
  spurious. We adopt $S/N=6.5$ as the threshold for our subsample of
  761 ``highly reliable'' GMCs.

\begin{figure}[h]
\begin{center}
\includegraphics[width=0.6 \textwidth]{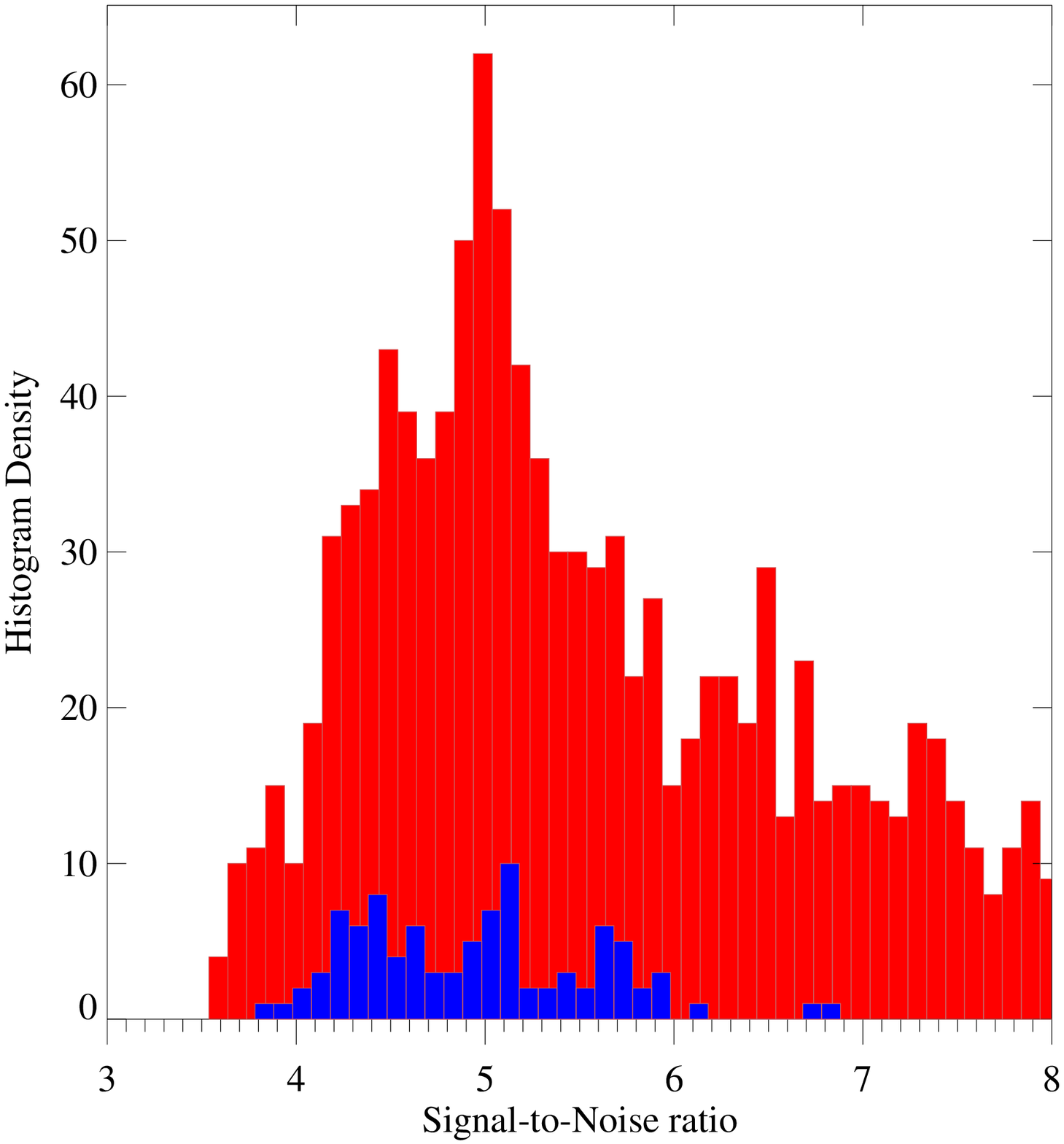} 	
\caption{\footnotesize Histograms of S/N distribution of cataloged
objects (red) and false positives (blue) eliminated via application of
the CLEAN mask. The histogram range is restricted to a $S/N=8$ to
emphasize the distribution of the removed false positives.}
\end{center}
\label{false}
\end{figure}

\begin{deluxetable}{ccccccccccccccc}
\tabletypesize{\scriptsize}
\rotate
\tablecaption{PAWS GMC catalog\label{gmcs_catalog}}
\tablewidth{0pt}
\setlength{\tabcolsep}{0.02in}
\renewcommand{\arraystretch}{1.2}
\tablehead{
\colhead{ID} & \colhead{RA (J2000)} & \colhead{Dec (J2000)} & \colhead{$\Delta V_{LSR}$} &
\colhead{$T_{max}$} & \colhead{$S/N$} & \colhead{$R$} & \colhead{$\sigma_{v}$} &
\colhead{$L_{CO}$} & \colhead{$M_{vir}$} &
\colhead{$\alpha$} & \colhead{PA}& \colhead{b/a}& \colhead{Reg} &\colhead{Flag}\\
& \colhead{$hh\,mm\,ss.ss$} & \colhead{$dd\,mm\,ss.ss$} & \colhead{km\,s$^{-1}$} & \colhead{K} &&
\colhead{pc} &
\colhead{km\,s$^{-1}$} & \colhead{$10^{5}$\,K\,km\,s$^{-1}$\,pc$^{2}$} & \colhead{$10^{5}$\,M$_{\odot}$} & 
&& \colhead{deg}&&\\
\colhead{(1)} & \colhead{(2)} & \colhead{(3)} & \colhead{(4)} & \colhead{(5)} & \colhead{(6)} &
\colhead{(7)} & 
\colhead{(8)} & \colhead{(9)} & \colhead{(10)} & \colhead{(11)} & \colhead{(12)} & \colhead{(13)} & 
\colhead{(14)} & \colhead{(15)}}
\startdata
$1$&$13^{h}30^{m}0.65^{s}$&$47^{\circ}
11'10.58"$&$-4.3$&$2.5$&$5.2$&$18\pm19$&$3.7\pm2.6$&$0.9\pm0.3$&$2.7\pm5.7$&$0.7$&$135$&$1.0$&$IA$&$0$\\
$2$&$13^{h}30^{m}0.87^{s}$&$47^{\circ}
10'56.15"$&$52.8$&$5.3$&$7.0$&$50\pm8$&$10.2\pm1.9$&$4.4\pm0.7$&$54.0\pm25.4$&$2.8$&$49$&$0
.9$&$IA$&$0$\\
$3$&$13^{h}30^{m}1.54^{s}$&$47^{\circ}
11'4.84"$&$60.5$&$4.6$&$5.1$&$32\pm0$&$10.8\pm4.5$&$2.1\pm0.8$&$38.5\pm31.9$&$4.3$&$152$&$0
.6$&$IA$&$1$\\
$4$&$13^{h}29^{m}58.01^{s}$&$47^{\circ}
11'6.34"$&$-2.4$&$1.3$&$3.8$&$32\pm0$&$5.1\pm3.9$&$0.7\pm0.6$&$8.5\pm13.0$&$2.8$&$179$&$0.2$&$SA$&$1$\\
$5$&$13^{h}29^{m}57.79^{s}$&$47^{\circ}
11'7.20"$&$3.3$&$2.1$&$5.8$&$40\pm21$&$9.6\pm3.7$&$1.6\pm0.6$&$38.0\pm33.7$&$5.4$&$8$&$0.9
$&$SA$&$0$\\
$6$&$13^{h}29^{m}58.14^{s}$&$47^{\circ}
11'6.34"$&$15.3$&$2.5$&$6.7$&$27\pm33$&$1.9\pm2.0$&$0.6\pm0.8$&$1.1\pm3.2$&$0.4$&$116$&$0.6
$&$SA$&$0$\\
$7$&$13^{h}29^{m}58.76^{s}$&$47^{\circ}
11'9.41"$&$13.4$&$2.2$&$5.8$&$32\pm0$&$3.1\pm3.6$&$0.3\pm0.2$&$3.2\pm7.6$&$2.2$&$11$&$0.4$
&$SA$&$1$\\
$8$&$13^{h}29^{m}58.36^{s}$&$47^{\circ}
11'10.50"$&$15.8$&$2.8$&$7.8$&$32\pm0$&$11.1\pm7.3$&$0.5\pm1.1$&$40.9\pm53.3$&$18.4$&$158$&
$0.5$&$SA$&$1$\\
$9$&$13^{h}29^{m}57.72^{s}$&$47^{\circ}
11'2.80"$&$24.0$&$4.1$&$9.9$&$118\pm14$&$7.3\pm1.1$&$10.8\pm3.1$&$65.5\pm24.8$&$1.4$&$163$&
$0.5$&$SA$&$0$\\
$10$&$13^{h}29^{m}58.24^{s}$&$47^{\circ}
11'9.36"$&$19.6$&$5.0$&$12.0$&$32\pm15$&$8.0\pm3.9$&$3.8\pm3.3$&$21.3\pm21.6$&$1.3$&$133$&$
0.6$&$SA$&$0$\\
...&...&...&...&...&...&...&...&...&...&...&...&...&...&...\\
...&...&...&...&...&...&...&...&...&...&...&...&...&...&...\\
...&...&...&...&...&...&...&...&...&...&...&...&...&...&...\\
$1507$&$13^{h}29^{m}46.33^{s}$&$47^{\circ}
12'40.28"$&$-0.4$&$2.9$&$5.6$&$32\pm0$&$10.1\pm3.5$&$0.9\pm0.2$&$33.7\pm23.4$&$9.2$&$148$&$
0.6$&$IA$&$1$\\
\enddata
\tablecomments{\footnotesize 
(1) cloud identification number (\emph{ID}), 
(2) Right Ascension (\emph{RA (J2000)}), 
(3) Declination (\emph{Dec (J2000)}), 
(4) Velocity with respect to the systematic velocity of NGC5194 ($\sim472$ km/s, \citealt{shetty07}), 
(5) Peak brightness temperature ($T_{max}$),
(6) Peak signal-to-noise ratio ($S/N$),
(7) Radius (\emph{R}), 
(8) Velocity dispersion ($\sigma_{v}$), 
(9) CO luminosity ($L_{CO}$), 
(10) Mass from virial theorem ($M_{vir}$),
(11) Virial parameter ($\alpha$), 
(12) Position angle of cloud major axis, measured from North through West (\emph{PA}), 
(13) Ratio between minor axis and major axis ($b/a$), 
(14) Region of M51 where a given cloud has been identified, i.e. center (\emph{CR}), spiral arms
(\emph{SA}), inter-arm (\emph{IA}),
(15) Flag$=0$ indicates the default measurement of the
cloud radius, Flag$=1$ indicates that the radius is substituted with an upper limit.}
\end{deluxetable}

\newpage

\section{Environmental dependence of the GMC properties in M51}\label{sec:props}

\noindent Previous observations of M51 have indicated
  that galactic environment is important for the organization and
  properties of the molecular gas. Recently, for example,
  \cite{koda09} showed that M51's spiral arms contain giant
  molecular associations (GMAs) with masses between $10^{7}-10^{8}$
  M$_{\odot}$, while the inter-arm region hosts only smaller clouds
  with masses less than $\sim10^{6}$ M$_{\odot}$. To test whether the
  physical properties of GMCs depend on environment in M51, we divide
  the PAWS FoV into seven distinct regions (see
  Section~\ref{sec:envs_def}). We analyze the global properties of the
  CO emission and the GMC ensemble in
  Section~\ref{sec:props_general}. Environmental trends in the GMC
  property distributions are examined in
  Section~\ref{sec:props_basic_deriv}.

\newpage  

\begin{figure}
\begin{center}
\includegraphics[width=1\textwidth]{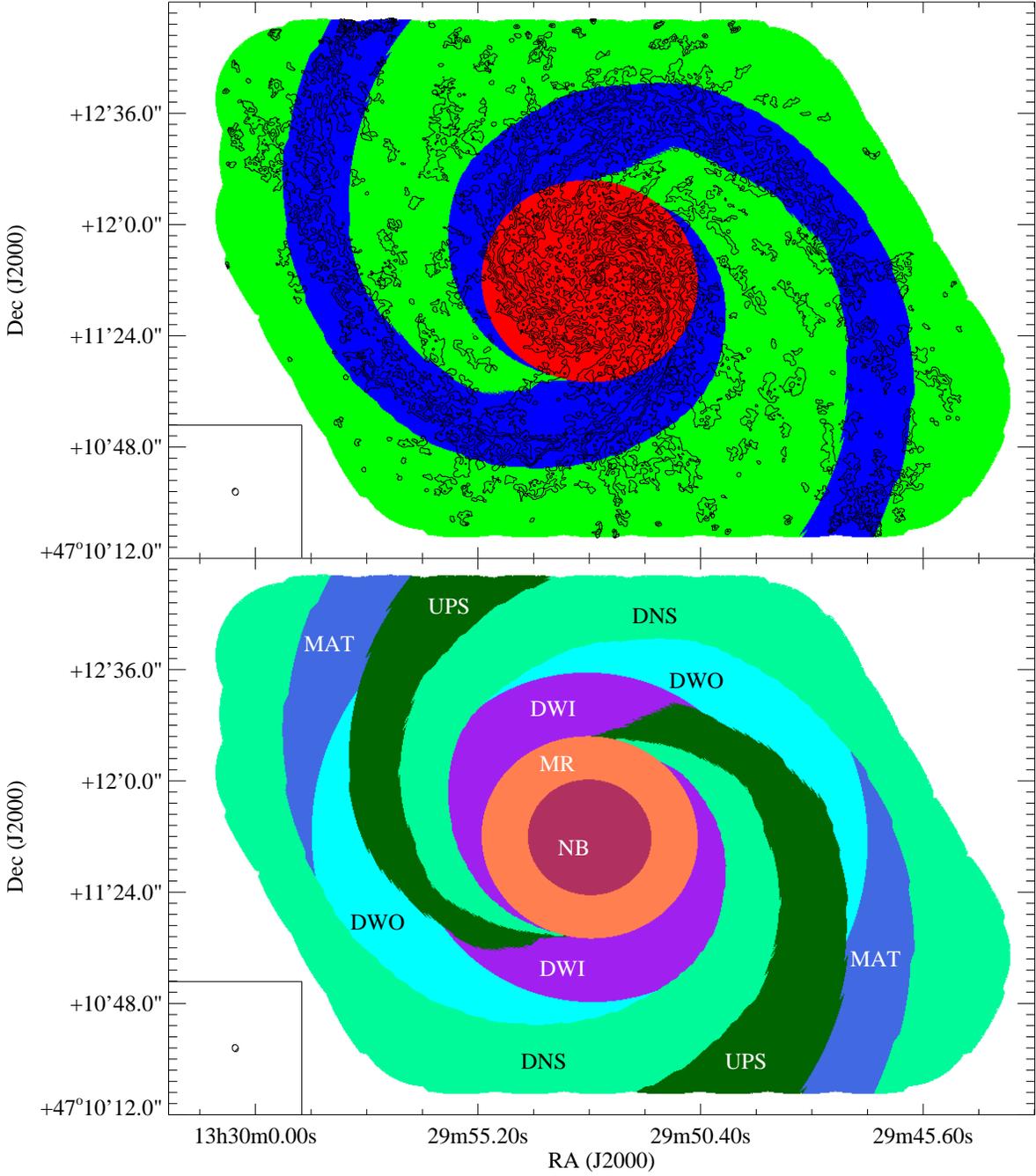}
\end{center}
\caption{\scriptsize Top: the three main regions in which the PAWS field of view is divided:
center in red, spiral arm in blue and inter-arm in green. Contours at 10, 50, 100, 200 and 400 K\,km\,s$^{-1}$ belong to the integrated
intensity map of islands. Bottom: M51 environmental mask. Nuclear bar (NB) and molecular ring (MR)
are indicated in dark red and orange, respectively. Inner density-wave
spiral arms (DWI) are indicated in purple,
outer density-wave spiral arms (DWO) in cyan, and material arms (MAT) in light
blue. Downstream with respect to the spiral arms (DNS) is shaded light green while upstream
is shaded dark green (UPS). These color codes will be kept throughout the paper. In the bottom left of both panels the beam ($\sim1"$ or 40\,pc) is shown.}
\label{mask}
\end{figure}

\clearpage
\newpage

\subsection{M51 environment definition}\label{sec:envs_def}

\noindent We use the stellar potential of M51 to divide the PAWS FoV
into seven distinct dynamical environments, each of which contains a
statistically significant GMC population. Initially, we distinguish
between the ``center'' ($R_{gal}\lesssim1.3$ kpc) and ``disk''
($1.3\lesssim R_{gal}\lesssim5$ kpc) regions within the PAWS FoV.  The
central region (CR) is further separated into i) a nuclear bar (NB)
region that is located within the corotation resonance of the bar, and
ii) the molecular ring (MR), which is a zone of zero torque created by
the combined dynamical effects of the spiral and nuclear bar.  The
``disk'' region is divided azimuthally into spiral arm (SA) and
inter-arm (IA) zones. Based on the direction of the gas flow within
the arms derived from the torque map (\citealt{meidt13}) and tracers
of massive star formation activity, we segment the spiral arm region
radially into: i) inner density-wave spiral arms (DWI), ii) outer
density-wave spiral arms (DWO) and iii) material arms (MAT). We divide
the inter-arm zones into downstream (DNS) and upstream (UPS) regions
relative to the spiral arms. The seven environments within the PAWS
FoV are illustrated in Fig.~\ref{mask}. We describe the construction
of our environmental mask in more detail in
Appendix~\ref{app:env_m51}.\\

\begin{figure}[!h]
\begin{center}
\includegraphics[width=0.95\textwidth]{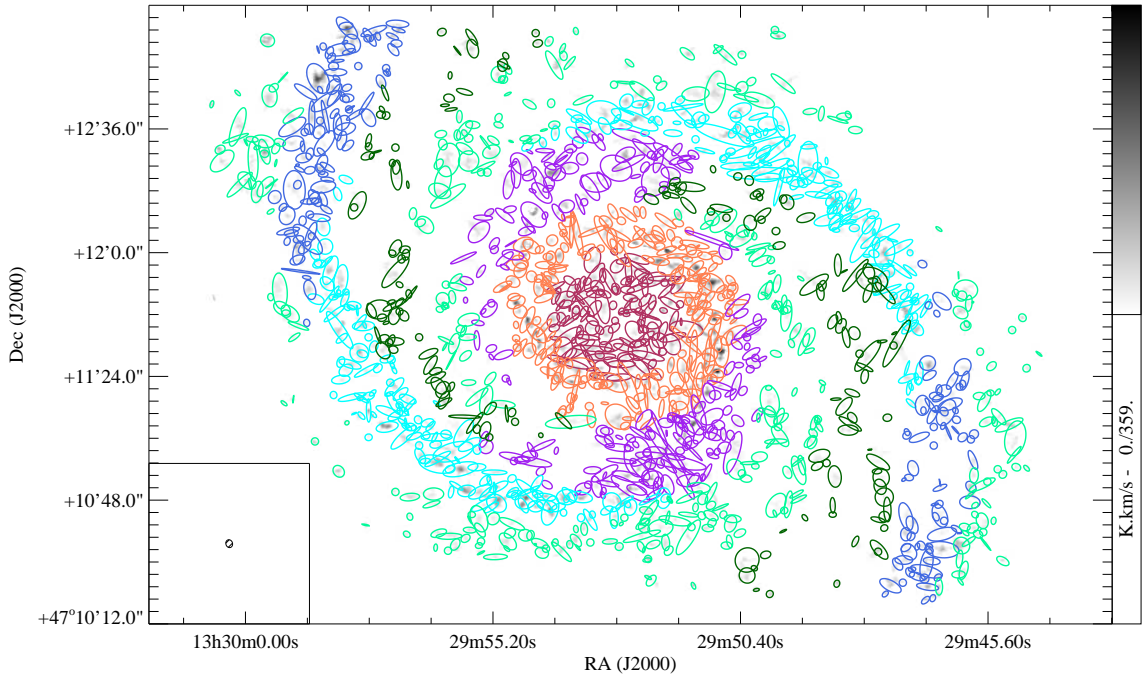}
\end{center}
\caption{\scriptsize The GMC distribution
in the PAWS field of M51 superimposed on the integrated
identified object CO intensity map (grey-scale). The sidebar indicates the color scale of the map in K\,km\,s$^{-1}$. The GMCs are represented as ellipses with the
extrapolated and deconvolved major and minor axes, oriented according to the measured position
angle. The clouds that appear overlapping are actually separated along the velocity axis. Colors
indicate the environment in which a given object has been identified following the color code of
Fig.~\ref{mask}. These color codes will be kept throughout the paper. In the bottom left of both panels the beam ($\sim1"$ or 40\,pc) is shown.}
\label{gmc_ellipse}
\end{figure}

\clearpage
\newpage

\subsection{Properties of CO emission and the GMC Ensemble in Different M51 Environments}\label{sec:props_general}

\noindent In Table~\ref{gmcs_flux}, we list several key properties of
the CO emission and GMC populations within the different galactic
environments. These tabulated properties include the total CO
luminosity, the fraction of the CO emission that is relatively bright
and hence included within the CPROPS ``working area'', and the total
number and number density of GMCs. One obvious difference between the
environments is the contribution of high S/N emission to the region's
total CO luminosity: emission belonging to the CPROPS working area
constitutes 80-90\% of the CO luminosity present in the spiral arm and
central regions, but only $\sim45\%$ of the inter-arm
emission. Another way to quantify this is via the average H$_{2}$ mass
surface density ($\Sigma_{H_{2}}$) calculated across each region. Assuming a constant
conversion factor ($X_{CO}=2\times10^{20}$cm$^{-2}$ (K km
s$^{-1}$)$^{-1}$), the center of M51 has the highest H$_{2}$ mass
surface density $\Sigma_{H_{2}} = 237$\,M$_{\odot}$\,pc$^{-2}$, while in the spiral arm and
in the inter-arm regions the $\Sigma_{H_{2}}$ is a factor 2 and 6
lower, respectively. Since the area of the inter-arm relative to the
spiral arm increases with galactocentric radius, this decline is
consistent with the radial decrease in the molecular mass surface
density reported by lower resolution CO studies of M51,
e.g. \cite{schuster07}. The number density of clouds, $N_{GMC}$,
shows a similar trend as $\Sigma_{H_{2}}$, decreasing from
72\,kpc$^{-2}$ in the central region to 45\,kpc$^{-2}$ in the spiral
arms and 19\,kpc$^{-2}$ in the inter-arm region.\\

\noindent Table~\ref{gmcs_flux} shows that the flux associated with
GMCs ($L_{CO}^{EX}$) is 54\% of the total flux in the PAWS data
cube $L_{CO}\approx91\times10^{7}$ \kkmspc.\footnote{In
    this paper we refer to the CO luminosity within the area observed
    by PAWS as the total CO luminosity. A detailed comparison of the
    flux measured by PAWS to equivalent measurements by the BIMA SoNG
    (\citealt{helfer03}) and CARMA-NRO (\citealt{koda09}) surveys is
    presented in \cite{pety13}. These authors find that the flux
    measurements agree within 10\%, which is consistent with the
    uncertainties in absolute flux calibration for millimeter data.} A
  significant fraction of the emission of the PAWS cube is thus not
  decomposed by CPROPS into GMCs. The remaining flux could be due to
  structures smaller than the beam or in the extended component
  identified by \cite{pety13}. We note that the CO luminosity
  contained in the identified objects ($L_{CO}^{NX}$) is only
  $\sim20\%$ of the total flux in the cube, i.e. more than half of the
  combined flux of GMCs is recovered through the extrapolation step of
  the CPROPS decomposition algorithm. We discuss this issue further in
  Section~\ref{sec:cat_reliability}.\\

\begin{deluxetable}{l|ccc|cc|cccccc}
\tabletypesize{\scriptsize}
\rotate
\setlength{\tabcolsep}{0.04in}
\renewcommand{\arraystretch}{1.2}
\tablewidth{0pt}
\tablecaption{Global Properties of M51's GMC population and environments\label{gmcs_flux}}
\tablehead{
\colhead{\textbf{Envir.}}&\multicolumn{3}{|c|}{\textbf{Whole region}}&\multicolumn{2}{|c|}{\textbf{Working Area}}&\multicolumn{6}{|c}{\textbf{GMC}}\\
&\colhead{\emph{$^{(1)}$A}}&\colhead{$^{(2)}L_{CO}$}&\multicolumn{1}{c|}{$^{(3)}\Sigma_{H_{2}}$}&\colhead{$^{(4)}$\emph{A}}&\multicolumn{1}{c|}{$^{(5)}L_{CO}$}&\colhead{$^{(6)}L_{CO}^{NX}$}&\colhead{$^{(7)}L_{CO}^{EX}$}&\colhead{$^{(8)}\%^{NX}$}&\colhead{$^{(9)}\%^{EX}$}&
\colhead{\emph{$^{(10)}$\#}}&\colhead{\emph{$^{(11)}$N$_{GMC}$}}\\
&\colhead{[kpc$^{2}$]}&\colhead{[$10^{7}$ K km s$^{-1}$ pc$^{2}$]}&\multicolumn{1}{c|}{[M$_{\odot}$ pc$^{-2}]$}&
\multicolumn{1}{|c}{[kpc$^{2}$]}&\multicolumn{1}{c|}{[$10^{7}$ K km s$^{-1}$ pc$^{2}$]}&\multicolumn{2}{|c}{[$10^{7}$ K km s$^{-1}$ pc$^{2}$]}&&&&\colhead{[kpc$^{-2}$]}\\
}
\startdata
\emph{Cube} &47.0&90.83&84.22&11.5&67.08&17.81&48.65&20&54&1507&32\\
\hline
\hline
\emph{CR} &4.7&25.47&237.02&1.2&22.85&4.71&14.48&18&57&335&73\\
\emph{SA} &14.6&43.44&129.94&2.3&35.10&8.16&23.22&21&59&657&45\\
\emph{IA} &27.8&21.88&34.37&1.0&9.12&4.93&10.93&19&42&514&19\\
\hline
\hline
\emph{NB} &1.5&7.48&213.11&2.7&6.49&1.43&4.18&19&56&126&82\\
\emph{NR} &3.2&17.99&248.62&5.5&16.35&3.28&10.30&18&57&209&66\\
\hline
\emph{DWI} &4.2&5.50&56.90&3.3&4.75&2.32&7.23&18&55&204&48\\
\emph{DWO} &5.3&10.54&87.09&1.0&9.16&3.69&10.72&20&58&274&52\\
\emph{MAT} &3.9&3.50&39.31&1.7&2.33&2.15&5.27&27&65&179&46\\
\hline
\emph{DNS} &20.7&8.21&17.25&1.8&6.81&3.57&7.66&20&43&350&17\\
\emph{UPS} &8.2&10.13&53.70&2.5&8.25&1.36&3.27&17&42&164&20\\
\enddata
\tablecomments{\footnotesize 
$^{(1)}$ area encompassed by M51's environments; 
$^{(2)}$ CO luminosity contained in the environment area; 
$^{(3)}$ H$_{2}$ mass surface density of the given environment;
$^{(4)}$ area encompassed by M51's environment working areas; 
$^{(5)}$ CO luminosity contained in the environment area within the working area; 
$^{(6)}$ and $^{(7)}$ CO luminosity associated with identified GMCs, before and after extrapolation,
respectively; 
$^{(8)}$ and $^{(9)}$ percentage CO luminosity contained in GMCs, before and after
extrapolation,
respectively, with respect to the total CO luminosity of the environment; 
$^{(10)}$ number of GMCs in a given environment; 
$^{(11)}$ number density of GMCs in a given environment.}
\end{deluxetable}

\newpage
 
\begin{figure}
\begin{center}
\includegraphics[width=0.85\textwidth]{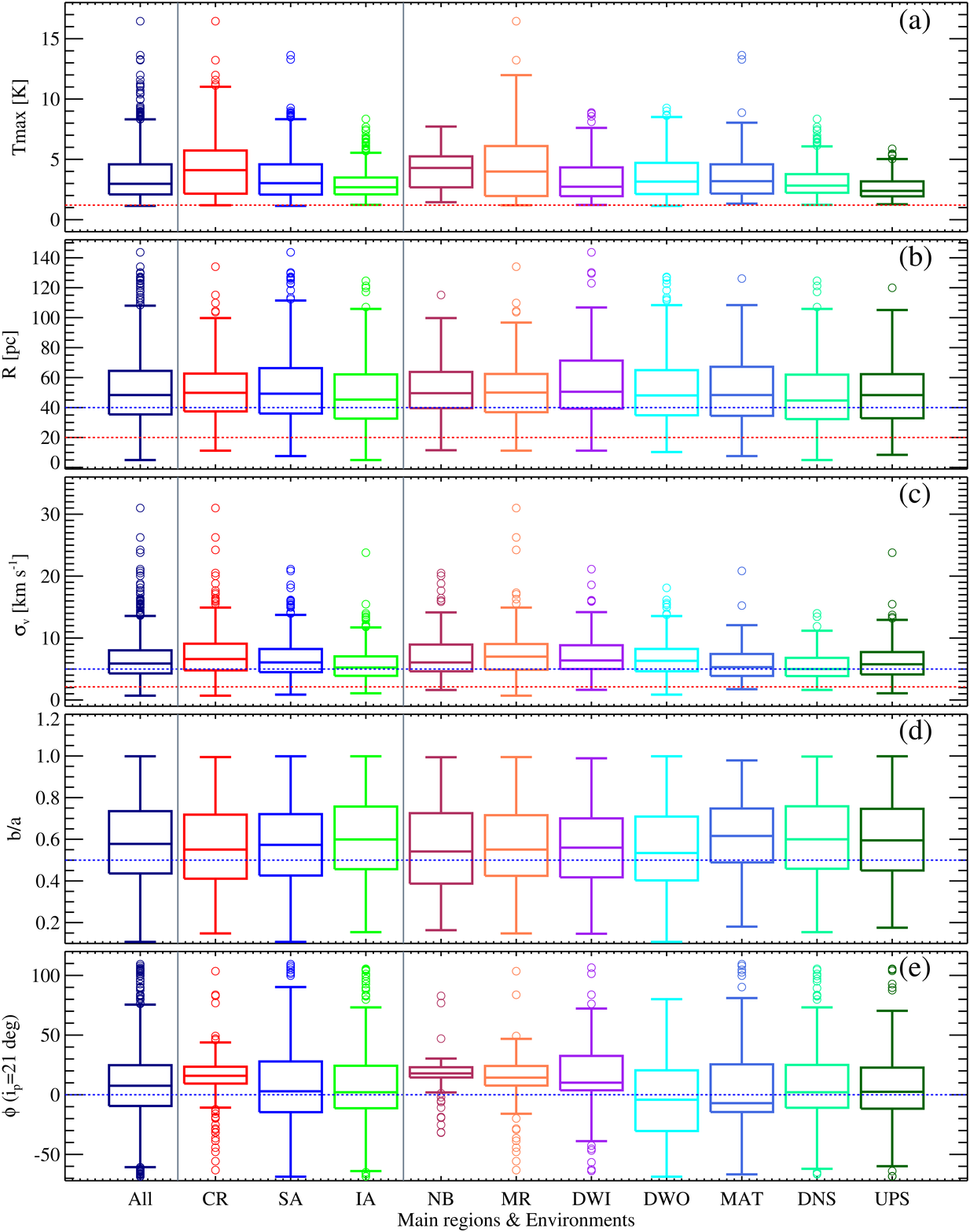}
\end{center}
\caption{\scriptsize Basic GMC properties (from the top to the bottom):
(a) peak brightness temperature $T_{max}$, 
(b) effective radius $R$, 
(c) velocity dispersion $\sigma_{v}$, 
(d) axis ratio $b/a$ and
(e) orientation $\phi$ shown in a ``box and whiskers''
representation for different M51 environments (from the left to the right: All -full sample;
3 main regions -center (CR), spiral arm (SA), inter-arm (IA) and 7 environments defined in
Fig.~\ref{mask} and Appendix~\ref{app:env_m51}). The
box middle band represents the median of the distribution. The box itself
contains $50\%$ of the data points. Each whisker that emerges from the box, coinciding with
$\sim25\%$
of the data points, corresponds roughly to $3\sigma$ of a normal distribution. The median of
velocity dispersion and brightness temperature is always higher in the central region (CR and MR,
NB) and in the density-wave spiral arms (DWI and DWO), compared to inter-arm environments (DNS,
UPS). Straight horizontal red lines indicate the resolution, instrumental or sensitivity
limits: 1.2 K for the peak brightness temperature, 20 pc for the radius, 2.12 km\,s$^{-1}$  for the
velocity dispersion. Reference lines at arbitrary values are
indicated in blue to help guide the eye. Circles represent the outliers of the distribution (see description at the beginning of Section~\ref{sec:props_general}).}
\label{basic_props_envs}
\end{figure}

\subsection{Variation of GMC physical properties with
environment}\label{sec:props_basic_deriv}

\noindent In this section, we examine whether the physical properties
of GMCs -- such as radius, velocity dispersion and mass -- vary with
galactic environment. To visualize the GMC property distributions, we
use a ``box and whiskers'' plot (e.g. \citealt{tukey77}) in
Figures~\ref{basic_props_envs} and~\ref{derived_props_envs}. This
representation is a useful tool to identify and illustrate differences
in the shape of non-Gaussian distributions. The box is
  delimited by two lines that indicate the lower $Q25$ and upper $Q75$
  quartiles of the distribution. The middle band represents the
  median. For a normal distribution, the interquartile range or
  distribution spread ($IQR \equiv Q75 - Q25$) corresponds to
  $1.35\sigma$, where $\sigma$ is the standard deviation. $0.5IQR$ 
  corresponds to $0.6745\sigma$ or to the median absolute
  deviation (MAD). The ends of the whiskers indicate the lowest and
  the highest data points that lie within 1.5 $\times IQR$ of the
  lower quartile (the bottom whisker, BW) and 1.5 $\times IQR$ of the
  upper quartile (the top whisker, TW). For a normal distribution, the
  range of values between TW (or BW) and the middle band roughly
  corresponds to $\pm3\sigma$. We define ``outliers'' as data points
  with values lower or greater than BW or TW, respectively
  (i.e. outside the $3\sigma$ range of a Gaussian distribution), and
  represent them as circles in the box and whiskers plots. The median
  and the lower and upper quartiles ($Q25$ and $Q75$, respectively) of
  the GMC property distributions are listed in
  Table~\ref{gmc_props_table}.\\

\noindent To test the statistical significance of differences between
the GMC property distributions, we use the two-sided
Kolmogorov-Smirnov (KS) test (e.g. \citealt{eadie71}) on both the full
and the ``highly reliable cloud'' samples.  The two-sided KS statistic
quantifies a distance between the empirical distribution functions of
two samples assuming as a null hypothesis that the samples are drawn
from the same parent distribution. This distance is directly connected
to the \emph{p-value}, the probability that two samples descend from
the same parent population. Traditionally, the null hypothesis is
rejected when the \emph{p-value} is smaller than a certain
significance level. We adopt the convention that there is a
significant difference between two samples if the \emph{p-value} is
lower than 0.001, while \emph{p-values} less than or equal to 0.05
indicate marginally significant differences.  We use a modified
version of the two-sided KS test that attempts to account for
measurement uncertainties (for details see Appendix~D).

\newpage

\begin{deluxetable}{lccccccccccc}
\tabletypesize{\scriptsize}
\rotate
\setlength{\tabcolsep}{0.04in}
\renewcommand{\arraystretch}{1.2}
\tablecaption{GMC properties in the different environments of M51\label{gmc_props_table}}
\tablehead{
\multicolumn{1}{l|}{ } & \multicolumn{10}{c}{\textbf{GMC Property}}\\
\multicolumn{1}{l|}{\textbf{Env.}} & \multicolumn{5}{c|}{\textbf{Basic}} &
\multicolumn{5}{c}{\textbf{Derived}} \\
\multicolumn{1}{l|}{} & \colhead{$T_{max}$} & \colhead{R} & \colhead{$\sigma_{v}$} & 
\colhead{b/a} & \colhead{$\phi$} & \multicolumn{1}{|c}{$M_{lum}$} &
\colhead{$M_{vir}$} &
\colhead{$\Sigma_{H_{2}}$} & \colhead{$c$} & \colhead{$\alpha$}\\
\multicolumn{1}{l|}{} & \colhead{[K]} & \colhead{[pc]} &
\colhead{[km\,s$^{-1}$]} & & \colhead{[deg]} & \multicolumn{1}{|c}{[$10^{5}$ M$_{\odot}$]} &
\colhead{[$10^{5}$ M$_{\odot}$]}
&
\colhead{[M$_{\odot}$\,pc$^{-2}$]} & \colhead{[km\,s$^{-1}$\,pc$^{-1/2}$]} & 
}
\tablewidth{0pt}
\startdata
\emph{All}&\multicolumn{1}{|c}{$3.0^{+4.6}_{-2.1}$}&\multicolumn{1}{|c}{$48.4^{+64.5}_{-35.4}$}
&\multicolumn{1}{|c}{$5.9^{+8.0}_{-4.3}$}&\multicolumn{1}{|c}{$0.6^{+0.7}_{-0.4}$}&\multicolumn{1}{
|c}{$7.6^{+24.8}_{-9.4}$}&\multicolumn{1}{|c}{$7.6^{+16.5}_{-3.4}$}&\multicolumn{1}{|c}{$19.6^{
+40.5}_{-9.4}$}&\multicolumn{1}{|c}{$177.4^{+298.5}_{-110.2}$}&\multicolumn{1}{|c}{$0.9^{+1.3}_{-0.7
}$}&\multicolumn{1}{|c}{$1.6^{+3.2}_{-0.9}$}\\
\hline
\hline
\emph{CR}&\multicolumn{1}{|c}{$4.1^{+5.7}_{-2.2}$}&\multicolumn{1}{|c}{$49.8^{+62.7}_{-37.4}$}
&\multicolumn{1}{|c}{$6.6^{+9.1}_{-4.8}$}&\multicolumn{1}{|c}{$0.6^{+0.7}_{-0.4}$}&\multicolumn{1}{
|c}{$15.9^{+23.5}_{-9.4}$}&\multicolumn{1}{|c}{$10.4^{+24.0}_{-3.9}$}&\multicolumn{1}{|c}{$25.1^{
+50.2}_{-12.5}$}&\multicolumn{1}{|c}{$212.4^{+368.2}_{-129.2}$}&\multicolumn{1}{|c}{$1.0^{+1.4}_{
-0.7}$}&\multicolumn{1}{|c}{$1.5^{+3.5}_{-0.9}$}\\
\emph{SA}&\multicolumn{1}{|c}{$3.0^{+4.6}_{-2.1}$}&\multicolumn{1}{|c}{$49.3^{+66.3}_{-36.0}$}
&\multicolumn{1}{|c}{$6.1^{+8.2}_{-4.5}$}&\multicolumn{1}{|c}{$0.6^{+0.7}_{-0.4}$}&\multicolumn{1}{
|c}{$2.9^{+27.9}_{-14.6}$}&\multicolumn{1}{|c}{$8.3^{+18.2}_{-3.6}$}&\multicolumn{1}{|c}{$21.7^{
+45.1}_{-10.7}$}&\multicolumn{1}{|c}{$185.3^{+304.1}_{-112.4}$}&\multicolumn{1}{|c}{$1.0^{+1.3}_{
-0.7}$}&\multicolumn{1}{|c}{$1.7^{+3.0}_{-0.9}$}\\
\emph{IA}&\multicolumn{1}{|c}{$2.7^{+3.5}_{-2.1}$}&\multicolumn{1}{|c}{$45.3^{+62.2}_{-32.6}$}
&\multicolumn{1}{|c}{$5.2^{+7.0}_{-3.9}$}&\multicolumn{1}{|c}{$0.6^{+0.8}_{-0.5}$}&\multicolumn{1}{
|c}{$2.1^{+24.3}_{-11.3}$}&\multicolumn{1}{|c}{$5.8^{+11.0}_{-3.1}$}&\multicolumn{1}{|c}{$14.8^{
+31.0}_{-6.9}$}&\multicolumn{1}{|c}{$143.4^{+228.1}_{-94.0}$}&\multicolumn{1}{|c}{$0.8^{+1.2}_{-0.6}
$}&\multicolumn{1}{|c}{$1.6^{+3.2}_{-0.8}$}\\
\hline
\hline
\emph{NB}&\multicolumn{1}{|c}{$4.3^{+5.2}_{-2.7}$}&\multicolumn{1}{|c}{$49.6^{+63.8}_{-39.6}$}
&\multicolumn{1}{|c}{$6.1^{+9.0}_{-4.6}$}&\multicolumn{1}{|c}{$0.5^{+0.7}_{-0.4}$}&\multicolumn{1}{
|c}{$17.9^{+23.0}_{-14.5}$}&\multicolumn{1}{|c}{$10.7^{+19.8}_{-5.7}$}&\multicolumn{1}{|c}{$20.7^{
+49.8}_{-11.5}$}&\multicolumn{1}{|c}{$184.3^{+291.1}_{-111.6}$}&\multicolumn{1}{|c}{$0.9^{+1.3}_{
-0.6}$}&\multicolumn{1}{|c}{$1.5^{+3.7}_{-0.9}$}\\
\emph{MR}&\multicolumn{1}{|c}{$4.0^{+6.1}_{-2.0}$}&\multicolumn{1}{|c}{$50.0^{+62.4}_{-36.9}$}
&\multicolumn{1}{|c}{$7.0^{+9.0}_{-4.9}$}&\multicolumn{1}{|c}{$0.6^{+0.7}_{-0.4}$}&\multicolumn{1}{
|c}{$14.4^{+24.1}_{-7.7}$}&\multicolumn{1}{|c}{$10.4^{+27.1}_{-3.5}$}&\multicolumn{1}{|c}{$26.8^{
+50.2}_{-13.6}$}&\multicolumn{1}{|c}{$227.4^{+387.6}_{-141.8}$}&\multicolumn{1}{|c}{$1.0^{+1.4}_{
-0.8}$}&\multicolumn{1}{|c}{$1.6^{+3.4}_{-0.9}$}\\
\hline
\emph{DWI}&\multicolumn{1}{|c}{$2.7^{+4.3}_{-1.9}$}&\multicolumn{1}{|c}{$50.5^{+71.3}_{-39.3}$}
&\multicolumn{1}{|c}{$6.4^{+8.8}_{-5.0}$}&\multicolumn{1}{|c}{$0.6^{+0.7}_{-0.4}$}&\multicolumn{1}{
|c}{$10.1^{+32.6}_{-3.8}$}&\multicolumn{1}{|c}{$8.5^{+16.5}_{-3.7}$}&\multicolumn{1}{|c}{$29.9^{
+52.0}_{-12.6}$}&\multicolumn{1}{|c}{$155.0^{+251.9}_{-110.2}$}&\multicolumn{1}{|c}{$1.0^{+1.3}_{
-0.7}$}&\multicolumn{1}{|c}{$2.1^{+3.6}_{-1.2}$}\\
\emph{DWO}&\multicolumn{1}{|c}{$3.2^{+4.7}_{-2.1}$}&\multicolumn{1}{|c}{$48.1^{+65.0}_{-34.8}$}
&\multicolumn{1}{|c}{$6.3^{+8.2}_{-4.6}$}&\multicolumn{1}{|c}{$0.5^{+0.7}_{-0.4}$}&\multicolumn{1}{
|c}{$-4.2^{+20.5}_{-30.4}$}&\multicolumn{1}{|c}{$8.6^{+22.8}_{-3.8}$}&\multicolumn{1}{|c}{$22.8^{
+42.3}_{-11.4}$}&\multicolumn{1}{|c}{$218.7^{+317.3}_{-123.5}$}&\multicolumn{1}{|c}{$1.0^{+1.3}_{
-0.8}$}&\multicolumn{1}{|c}{$1.7^{+2.7}_{-1.0}$}\\
\emph{MAT}&\multicolumn{1}{|c}{$3.2^{+4.6}_{-2.2}$}&\multicolumn{1}{|c}{$48.3^{+67.2}_{-34.5}$}
&\multicolumn{1}{|c}{$5.3^{+7.4}_{-3.9}$}&\multicolumn{1}{|c}{$0.6^{+0.7}_{-0.5}$}&\multicolumn{1}{
|c}{$-7.0^{+25.5}_{-14.5}$}&\multicolumn{1}{|c}{$7.1^{+15.0}_{-3.1}$}&\multicolumn{1}{|c}{$15.0^{
+31.6}_{-8.7}$}&\multicolumn{1}{|c}{$180.1^{+319.3}_{-92.3}$}&\multicolumn{1}{|c}{$0.8^{+1.3}_{-0.6}
$}&\multicolumn{1}{|c}{$1.5^{+2.5}_{-0.8}$}\\
\hline
\emph{DNS}&\multicolumn{1}{|c}{$2.8^{+3.8}_{-2.2}$}&\multicolumn{1}{|c}{$44.7^{+62.0}_{-32.3}$}
&\multicolumn{1}{|c}{$5.0^{+6.8}_{-3.9}$}&\multicolumn{1}{|c}{$0.6^{+0.8}_{-0.5}$}&\multicolumn{1}{
|c}{$2.1^{+25.0}_{-10.9}$}&\multicolumn{1}{|c}{$5.9^{+11.9}_{-3.1}$}&\multicolumn{1}{|c}{$12.8^{
+27.7}_{-6.7}$}&\multicolumn{1}{|c}{$147.0^{+235.0}_{-94.5}$}&\multicolumn{1}{|c}{$0.8^{+1.1}_{-0.6}
$}&\multicolumn{1}{|c}{$1.5^{+2.5}_{-0.8}$}\\
\emph{UPS}&\multicolumn{1}{|c}{$2.4^{+3.2}_{-1.9}$}&\multicolumn{1}{|c}{$48.3^{+62.3}_{-32.8}$}
&\multicolumn{1}{|c}{$5.8^{+7.7}_{-4.1}$}&\multicolumn{1}{|c}{$0.6^{+0.7}_{-0.5}$}&\multicolumn{1}{
|c}{$2.5^{+22.7}_{-11.7}$}&\multicolumn{1}{|c}{$5.3^{+10.3}_{-3.1}$}&\multicolumn{1}{|c}{$17.7^{
+37.4}_{-7.5}$}&\multicolumn{1}{|c}{$139.1^{+215.9}_{-92.5}$}&\multicolumn{1}{|c}{$0.9^{+1.3}_{-0.6}
$}&\multicolumn{1}{|c}{$1.9^{+4.3}_{-0.8}$}\\
\hline
\enddata
\tablecomments{\footnotesize Median, lower quartile ($Q25$) and upper quartile ($Q75$) of the
distributions. For Gaussian distributions a quartile corresponds to $0.6745\sigma$ or
to the median absolute deviation (MAD).}
\end{deluxetable}

\clearpage
\newpage

\subsubsection{Basic GMC properties}\label{sec:props_basic}

\noindent In Fig.~\ref{basic_props_envs}, we plot the distribution of
basic GMC properties within each of our environments. The results of
the KS tests that were used to assess whether the distributions exhibit
significant differences are reported in
Appendix~D. Fig.~\ref{basic_props_envs}a and
Fig.~\ref{basic_props_envs}c show that the distributions of GMC peak
brightness temperature T$_{max}$ and velocity dispersion $\sigma_{v}$
exhibit the most significant environmental variations: both properties
tend to decrease from the center to the spiral arm to the inter-arm
region. In the spiral arms and central region, GMCs span a large range
of T$_{max}$ and $\sigma_{v}$ values, while the inter-arm region lacks
GMCs with high T$_{max}$ and $\sigma_{v}$. There is also a subtle
difference between the peak brightness of inter-arm GMCs, such that
upstream GMCs tend to have lower T$_{max}$ than downstream clouds. The KS tests
generally confirm these findings.\\

\noindent Galactic environment appears to have at most a modest impact
on the size and elongation of GMCs in M51
(Fig.~\ref{basic_props_envs}b and Fig~\ref{basic_props_envs}d). GMCs
in M51 are generally elongated with an \emph{axis ratio} $b/a$ around
$\sim0.6$\footnote{It is worth noting that the typical GMC axis ratio
  ($\sim0.5$) is significantly lower than the beam axis ratio
  ($\sim0.84$), i.e. the clouds have a genuine tendency to be
  elongated rather than round.}. However, clouds in the material arm
and inter-arm regions have a slightly higher $b/a$ and visually appear
more round. By contrast, the \emph{cloud orientation}, $\phi$, shows a
clear connection to galactic structure in
M51. Fig.~\ref{basic_props_envs}e shows that $\langle\phi\rangle$ is
generally close to $0^{\circ}$ in the spiral arm and inter-arm
regions, confirming that the GMC orientation follows the spiral
geometry. Clouds in the central region show a larger deviation from
the spiral arm model, which is expected since the molecular ring is
not a direct extension of the spiral arms. Nevertheless, the width of
the $\phi$ distributions in all environments is fairly large. One
possible explanation is that the CO spiral arms are not perfect
logarithmic spirals.  Although they are well-approximated by a double
logarithmic spiral with $i_{p}=21^{\circ}\pm5^{\circ}$ for
galactocentric radii $1.9<R_{gal}<5.5$ kpc (Patrikeev et al.  2006)
several breaks are evident in a polar representation (see. Fig. 3 in
\citealt{schinnerer13}).  Another source of scatter might be due to
GMCs located in the spurs that are orthogonal to the spiral arms
(especially evident along the northern arm, see
Figure~\ref{gmc_ellipse}).\\

\subsubsection{Derived GMC properties}\label{sec:props_deriv}

\begin{figure}
\begin{center}
\includegraphics[width=0.85\textwidth]{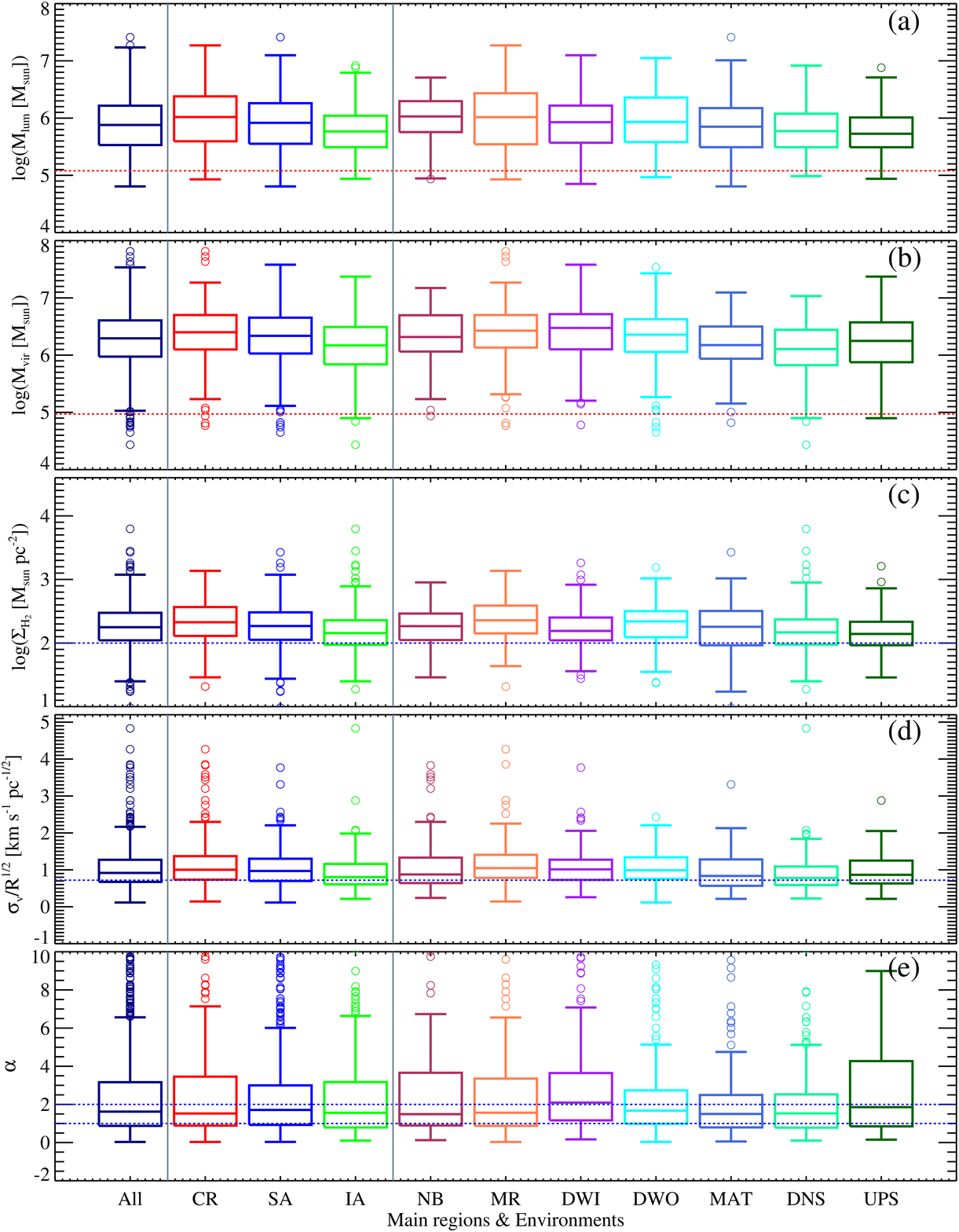}
\end{center}
\caption{\scriptsize Derived GMC properties (from top to bottom): 
(a) mass derived from CO luminosity $M_{lum}$ and 
(b) using the virial theorem $M_{vir}$, 
(c) H$_{2}$ surface density $\Sigma_{H_{2}}$, 
(d) scaling coefficient $\sigma_{v}/R^{1/2}$ and 
(e) virial parameter $\alpha$
shown in a ``box
and whiskers'' representation (see Fig.~\ref{basic_props_envs} for details) for different M51
environments (from the left to the right: All -full sample;
3 main regions -center (CR), spiral arm (SA), inter-arm (IA) and the 7 environments defined in
Fig.~\ref{mask} and Appendix~\ref{app:env_m51}). In general masses, H$_{2}$ mass
surface densities and scaling coefficients are higher in the center and in the spiral arm region
than in the inter-arm environments. The cloud population in every environment is, in general, self-gravitating, 
however a number of objects appears unbound ($\alpha>2$).
Straight horizontal red lines indicate the sensitivity or resolution limits: $1.2\times10^{5}$
$M_{\odot}$ for the luminosity mass and $10^{5}$
$M_{\odot}$ for the virial mass. For surface density and scaling coefficient the blue
lines
show values observed in the Galaxy: $100$ M$_{\odot}$ pc$^{-2}$ (\citealt{heyer09}) and $0.72$ km
s$^{-1}$
pc$^{-1/2}$ (S87), respectively.
Horizontal blue lines in the virial parameter panel indicate the limit for the virialized ($\alpha=1$) and self-gravitating objects ($\alpha=2$) (see text for details).}
\label{derived_props_envs}
\end{figure}

\noindent In Figure~\ref{derived_props_envs}, we plot the
distributions of GMC mass, as inferred from both the CO luminosity and
the virial theorem, H$_{2}$ mass surface density, scaling coefficient
and virial parameter for each of the M51 environments. The differences
in the brightness and velocity dispersion of GMCs that we detected in
Fig.~\ref{basic_props_envs} are likely to produce variations in the
distributions of cloud properties that are estimated using a
combination of these parameters. This is what we observe:
Fig.~\ref{derived_props_envs}a shows the GMC mass inferred from the CO
luminosity $M_{lum}$ declines from the central and density-wave spiral
arm regions to the material arm and inter-arm regions. This is
expected since $M_{lum}\propto L_{CO}\propto\langle T\rangle
R^{2}\sigma_{v}$.\footnote{A parametric description of the CO
  luminosity is legitimate, although CPROPS calculates $L_{CO}$ by
  summing the emission from all pixels that constitute one cloud
  asdescribed in Section~\ref{sec:constr_props}.} In broad terms, the
mass derived from the virial theorem exhibits a similar trend (see
Fig.~\ref{derived_props_envs}b), although by definition it is
dependent only on $\sigma_{v}$ and $R$. We note that the average virial mass
for GMCs in the PAWS catalog is $\sim2\times$ greater than the average
value of $M_{lum}$, derived assuming $X_{CO}=2\times10^{20}$cm$^{-2}$
(K km s$^{-1}$)$^{-1}$.\\

\noindent Fig.~\ref{derived_props_envs}c shows that the average GMC
mass surface density $\langle\Sigma_{H_{2}} \rangle$ is highest in the
central zone (212\,M$_{\odot}$\,pc$^{-2}$), and lower in spiral arm
(185\,M$_{\odot}$\,pc$^{-2}$) and the inter-arm region
(143\,M$_{\odot}$\,pc$^{-2}$). Across the entire PAWS FoV, the median
H$_{2}$ mass surface density is $\Sigma_{H_{2}}\approx180$
M$_{\odot}$\,pc$^{-2}$, almost twice the average value observed for
GMCs in the inner Milky Way ($\sim100$ M$_{\odot}$\,pc$^{-2}$,
\citealt{heyer09}). We note that the PAWS and Galactic values
  are not strictly comparable: the Galactic structures described by
  \cite{heyer09} are typically smaller than the GMCs in M51, and are
  observed at high spatial resolution (i.e. the telescope beam is much
  smaller than the angular size of the observed GMCs). The filling
  factor of CO emission within the PAWS beam, by contrast, is likely
  to be less than unity since the typical peak brightness is only
  T$_{max}\approx4$\,K. The difference between the typical
  mass surface densities of the M51 and Milky Way GMCs is therefore
  probably a lower limit, with high resolution observations likely
  to yield even higher mass surface densities for M51 cloud
  structures.\\

\noindent Fig.~\ref{derived_props_envs}e shows that the median value
of the virial parameter is $\sim1.6$ across all M51 environments, with
values for individual GMCs ranging between 1 and 8. This suggests that
the GMC population in M51 is, on average, self-gravitating, although
$\sim30$\% of the clouds have $\alpha>2$.  The fraction of clouds with
$\alpha>2$ is higher for the upstream subsample than for the
downstream subsample of GMCs.  Fig.~\ref{derived_props_envs}d shows
that the average scaling coefficient $c
=0.90$\,km\,s$^{-1}$\,pc$^{-1/2}$ of the size-linewidth relation is
also roughly constant across the different environments. The median
value $\langle c\rangle\approx0.90$\,km\,s$^{-1}$\,pc$^{-1/2}$ is
always higher than the Galactic value of $0.72$ km s$^{-1}$
pc$^{-1/2}$ (S87), indicating that GMCs in M51 tend to have higher
velocity dispersions than GMCs with comparable size in the Milky
Way.\\

\newpage

\subsubsection{Radial Trends in GMC Properties}\label{sec:radialtrends}

\noindent Our investigation differs from several previous
  surveys of molecular gas across the disk of external galaxies, which
  have tended to analyze the properties of the molecular gas and/or
  GMCs as a function of galactocentric radius
  (e.g. \citealt{hitschfeld09}, \citealt{gratier12}). In contrast to
  these CO surveys, PAWS is restricted to the inner disk of M51
  ($R_{gal}\lesssim5$ kpc), and many environmental parameters that
  could produce a change in the GMC properties show only modest
  variations. For example, the molecular gas fraction
  $M_{H_{2}}/(M_{H_{2}}+M_{H})$ is $\sim80$\% across the FoV
  (\citealt{leroy08}, but see also \citealt{schuster07},
  \citealt{koda09}), while the dust-to-gas ratio and ambient
  interstellar radiation field are roughly constant across our FoV
  (\citealt{cooper12}, \citealt{munoz11}).\\

\noindent Nevertheless, for comparison with previous studies, we
examined whether the GMC properties exhibit trends with galactocentric
radius. We divided the PAWS FoV into 5 radial bins (2 covering the
central region, 3 for the disk) of $\sim2$\,kpc width, each containing
$\sim300$ objects, and compared the statistics of the cloud property
distributions in the different radial bins. As seen in
  Fig.~4 and 5, clouds in the central region
  tend to have higher peak brightness temperatures, velocity
  dispersions and CO luminosities compared to clouds at larger
  radii. Within the bins covering the disk region, however, we see no
  evidence for variations in the average physical properties of the
  GMCs with galactocentric radius. Due to the shape of the PAWS FoV,
each radial disk bin contains an almost equal number of spiral arm and
inter-arm GMCs. We conclude that this uniform mixture of arm and
inter-arm clouds suppresses the environmental variations that we
described above when we examine the cloud properties as a function of
galactocentric radius beyond the central zone. In light of our results
for the GMCs in PAWS, it would be interesting to examine whether the
radial trends reported by previous studies reflect a combination of
variations between the properties of clouds in the arm and inter-arm
regions, as well as variations along the spiral arms.

\begin{deluxetable}{lccccccc}
\tabletypesize{\scriptsize}
\setlength{\tabcolsep}{0.04in}
\renewcommand{\arraystretch}{1.2}
\tablecaption{Median of corrections applied to measurements of GMC properties\label{tab:corr_table}}
\tablehead{
\multicolumn{1}{l|}{\textbf{Envir.}} & \multicolumn{3}{c|}{\textbf{Sensitivity}} &
\multicolumn{2}{c|}{\textbf{Resolution}} &
\multicolumn{2}{c}{\textbf{Global}} \\
\multicolumn{1}{l|}{} & \colhead{$R^{ext}/R^{obs}$} & \colhead{$\sigma_{v}^{ext}/\sigma_{v}^{obs}$}
&
\multicolumn{1}{c|}{$L_{CO}^{ext}/L_{CO}^{obs}$}&$R^{dec}/R^{obs}$ &
\multicolumn{1}{c|}{$\sigma_{v}^{dec}/\sigma_{v}^{obs}$} &
\colhead{$R^{corr}/R^{obs}$} & \colhead{$\sigma_{v}^{corr}/\sigma_{v}^{obs}$}  }
\tablewidth{0pt}
\startdata
\multicolumn{1}{l|}{\emph{All}}&1.6&1.6&\multicolumn{1}{c|}{2.5}&0.7&\multicolumn{1}{c|}{0.8}
&1.3&1.5\\
\hline
\hline
\multicolumn{1}{l|}{\emph{CR}}&1.8&1.8&\multicolumn{1}{c|}{2.8}&0.7&\multicolumn{1}{c|}{0.8}
&1.5&1.7\\
\multicolumn{1}{l|}{\emph{SA}}&1.6&1.7&\multicolumn{1}{c|}{2.6}&0.7&\multicolumn{1}{c|}{0.9}
&1.3&1.6\\
\multicolumn{1}{l|}{\emph{IA}}&1.4&1.4&\multicolumn{1}{c|}{2.1}&0.7&\multicolumn{1}{c|}{0.8}
&1.1&1.3\\
\hline
\hline
\multicolumn{1}{l|}{\emph{NB}}&1.8&1.8&\multicolumn{1}{c|}{2.8}&0.7&\multicolumn{1}{c|}{0.8}
&1.5&1.7\\
\multicolumn{1}{l|}{\emph{MR}}&1.6&1.6&\multicolumn{1}{c|}{2.5}&0.7&\multicolumn{1}{c|}{0.9}
&1.4&1.5\\
\hline
\multicolumn{1}{l|}{\emph{DWI}}&1.6&1.6&\multicolumn{1}{c|}{2.5}&0.7&\multicolumn{1}{c|}{0.8}
&1.2&1.5\\
\multicolumn{1}{l|}{\emph{DWO}}&1.4&1.4&\multicolumn{1}{c|}{2.2}&0.7&\multicolumn{1}{c|}{0.8}
&1.2&1.4\\
\multicolumn{1}{l|}{\emph{MAT}}&1.5&1.4&\multicolumn{1}{c|}{2.2}&0.7&\multicolumn{1}{c|}{0.8}
&1.2&1.3\\
\hline
\multicolumn{1}{l|}{\emph{DNS}}&1.8&1.8&\multicolumn{1}{c|}{2.8}&0.7&\multicolumn{1}{c|}{0.8}
&1.5&1.7\\
\multicolumn{1}{l|}{\emph{UPS}}&1.6&1.6&\multicolumn{1}{c|}{2.5}&0.7&\multicolumn{1}{c|}{0.9}
&1.4&1.5\\
\hline
\enddata
\tablecomments{\footnotesize Median of the sensitivity, resolution and global corrections applied to
the observed values of the GMC properties as a function of environment.}
\end{deluxetable}

\clearpage
\newpage 

\subsection{The effect of CPROPS bias corrections on GMC property measurements}\label{sec:cat_reliability}

\noindent As noted in Section~\ref{sec:props_general}, the flux
contained in the cataloged GMCs is nearly three times greater than
the flux that is directly measured within the objects that are
initially identified by CPROPS. Here, we assess the reliability of the
cloud property measurements in our catalog, paying particular
attention to whether the environmental trends that we described above
could result from the CPROPS extrapolation and deconvolution
corrections.

\subsubsection{Dependence of resolution and sensitivity correction on environment}\label{sec:corr}

\noindent In Table~\ref{tab:corr_table}, we list the median ratio of
the corrected and uncorrected cloud properties within the different
M51 environments. The properties related to the identified objects are
indicated with the superscript \emph{obs}, the superscript \emph{ext}
denotes the extrapolated (but not deconvolved) GMC properties, while
\emph{dec} stands for deconvolution from the beam or the channel width
(without extrapolation). The superscript \emph{corr} denotes cloud
properties corrected for both resolution and sensitivity bias,
and corresponds to the cloud property values listed in the catalog.

\noindent The \emph{resolution correction} (i.e. deconvolution for
beam or channel width) is approximately constant with environment,
decreasing the effective radius and velocity dispersion of GMCs across
the PAWS FoV by 20-30\% on average. \emph{The sensitivity correction}
(i.e. extrapolation), by contrast, varies with environment. Compared
to the extrapolated radius $R^{ext}$, the observed radius $R^{obs}$ is
underestimated by $\sim80\%$ in the central region, $\sim60\%$ in the
spiral arms and $\sim40\%$ in the inter-arm region. The sensitivity
correction yields a similar trend for the velocity dispersion
measurements. The CO luminosity is even more dependent on
extrapolation than the radius and velocity dispersion measurements:
$L_{CO}^{ext}$ is typically a factor of $\sim1.5$ to 2 higher than its
uncorrected value for clouds in the central and spiral arm regions,
and a factor of $\sim1.3$ higher in the inter-arm region.\\

\begin{figure}[h]
\begin{center}
\includegraphics[width=1\textwidth]{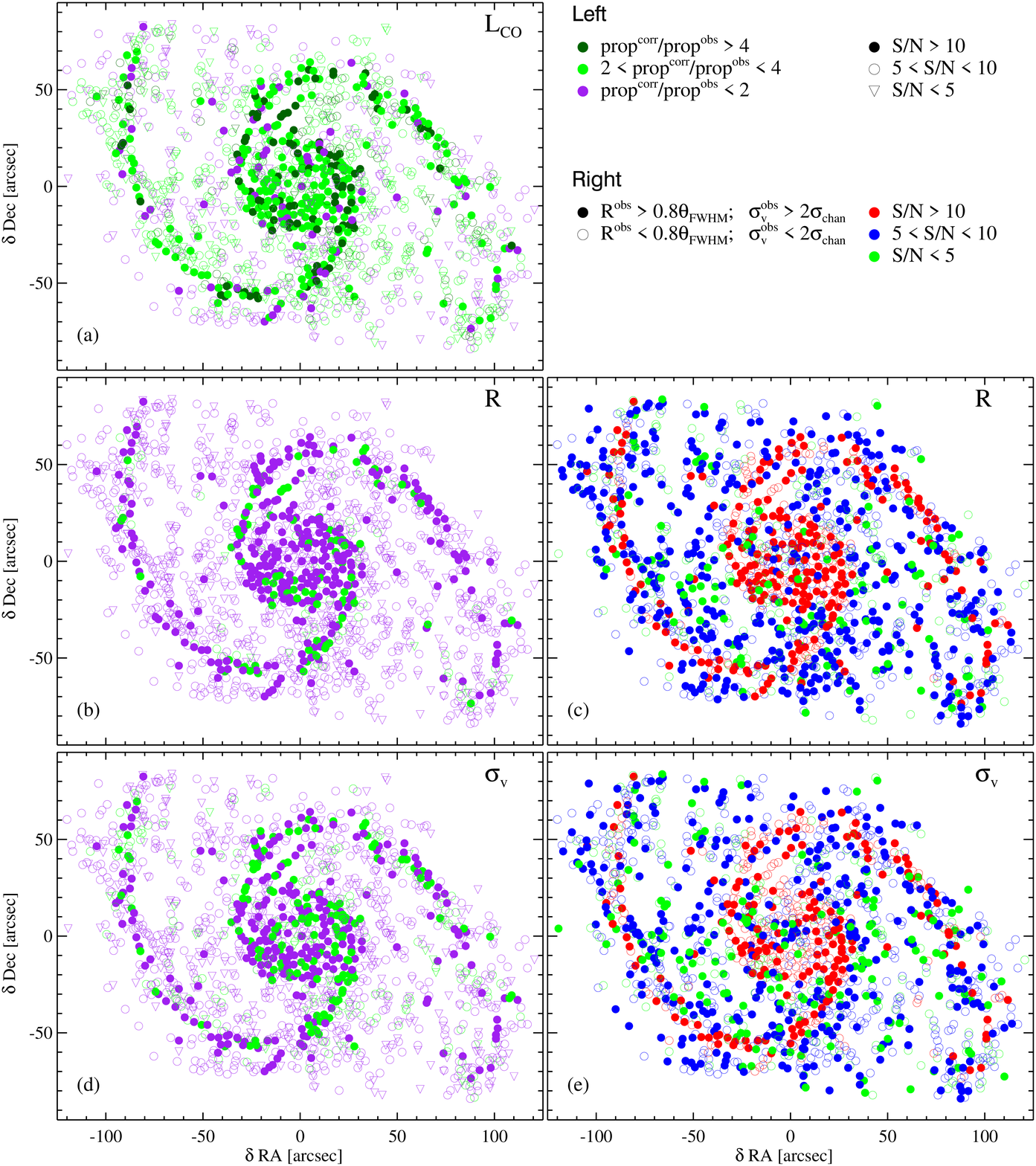}
\end{center}
\caption{\scriptsize \emph{Left:} Spatial illustration of the global correction applied to (a) 
the CO luminosity, (b) effective radius and (d) velocity dispersion measurements of GMCs
as a function of signal-to-noise ($S/N$). The superscript \emph{corr} refers to GMC
properties corrected for both sensitivity and resolution biases, while the superscript \emph{obs} to the properties of identified objects. \emph{Right:} Spatial illustration of the reliability of (c) the
effective radius and (e) velocity dispersion measurement as a function of $S/N$. A cloud
is considered fully resolved by CPROPS if
$R^{obs}>0.8\theta_{FWHM}$ and $\sigma_{v}^{obs}>2\sigma_{v}^{chan}$, where $R^{obs}$ and
$\sigma_{v}^{obs}$ represent the effective radius and the velocity dispersion of the identified
objects, respectively; while $\theta_{FWHM}$ and $\sigma_{v}^{chan}$ are the beam FWHM and the
channel width. Axis coordinates refer to the galactic center 13h\,29m\,52.7087s ;+47$^{\circ}$\,11′\,42.789” (\citealt{hagiwara07}).}
\label{extr_rel_vis}
\end{figure}

\noindent The combined effect of the CPROPS corrections on the cloud radius 
and velocity dispersion is summarized in the final
two columns of Table~\ref{tab:corr_table} and illustrated in
Fig.~\ref{extr_rel_vis}. The correction is higher in the central
region and in the density-wave spiral arm where $R^{corr}$ is around
$30-50\%$ higher than $R^{obs}$. In the inter-arm region, the
corrected radius is only $\sim10\%$ higher than the uncorrected
one. The CPROPS corrections have a larger impact on the velocity
dispersion: in the central and spiral arm regions, the corrected
$\sigma^{corr}_{v}$ is $60-70\%$ higher than the uncorrected
measurement. In the inter-arm region, $\sigma^{corr}_{v}$ is
$\sim40\%$ higher than the uncorrected velocity dispersion.\\

\clearpage
\newpage

\noindent The environmental dependence of the sensitivity correction
becomes easy to understand if we consider the method that CPROPS uses
to perform the extrapolation. An identified object is defined as a set
of (x,y,v) pixels with brightness temperature $T>T_{edge}^{min}$, where
$T_{edge}^{min}$ represents the cloud boundary above a certain
signal-to-noise level. The unextrapolated properties derived for the
identified objects are then a function of the cloud boundary, whereas
the estimate of the properties at $T\equiv0$ K (extrapolation for
perfect sensitivity) is performed using a weighted linear -- or, for
the flux, quadratic -- least-squares fit that takes into account the
brightness temperature profile within the cloud. Thus the difference
between the cloud property values before and after the sensitivity
correction (extrapolation) is determined by the magnitude of the
brightness temperature gradient within the cloud and consequently by
the value of $T_{edge}^{min}$. \\

\begin{figure}
\begin{center}
\includegraphics[width=0.95\textwidth]{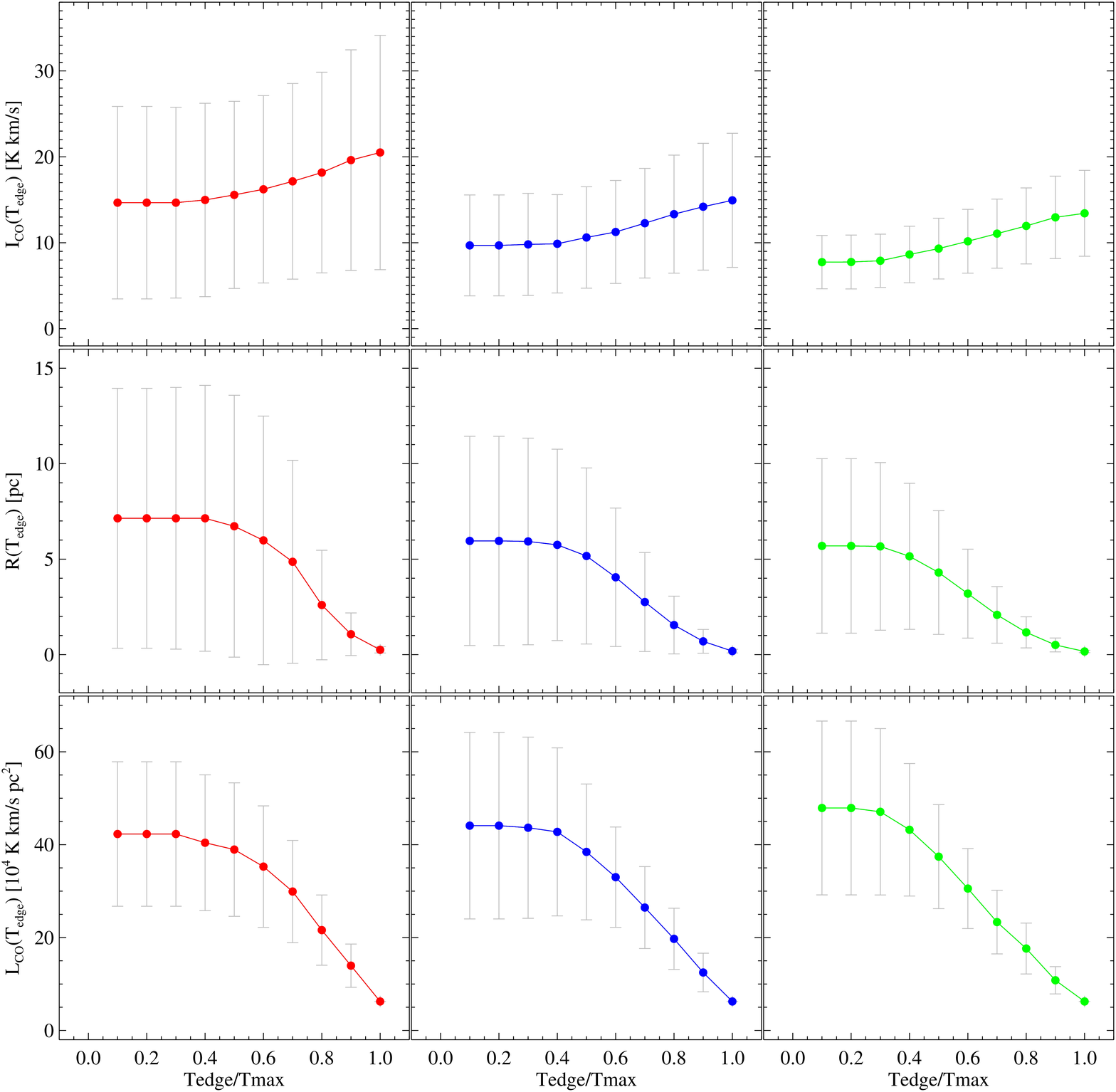}
\end{center}
\caption{\footnotesize Median of cloud profiles relative to surface brightness $I_{CO}$ (top), effective
radius $R$ (middle), and CO luminosity $L_{CO}$ (bottom) for the three main region (from left to
right: central,
spiral arm and inter-arm region). Error bars indicate the median absolute deviation of the
distributions.}
\label{extrapol_check}
\end{figure}

\noindent To test whether the cloud brightness temperature gradient
varies with environment, we analyzed the full cloud sample in the
three main regions (i.e. M51's center, spiral arms, and inter-arm). We
fixed 10 $T_{edge}$ levels corresponding to $10\%-20\%-...100\%$ of
the peak temperature of a cloud and we calculated the radius, the CO
luminosity and CO surface brightness of the object at each level. The
radius is estimated as:
\begin{equation}
 R=\sqrt{\frac{A}{\pi}},
\end{equation}
\noindent where $A$ is the area of the cloud (in pixels) at a given
$T_{edge}$. Figure~\ref{extrapol_check} shows the result as a median
of the property distribution at a given $T_{edge}/T_{max}$ value. The
cloud radius profiles show similar slopes in all three
environments. The CO luminosity profiles, however, appear steeper in
the central region. The surface brightness profiles $I_{CO}$
  also differ between the three main regions. The central region
  profile is the steepest, and the inter-arm profile is the most
  shallow. These differences indicate that the brightness temperature
  gradient inside the clouds is varying between the different regions,
  which explains why the magnitude of the sensitivity correction
  depends on environment.\\

\noindent The difference between the extrapolated and uncorrected
properties is also proportional to the value of $T_{edge}^{min}$. We
can assess the effect of $T_{edge}^{min}$ by examining the brightness
temperature distributions of the watershed (i.e. undecomposed emission
within the CPROPS working area) in the different environments. In the
central and spiral arm regions, where the difference between
extrapolated and unextrapolated properties is higher, large areas have
brightness temperatures $>4$\,K. In the inter-arm region, where
the difference between corrected and uncorrected properties is lower,
the watershed mostly has brightness temperatures $<2$\,K.

\newpage

\subsubsection{Reliability of extrapolated property measurements}\label{sec:props_reliab}

\noindent CPROPS obtains measurements of GMC properties only if
certain requirements on the sensitivity and resolution are satisfied
(RL06). Here we take a conservative approach, 
examining the properties of the identified objects in
order to determine whether the final corrected measurements can be
considered reliable.\\

\noindent As discussed by RL06, the sensitivity correction of CPROPS
will yield the \emph{effective radius} of a cloud with an error below
$10\%$ if the signal-to-noise $S/N$ is greater than 10. The algorithm
performs well even for barely resolved objects, i.e. for clouds with
$R^{obs}>0.8\theta_{FWHM}$, where $\theta_{FWHM}$ is the full width at
half maximum size of the beam. For clouds with $5<S/N<10$, the
measured radius may be underestimated by up to $20\%$. The accuracy of
the corrected radius measurements deteriorates for faint clouds
($S/N<5$), and when an object is unresolved.\\

\noindent Fig.~\ref{extr_rel_vis} shows the spatial distribution of
M51 clouds as a function of the signal-to-noise and the observed
radius relative to the beam size. The identified clouds with $S/N>10$
constitute $\sim25\%$ of the catalog. These clouds are typically
located in the ridge line of the spiral arms and in the central
region. More than $50\%$ of the objects have a $S/N$ between 5 and 10
and the remaining $25\%$ of clouds have $S/N<5$. These faint clouds
are distributed across the PAWS field.  The objects with a peak
signal-to-noise above 5 that satisfy the resolution requirement of
CPROPS ($R^{obs}>0.8\theta_{FWHM}$) are $40\%$ of the total, while the
objects with an observed radius below this limit that show the same
range of $S/N$ are more than $\sim35\%$ of the catalog and could
suffer a 10\% underestimation of their actual radii. Thus $65\%$ of
the clouds have a radius measurement that can be considered
reliable. According to Fig.~\ref{extr_rel_vis}, the bright clouds with
the most reliable radius measurements tend to be located in
environments where extrapolation correction for the cloud radius is largest.\\

\noindent The CPROPS performance requirements for the cloud
\emph{velocity dispersion} determination are less demanding
(RL06). The extrapolation works well -- independently of the cloud
$S/N$ -- if the line width of the identified object is at least twice
the channel width. Fig.~\ref{extr_rel_vis} shows a map of the clouds
as a function of the velocity dispersion with respect to the channel
RMS. The identified clouds with $\sigma_{v}^{obs}/\sigma_{chan}>2$ are
$\sim40\%$ of the total. Of the remaining objects, $\sim15\%$ have a
signal-to-noise peak greater than 10. In this case, according to RL06,
the overestimation of the actual velocity dispersion of the cloud is
around $20\%$. The spatial distributions of these two classes of
clouds are quite uniform and do not depend on environment. In the PAWS
catalog, we therefore have a large number of clouds for which the
cloud velocity dispersion may be overestimated. This is especially in
the inter-arm, where the signal-to-noise is typically lower. This
reinforces our conclusion that GMCs in the spiral arm and the central
regions tend to have a higher velocity dispersions than inter-arm
GMCs, since the former have higher $S/N$ ratios and hence more
accurate velocity dispersion measurements. Nevertheless 
this does not influence the conclusions on the unboundness of the clouds, 
since the objects with an intrinsically low velocity dispersion represent 
only the 5\% of the 394 clouds with $\alpha>2$.\\

\noindent The difference between the GMC flux after
  extrapolation and the flux measured directly within the identified
  objects is high (Table~\ref{gmcs_flux}). Indeed the average
  corrected CO luminosity of the GMC is 2.5$\times$ greater than the
  unextrapolated value (Table~\ref{tab:corr_table}). Although this is
  consistent with the results obtained on IC10 in RL06, it represents
  a significant addition to the flux of our identified GMCs and
  therefore merits further examination.\\

\noindent While the original CPROPS paper (RL06) provides guidelines
for checking whether extrapolated measurements of the cloud radius and
velocity dispersion can be considered reliable, this is not the case
for extrapolated measurements of the CO luminosity.  Nevertheless we
can draw some conclusions based on a comparison between the
extrapolated and the observed flux within GMCs (see
Section~\ref{sec:props_general}) and the extended component discussed
in \cite{pety13}. Although GMCs are often considered to account for
nearly all the CO emission in normal galactic disks ($\sim85\%$,
\citealt{sanders85}), roughly half of the CO flux in M51 arises from a
diffuse thick disk of molecular gas (see \citealt{pety13} for a
detailed discussion of its properties).  The fact that GMCs (after
extrapolation) contribute 54\% of the total CO flux in the PAWS FoV
would seem compatible with the existence of a diffuse, extended
component that is responsible for a comparable fraction of the total
CO luminosity. If, instead, the CO luminosities of GMCs were closer to
their unextrapolated values, $\sim30$\% of the CO emission within the
PAW FoV must be attributed to an ill-defined ``watershed''. Much of
this undecomposed ``watershed'' emission reaches temperatures above 4
K, characteristic of compact structures in the Galaxy
(\citealt{sawada12}). While this flux could be associated with
entities smaller than the beam, it is also possible that the watershed
is actually part of the GMCs.  Presumably, this part of the emission
could not be properly attributed to clouds by the identification
algorithm, given the low contrast between cloud and intra-cloud
emission. We might therefore assume the initially identified objects
as ``bright cores'' of more extended structures that we recover only
through the extrapolation correction. 

\noindent Overall, our examination of the effects of the sensitivity
and resolution corrections on the measured cloud properties highlights
the limitations of the CPROPS method in decomposing physically reliable
objects in highly crowded and low contrast environments. Although other methods, like the
``patchwork'' separation performed by CLUMPFIND, are able to attribute
all the measured flux to discrete objects, the resulting separation is
ambiguous when GMCs do not have well-defined boundaries, as in the
case of the cloud population in M51.\\

\section{Scaling relations}\label{sec:larson}

\noindent Having reviewed the physical properties of GMCs in different
regions of M51, we now examine whether the clouds obey the scaling
relations commonly referred to as ``Larson's laws''
(\citealt{larson81}).  The first Larson's law, or \emph{size-velocity
  dispersion relation}, states that $\sigma_{v}\propto R^{0.5}$ (S87);
it is considered to be a manifestation of turbulence inside the cloud
or of virial equilibrium (see \citealt{kritsuk11}). The second
Larson's law asserts that GMCs are roughly self-gravitating. The third
law describes an inverse correlation between the size of a cloud and
its density, implying that all GMCs have approximately constant
surface density.\\

\noindent To estimate the degree of correlation between GMC properties
we calculate the Spearman's rank correlation coefficient
(\citealt{spearman04}). This coefficient, $r_{s}$, assesses how well
the relationship between two variables can be described by a monotonic
function. If there are no repeated data values, +1 indicates a perfect
monotonically increasing function. We consider the properties to be
strongly correlated if $r_{s}\geq0.8$, and moderately correlated if
$0.5<r_{s}<0.8$. For the scaling relations shown in
Fig.~\ref{mlco_envs} and Fig.~\ref{lars2_envs}, the corresponding
$r_{s}$ values are indicated in the bottom corner of each panel.\\

\noindent To fit any correlations that we detect, we use the IDL
implementation distributed by Erik Rosolowsky of the ``BCES''
(bivariate, correlated errors with intrinsic scatter) method described
by \citealt{akritas96}. The BCES bisector estimator takes into account
the uncertainty associated with each cloud property measurement. In
our estimate for the best-fitting relation, we use only the ``highly
reliable sample'' of clouds of the catalog, i.e. GMCs with $S/N>6.5$
(see Section~\ref{sec:cat}), and we assume that the measurement
uncertainties are uncorrelated.\\

\subsection{First Larson's law: size-velocity dispersion relation}

\begin{figure}[h!]
\begin{center}
\includegraphics[width=0.85\textwidth]{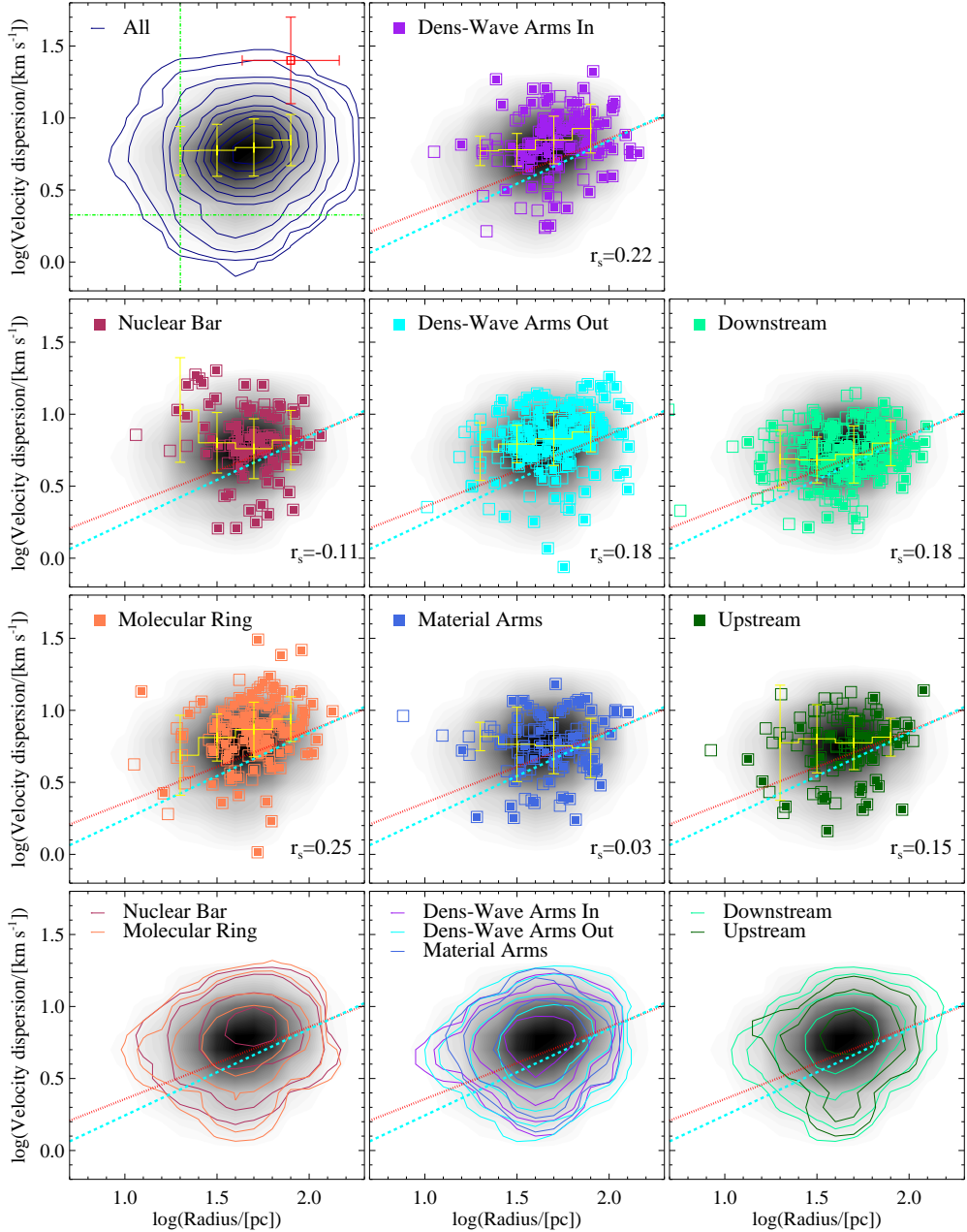}
\end{center}
\caption{\scriptsize Size-velocity dispersion relation (first Larson's law) for GMCs in M51 within
the various environments. Every
column refers to a different region
(from left
to right: spiral arm, inter-arm and central region). Data points corresponding to clouds with
$S/N>6.5$ are highlighted with filled symbols. The shaded area
shows the density distribution of the full catalog. Red dotted lines indicate the Galactic fit
($\sigma_{v}(km/s)=0.72R(pc)^{0.5}$, S87) and cyan dashed lines the extragalactic fit
($\sigma_{v}(km/s)=0.44R(pc)^{0.6}$, B08). In the
bottom right corner of each panel the
Spearman's correlation rank is given. The histogram in yellow illustrates the median and the MAD
of log($\sigma_{v}$/[km/s]) in bins of 0.2 dex for log(R/[pc])$\in(1.0-2.0)$. Then  bottom row
shows a contour representation of all GMCs
with $S/N>6.5$ within the various environments.
In the top left panel the
contours show the distribution of the full sample of ``highly reliable clouds'' (with $S/N > 6.5$).
Green horizontal and vertical
lines indicate the nominal resolution limit: 20 pc (CLEAN beam radius) and 2.12 km/s (channel
velocity dispersion). The average error bars are reported in red in the top right corner of the top
right panel.}
\label{lars1_envs}
\end{figure}

\noindent The relationship between the size and velocity dispersion of
GMCs in the PAWS catalog is shown in Fig.~\ref{lars1_envs}.  For all
environments, there is a high degree of scatter and the $r_{s}$ values
indicate that the size and linewidth of the M51 GMCs are, at best,
weakly correlated. If we restrict our comparison to GMCs with high
signal-to-noise ($S/N>6.5$), then a linear trend between $R$ and
$\sigma_{v}$ becomes apparent for some environments, although the
correlation is still very weak ($r_{s} \leq 0.25$). In the bottom row
of Fig.~\ref{lars1_envs}, we use contours to indicate the region of
the size-velocity dispersion space occupied by GMCs in different M51
environments. Compared to spiral arm environments, the inter-arm
region lacks clouds with high $\sigma_{v}$, while GMCs in the central
region seem shifted slightly towards higher values of $R$ and
$\sigma_{v}$. It is worth to note also that the majority of the data
points lies above the Galactic (S87) and extragalactic (B08) fits, in
particular in the case of the center and spiral arm samples. This
shows that GMCs in M51 have a higher velocity dispersion compared with
similar size clouds in the Milky Way or Local Group galaxies.\\

\subsection{Second Larson's law: virial mass-luminosity relation}\label{sec:lars2}

\begin{figure}[h!]
\begin{center}
\includegraphics[width=0.85\textwidth]{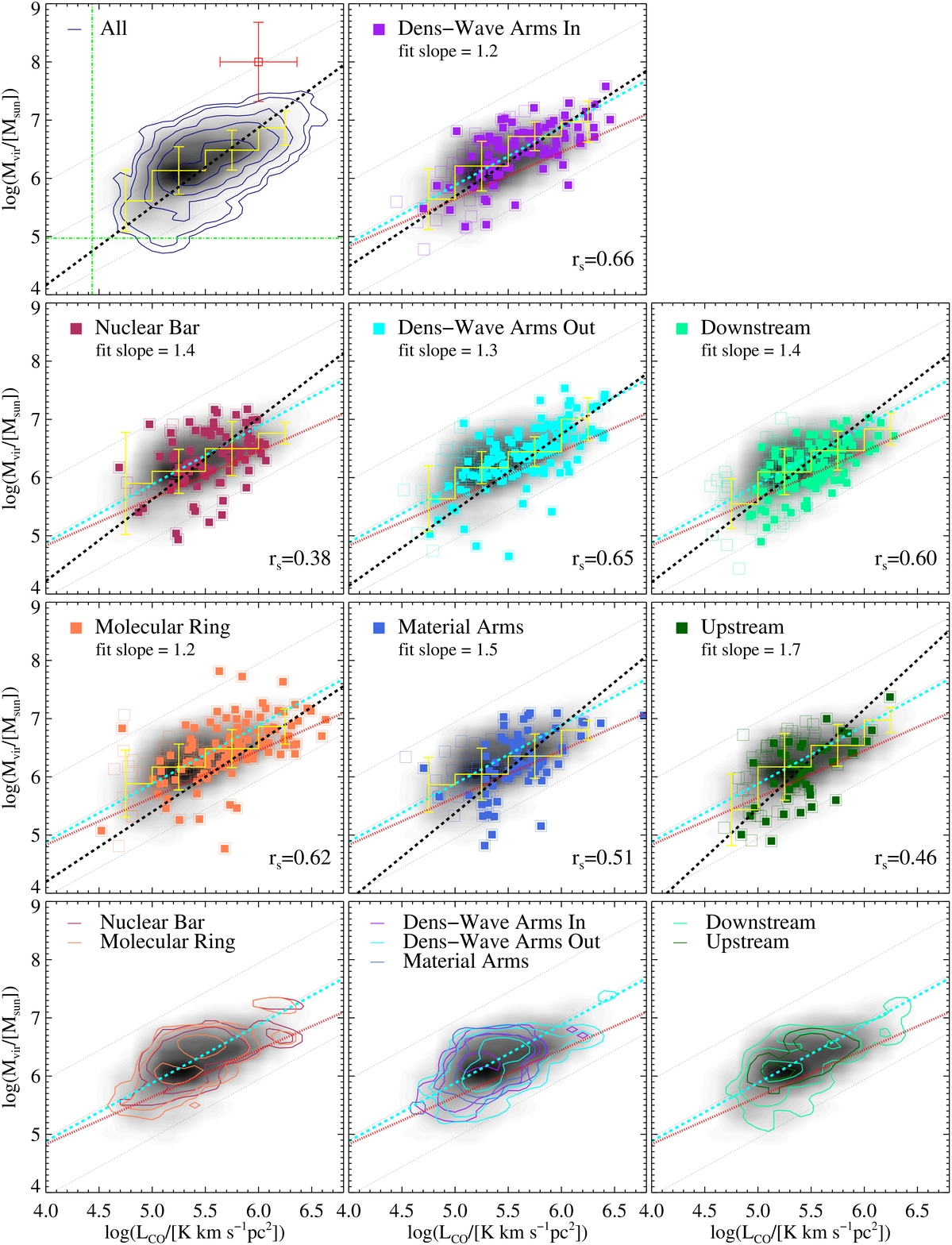}
\end{center}
\caption{\scriptsize Virial mass-luminosity relation (second Larson's
  law) for GMCs in M51 for the various environments. Every column
  refers to a different region (from left to right: spiral arm,
  inter-arm and central region). Data points corresponding to clouds
  with $S/N>6.5$ are highlighted with filled symbols. The shaded area
  shows the density distribution of the full catalog. Red dotted lines
  indicate the Galactic fit
  ($M_{vir}$(M$_{\odot})=39L_{CO}^{0.81}$(K\,km\,s$^{-1}$\,pc$^{2}$),
  S87), cyan dashed lines the extragalactic fit
  ($M_{vir}$(M$_{\odot})=7.6L_{CO}^{1.00}$(K\,km\,s$^{-1}$\,pc$^{2}$),
  B08) and black dotted lines the fits for the different
  environments. The slopes of our fits are indicated in the figure
  panels. Dashed grey lines indicate different $X_{CO}$ values, from
  bottom to top $X_{CO}=$ $4\times10^{19}$, $4\times10^{20}$, and
  $4\times10^{21}$ cm$^{-2}$\,K$^{-1}$\,km$^{-1}$\,s.  Spearman's
  correlation rank is indicated in the bottom right of each panel. The
  histogram in yellow illustrates the median and the MAD of
  log($M_{vir}/$[M$_{\odot}$]) in bins of 0.5 dex for log($L_{CO}/$[K
    km s$^{-1}$ pc$^{-2}$])$\in(4.5-6.5)$. The bottom row shows a
  contour representation of the GMCs with $S/N>6.5$ within the various
  environments. In the top left panel the contours show the
  distribution of the full sample of ``high reliable clouds'' (with
  $S/N > 6.5$). Green lines indicate resolution limit:
  $2.7\times10^{4}$ K km s$^{-1}$ pc$^{-2}$ for CO luminosity and
  $9.3\times10^{4}$ $M_{\odot}$ for the virial mass. The average error
  bars are reported in red in the top right corner of the top right
  panel.}
\label{mlco_envs}
\end{figure}

\noindent In Fig.~\ref{mlco_envs}, we plot the virial mass of the M51
GMCs as a function of their CO luminosity. We note that both virial
mass and CO luminosity depend on a combination of $R$ and
$\sigma_{v}$, i.e. $M_{vir}\propto \sigma_{v}^{2} R$ and
$L_{CO}\propto\langle T \rangle R^{2} \sigma_{v}$, so a significant
degree of correlation between these quantities is
expected. Fig.~\ref{mlco_envs} shows that GMCs in M51 are scattered
around the extragalactic relation obtained by B08
($M_{vir}$(M$_{\odot}$)=$7.6L_{CO}^{1.00}$(K\,km\,s$^{-1}$\,pc$^{2}$)), although the
peak-to-peak variations in $M_{vir}/L_{CO}$ span up to $\sim2$ orders
of magnitude. The best-fitting mass-luminosity relations that we
obtain for the different M51 GMC populations are steeper than the B08
relation by $\sim0.2$ to 0.5\,dex. We note that the slope of the
mass-luminosity relation varies with environment, increasing from
$\sim1.3$ in the spiral arm and central regions to $\sim1.5$ in the
inter-arm region. This increment is likely driven by differences in
luminosity and velocity dispersion observed within the environments.
Nevertheless, the clouds appear roughly distributed around a
$X_{CO}=4\times10^{20}$ cm$^{-2}$ (K km s$^{-1}$)$^{-1}$, consistent
with the average value that has been observed for other nearby
galaxies (e.g. \citealt{blitz07}, B08).\\
 
\begin{figure}[h!]
\begin{center}
\includegraphics[width=0.85\textwidth]{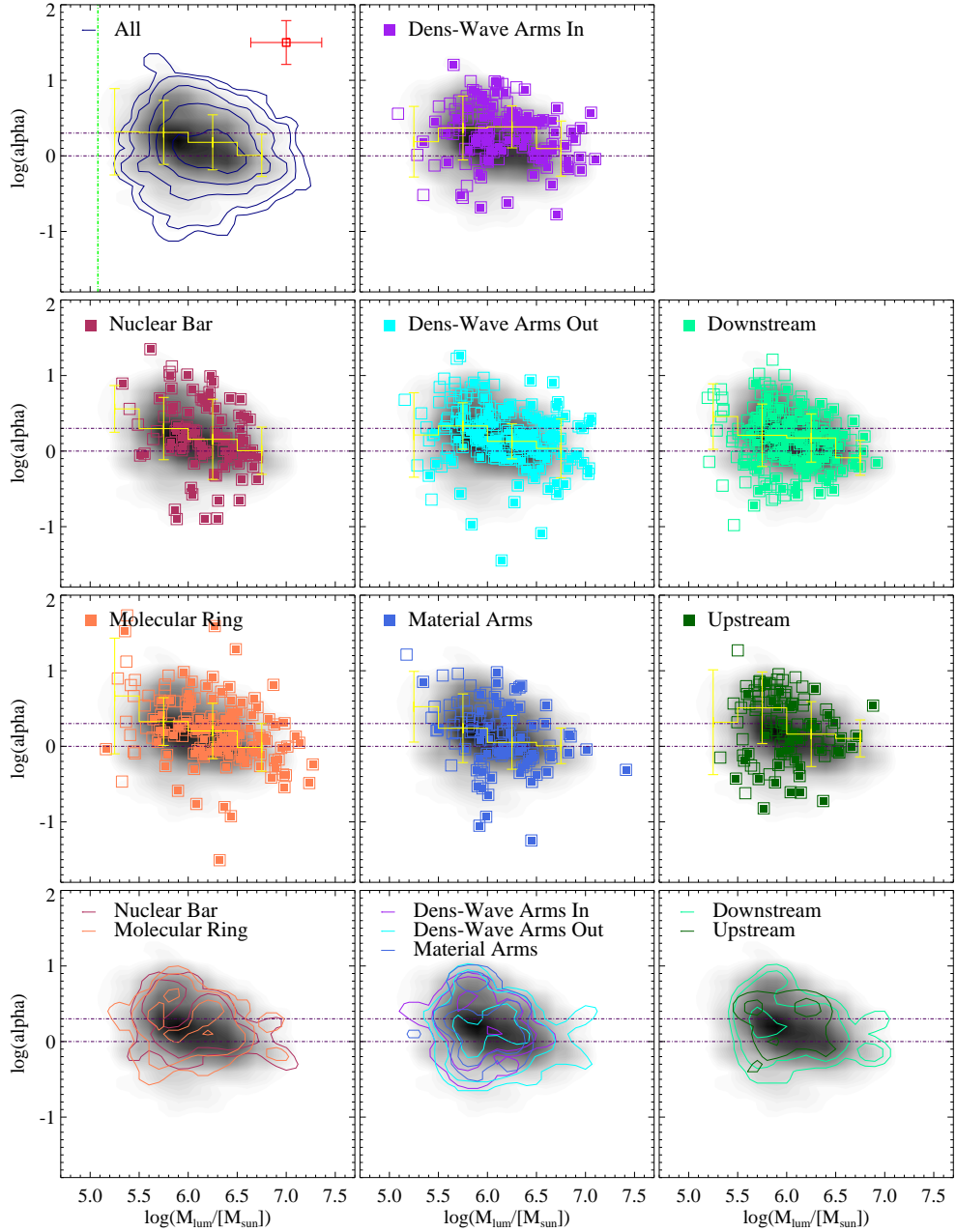}
\end{center}
\caption{\scriptsize Mass-virial parameter relation for GMCs in
the various M51 environments. Every column refers to a different region (from left to right:
spiral arm, inter-arm and central region). Data points corresponding
to clouds with $S/N>6.5$ are highlighted with filled symbols. The
shaded area shows the density distribution of the full catalog. The histogram in
yellow illustrates the median and the MAD of log($\alpha$) in bins of 0.5 dex for
log($M_{lum}$/[M$_{\odot}]\in(5.0-7.0)$. The
bottom row shows a contour representation of the GMCs with $S/N>6.5$ within the various
environments. In the top left panel the contours show the
distribution of the full sample of ``highly reliable clouds'' (with $S/N >
6.5$). Purple horizontal dashed lines indicate the limit between self-gravitating and pressure
confined clouds ($\alpha=1$) and unbound clouds ($\alpha=2$). Green line indicates our nominal
sensitivity limit:
$2.7\times10^{4}$ K km s$^{-1}$ pc$^{-2}$ for CO luminosity. The average
error bars are reported in red in the top right corner.}
\label{alpha_envs}
\end{figure}

 The analysis of the distribution of the virial parameter of
  Section~\ref{sec:props_deriv} has shown that clouds in M51 are in
  general self-gravitating. Here we check if $\alpha$ is correlated
  with the cloud mass. In Fig.~\ref{alpha_envs}, we plot
  $\alpha$ as a function of $M_{lum}$ finding that although GMCs with
  $\alpha>2$ are present across our entire observed mass range, the
  average value of $\alpha$ tends to decrease for high mass
  clouds. This plot should be interpreted with care, since the axes
  are correlated ($M_{lum}$ appears in the denominator of the virial
  parameter definition). Nevertheless, since there are low- to
  intermediate-mass clouds with high signal-to-noise and large virial
  parameters ($\alpha>2$), Fig.~\ref{alpha_envs} suggests that overall
  the high mass clouds in M51 tend to be more strongly bound than low
  mass clouds.

\subsection{Third Larson's law: Luminosity-size relation} 

\begin{figure}[h!]
\begin{center}
\includegraphics[width=0.85\textwidth]{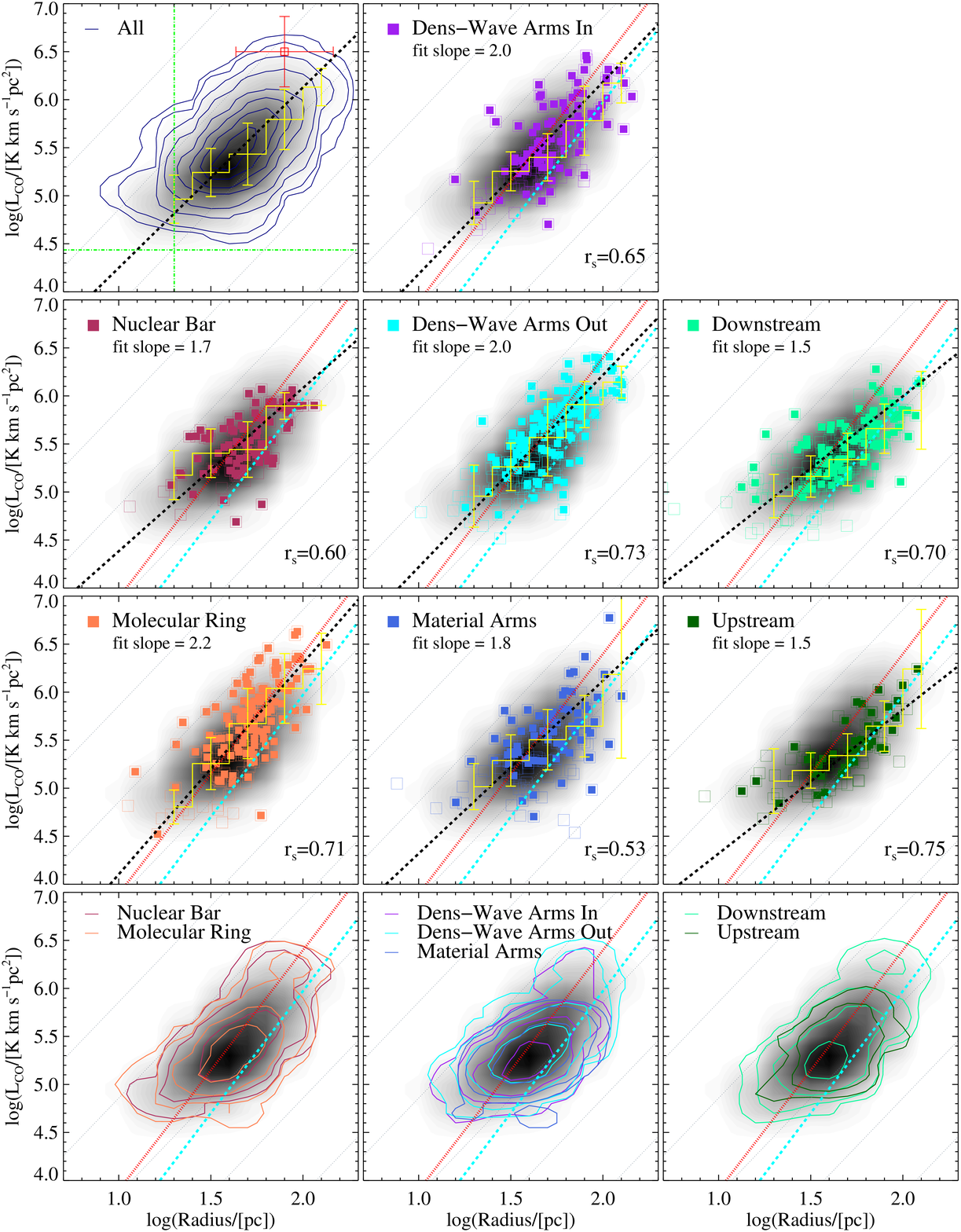}
\end{center}
\caption{\scriptsize Luminosity-size relation (third Larson's law) for
GMCs in the various M51 environments. Every column refers to a different region (from left to
right: spiral arm, inter-arm and central region). Data points
corresponding to clouds with $S/N>6.5$ are highlighted with filled
symbols. The shaded area shows the density distribution of the full
catalog. Red dotted lines indicate the Galactic fit ($L_{CO}$(K\,km\,s$^{-1}$\,pc$^{2}$)=$25R^{5}$(pc), S87),
cyan dashed
lines the extragalactic fit ($L_{CO}$(K\,km\,s$^{-1}$\,pc$^{2}$)=$7.8R^{2.54}$(pc),
B08) and black dotted lines the fits for the
different environments, which slopes are directly indicated in the figure panels. Dashed grey lines
indicate different H$_{2}$ surface density values, from bottom to
top $\Sigma_{H_{2}}=$ 1, 10, 100, 10$^{3}$, and 10$^{4}$
M$_{\odot}$\,pc$^{-2}$. At the bottom of the panels the Spearman's
correlation rank is indicated. The histogram in yellow illustrates the median and the MAD
of log($L_{CO}/$[K\,km\,s$^{-1}$\,pc$^{2}$]) in bins of 0.2 dex for log(R/[pc])$\in(1.2-2.0)$. The
bottom row shows a contour
representation of the various environments.  In the top left panel
the contours show the distribution of the full sample of reliable
clouds (with $S/N > 6.5$). Green horizontal and vertical lines
indicate the nominal sensitivity and resolution limits:
$2.7\times10^{4}$\,K\,km\,s$^{-1}$\,pc$^{-2}$ for CO luminosity and 20 pc for the
radius, respectively. The average error bars are reported in red in
the top right corner of the top right panel.}
\label{lars2_envs}
\end{figure}

\noindent Fig.~\ref{lars2_envs} shows that the size and CO luminosity
of M51 GMCs are strongly correlated, with $0.5<r_{s}<0.8$. This is not
surprising since $L_{CO}\propto\langle T\rangle R^{2}\sigma_{v}$. The
bottom row of Fig.~\ref{lars2_envs} shows that the relationship
between $R$ and $L_{CO}$ is steeper in the central and spiral arm
regions than in the inter-arm region. This is confirmed by the results
of a linear regression fit: the slope of the best-fitting power law
flattens from 2.4 for GMCs in the molecular ring, to $\sim2$ for
clouds in the density wave spiral arms, to $<1.5$ for the inter-arm
environments. The origin of such effect is likely to be the different
CO emission properties within the different M51 environments (such as
the geometry, CO filling factor and/or density distribution, see also
Hughes et al. 2013b) but further investigation into its physical
significance is required. Nevertheless, the change in slope of the fit
appears to be real, given the fact that all environments span a
similar range of GMC radii but contain clouds with very different
luminosity.  Assuming a uniform $X_{CO}$ factor throughout the PAWS
field, the linear regression illustrates why the median H$_{2}$ mass
surface density varies with environment: large GMCs located in
molecular ring and density-wave spiral arms contain more high
brightness CO emission than clouds of an equivalent size in the
inter-arm region.\\
 
\subsection{CPROPS bias corrections and scaling relations}\label{sec:corr_scalrel}

\begin{figure}[h!]
\begin{center}
\includegraphics[width=1\textwidth]{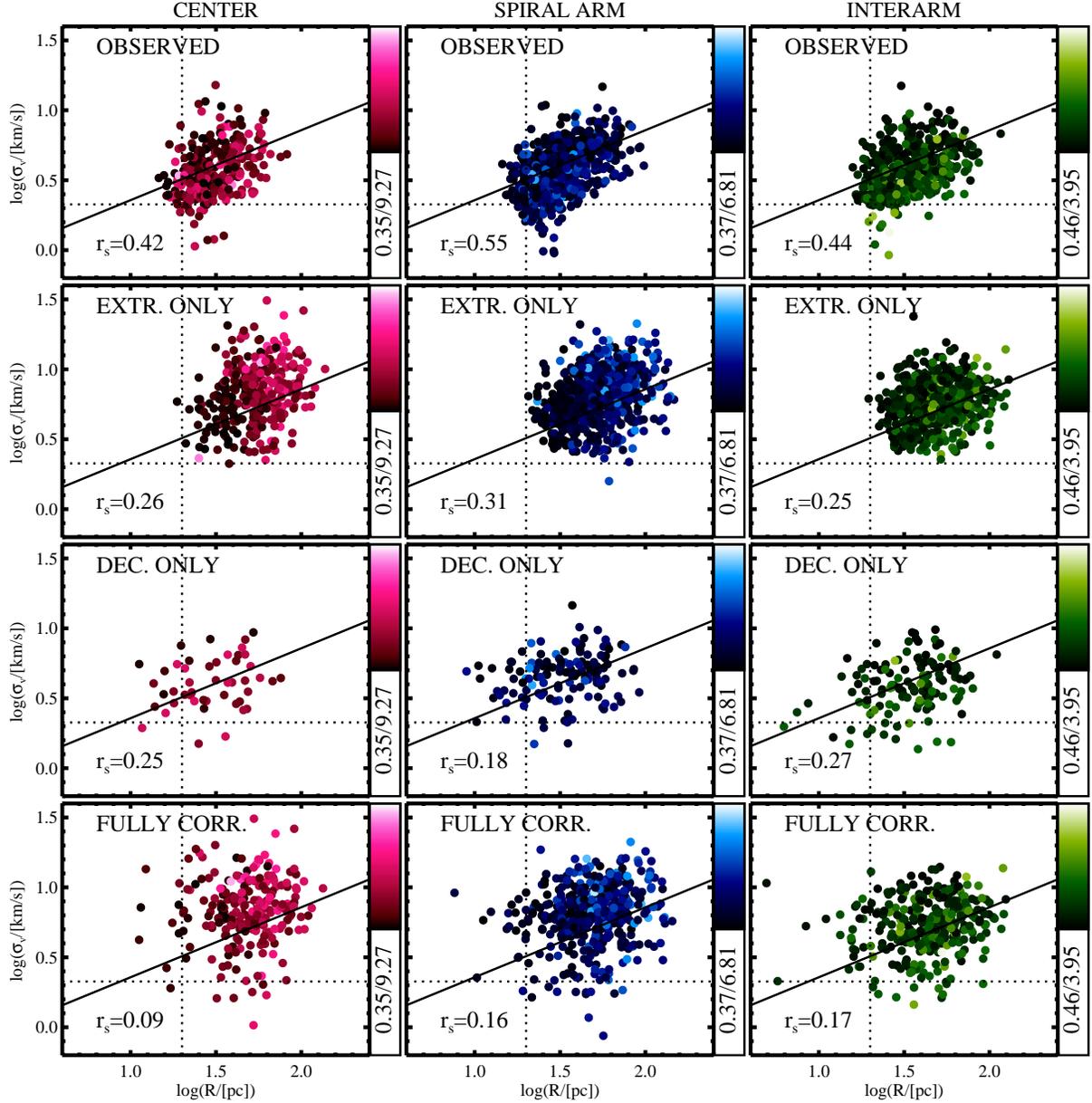}
\end{center}
\qquad
\caption{\footnotesize Comparison of the Larson's laws for observed
  (top row), extrapolated only (second row), deconvolved only (third
  row) and fully corrected (extrapolated and deconvolved, bottom row)
  properties of the full GMC catalog. The three columns present
  the central (left), arm (middle) and inter-arm (right) GMC
  populations. Spearman's rank correlation coefficients ($r_{s}$) of
  the full catalog are indicated at the bottom left of each
  panel. Straight dotted lines indicate resolution limits: 2.12
  km\,s$^{-1}$ channel ``velocity dispersion'' and 20\,pc beam
  ``radius''. The full black line represents the Galactic fit by S87.}
\label{fig:lars_nx_ex}
\end{figure}

\noindent Although Larson's Laws have regularly been used as
  yardstick for comparing GMC populations, a number of previous
  studies have demonstrated that the method used to identify clouds
  and measure their properties has a large impact on the appearance of
  the Larson-type scaling relations \citep[e.g.][]{wong11}. In
  Section~\ref{sec:cat_reliability}, we argued that the
  \textsc{CPROPS} bias corrections are important for recovering a
  reliable estimate for the properties of GMCs within the PAWS
  field. In Fig.~\ref{fig:lars_nx_ex}, we plot the size-linewidth
  relation for the PAWS clouds in the three main environments, using
  measurements with and without the resolution and sensitivity
  corrections applied. It is clear that the uncorrected properties
  (top row) exhibit the most robust correlations. Taken individually,
  the corrections for sensitivity (i.e. extrapolation, second row) and
  resolution (i.e. deconvolution, third row) appear to introduce a
  comparable level of scatter into the size-linewidth relation,
  decreasing the Spearman rank correlation coefficient by a factor of
  $\sim2$ with respect to the relation exhibited by the uncorrected
  properties. It is important to recall, however, that the observed
  objects are not uniformly defined across the PAWS field: the CO
  brightness at the cloud boundary tends to be higher for objects in
  the spiral arm region ($\langle T_{edge} \rangle \in [0.4,6.8]$\,K,
  middle column) than for the inter-arm ($\langle T_{edge} \rangle \in
  [0.5,4.0]$\,K, right column). The top row of
  Fig.~\ref{fig:lars_nx_ex} shows that these differences in the
  definition of the cloud lead to some segregation of the data points
  within the size-linewidth plot, i.e. objects with low brightness
  boundaries (darker points) tend to have larger linewidths relative
  to their size than objects with boundaries at a higher brightness
  threshold (lighter points). In summary, our analysis re-inforces
  conclusions from previous observational studies that the methods
  used to identify GMCs and measure their properties exerts a
  significant influence over the existence and slope of a
  size-linewidth relation, and that decomposition methods that use a
  fixed brightness threshold to define cloud boundaries seem to yield
  stronger size-linewidth relations. This should be kept in mind by
  studies that collate literature values to, e.g., compare the
  physical properties of extragalactic GMC populations, or validate
  physical models for the origin of the first Larson Law.

\newpage  
  
\newpage

\newpage

\newpage

\newpage

\clearpage
\newpage

\section{GMC Mass spectra}\label{sec:masspect}

\subsection{Construction and general properties}

\noindent The GMC luminosity distribution depicts how the CO flux is
organized into clouds of different luminosity within a galaxy
(e.g. \citealt{rosolowsky05}).  In this section, we frame our
discussion in terms of the GMC mass spectrum, which equivalently
describes how molecular gas is organized into cloud structures of
different mass, assuming that CO emission is a reliable tracer of
H$_{2}$. We convert the CO luminosity to H$_{2}$ mass assuming a
constant Galactic conversion factor $X_{CO}=2\times10^{20}$ cm$^{-2}$
(K km s$^{-1}$)$^{-1}$, and including the mass contribution of helium,
thus $M_{lum}=4.4L_{CO}$ (eq.~\ref{mlum}). \\

\noindent The GMC mass spectrum is usually expressed in differential
form and modeled as a power law:

\begin{equation}\label{diffmass}
f(M)=\frac{dN}{dM} \propto M^{\gamma}
\end{equation}

\noindent The integral of this expression yields the cumulative mass
  distribution, i.e. the number of clouds $N$ with masses $M$ greater
than a reference mass $M_{0}$ as a function of that reference mass:

\begin{equation}\label{cummass}
    N(M'>M)=\left[\left(\frac{M}{M_{0}}\right)^{\gamma+1}\right].
\end{equation}

\noindent The index $\gamma$ describes how the mass is distributed: for values
$\gamma>-2$, the gas is preferentially contained in
massive structures, while for values $\gamma<-2$, small clouds
dominate the molecular mass budget.\\

\noindent Several studies have reported that the mass spectrum
steepens at high cloud masses (e.g. \citealt{fukui01},
\citealt{rosolowsky07}, \citealt{gratier12}). In this case, it can be
useful to model the mass spectra using a truncated power-law
(\citealt{williams97}):

\begin{equation}\label{trunclaw}
    N(M'>M)=N_{0}\left[\left(\frac{M}{M_{0}}\right)^{\gamma+1}-1\right],
\end{equation}

\noindent where $M_{0}$ is the maximum mass in the distribution and
$N_{0}$ is the number of clouds more massive than
$2^{1/(\gamma+1)}M_{0}$, the mass where the distribution deviates from
a simple power-law (i.e. the truncation mass).\\

\begin{figure}[!h]
\begin{center}$
\begin{array}{cc}
\includegraphics[width=0.4\textwidth]{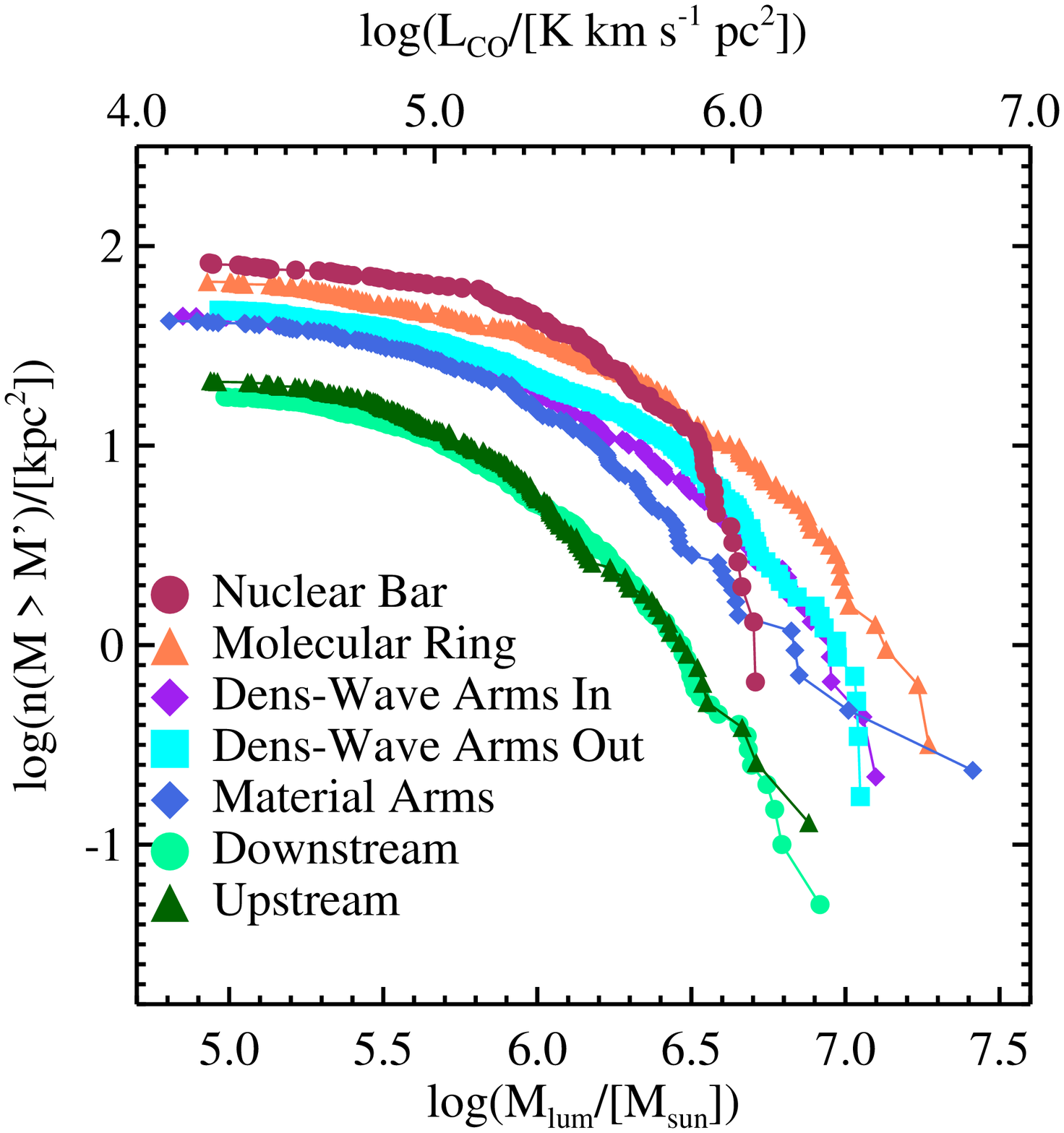} &
\includegraphics[width=0.4\textwidth]{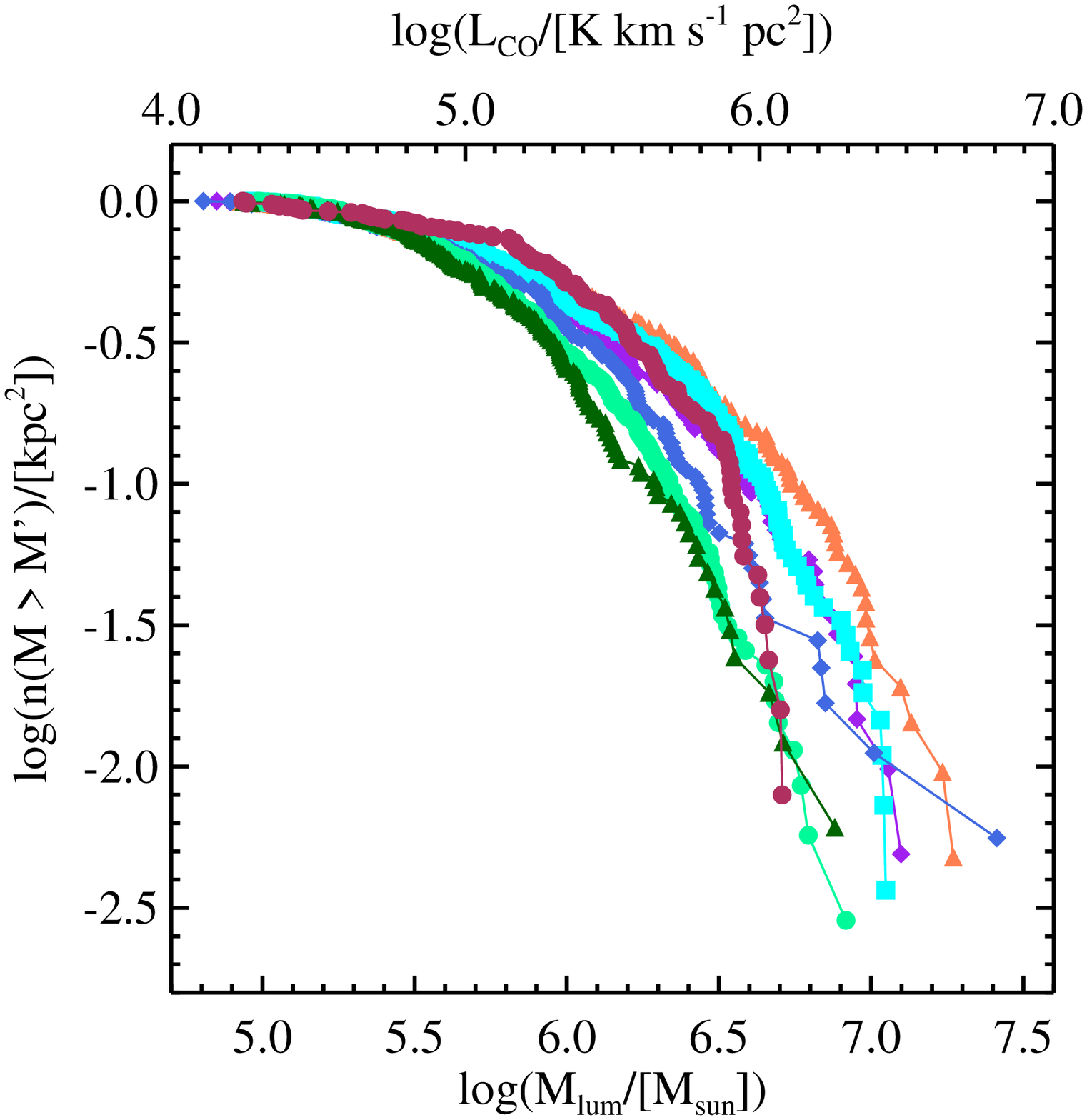} \\ \
\end{array}$
\end{center}
\caption{\footnotesize Cumulative mass spectra for GMCs in the
  different environments of M51 normalized by the area covered by the
  environments in kpc$^{2}$ ({\it left}; see Fig.~\ref{GMC_spect2} for
  exact area) and to the total number of clouds for each environment
  ({\it right}).  The distributions clearly exhibit both a vertical
  offset in the left panel (i.e. a different number density of GMCs)
  and a horizontal offset (i.e. a different maximum cloud mass), as
  well as the different distribution shapes. The equivalent CO
  luminosity is indicated on the top axis.}
\label{GMC_spect}
\end{figure}

\noindent Fig.~\ref{GMC_spect} shows the cumulative $M_{lum}$
distributions for GMCs in different M51 environments. The equivalent
values of CO luminosity are indicated on the top x-axis. In the left
panel, the distributions are normalized by the projected area (in
kpc$^{2}$) of the different environments (listed in
Table~\ref{gmcs_flux}, and indicated in the top-right corner of the
panels in Fig.~\ref{GMC_spect2}). Using this normalization, the
vertical offsets between the different mass distributions reflect true
variations in the number surface density of GMCs: as noted in
Section~\ref{sec:props_general}, the number density of GMCs is higher
in the center than the spiral arms, and higher in the spiral arms than
the inter-arm region. The right panel of Fig.~\ref{GMC_spect} shows
the same GMC mass distributions, this time normalized by the total
number of GMCs in each environment to facilitate a comparison of the
distribution shapes. \\

\noindent The top-left panel of Fig.~\ref{GMC_spect2} shows that
  the overall mass distribution of GMCs within the PAWS field steepens
  continuously with increasing mass. Comparing this global
  distribution with those in the other panels of Fig.~\ref{GMC_spect2}
  suggests that the non-power-law shape of the overall distribution is
  due to combining the intrinsically diverse GMCs mass distributions
  that characterize different galactic environments.  The GMC mass
  distribution in the inter-arm and material arm environments, for
  example, can be adequately represented by simple or truncated
  power-laws across the range of cloud masses probed by PAWS, and are
  hence more similar to the GMC mass distributions that have been
  previously observed for M33 and the LMC (\citealt{wong11},
  \citealt{gratier12}). Across most of the observed mass range, the slope
  of the mass distribution is shallower in the molecular ring and the
  density-wave spiral arms than in the inter-arm, while the mass
  distribution in the material arms has a slope that is intermediate
  between these extremes. Extremely high mass objects
  ($M_{lum}>10^{7}$ $M_{\odot}$) are only observed in the molecular
  ring and spiral arms. The inter-arm region contains very few clouds
  with masses greater than $10^{6.5}$ $M_{\odot}$, although the mass
  distribution of downstream GMCs reaches slightly higher cloud masses
  than the upstream cloud distribution. The nuclear bar has a high
  number density of clouds, and shows evidence for a very strong
  truncation at $10^{6.5}$ $M_{\odot}$.\\

\subsection{Variation in the GMC Mass Distribution with Environment}\label{sec:slope}

\begin{figure}
\begin{center}
\includegraphics[width=1\textwidth]{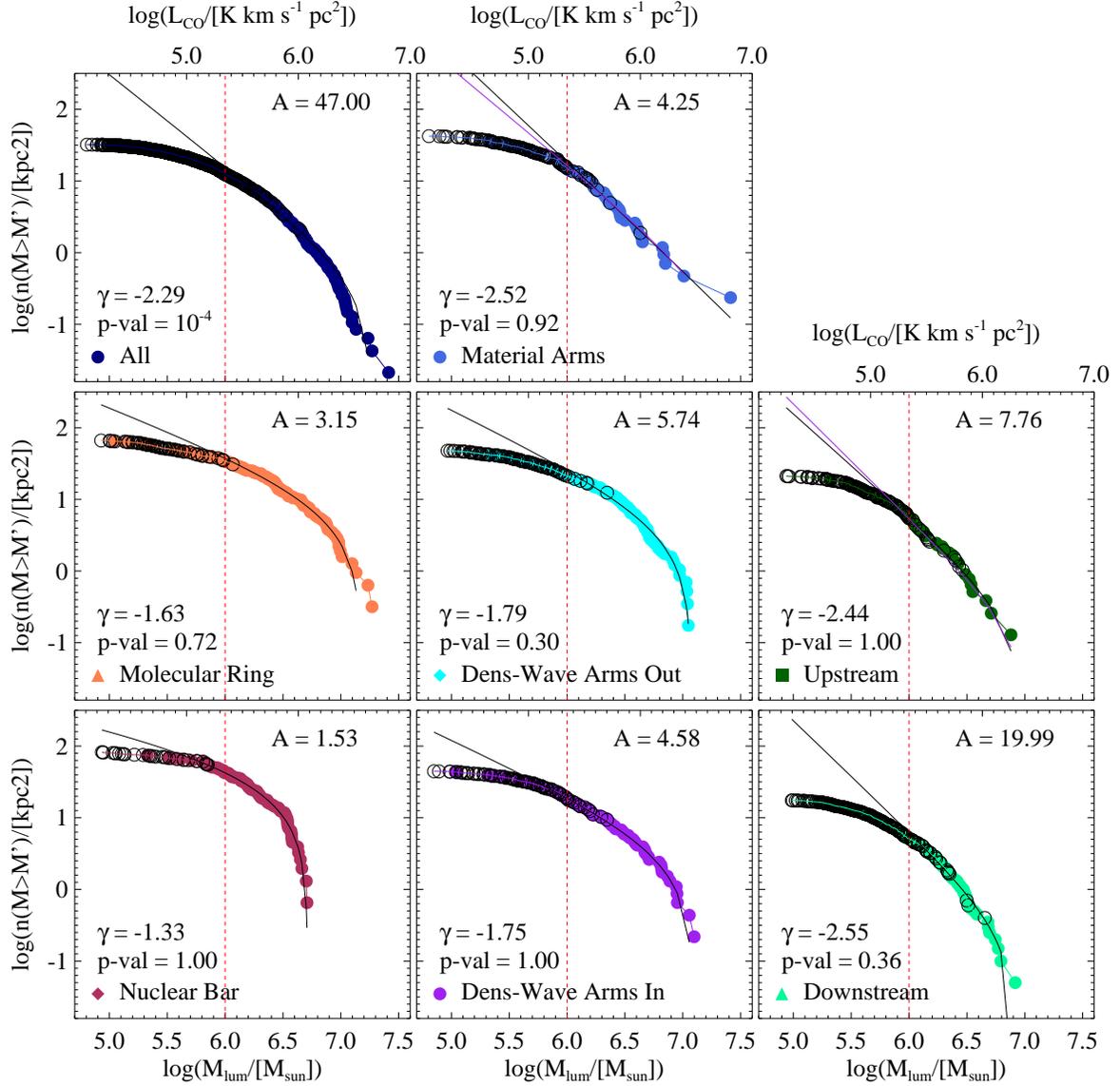}
\end{center}
\caption{\footnotesize Cumulative mass spectra for GMCs in the
different environments (from left to right: central, spiral arm, inter-arm regions with the full
catalog shown in the top left
panel). Colored full circles indicates clouds within the ``highly reliable sample'', while empty black circles clouds with $S/N<6.5$. 
Solid black lines represent the truncated power-law fits while the purple line indicates
the power-law fits for distributions that show resemblance with simple power-law. The red vertical
dashed line indicates the lower mass limit of the fit ($10^{6}$ M$\odot$). In the
top-right corner of each panel the normalization area ``A'' ( in kpc$^{2}$) is given, while on the
lower-left corner the value of the slope ($\gamma$) and of the KS test p-value (p-val) are indicated. For reference, the top axis provides the equivalent CO luminosity.}
\label{GMC_spect2}
\end{figure}

\noindent In the Milky Way and other Local Group galaxies, GMC mass
distributions tend to be adequately represented by simple power-laws
(e.g. \citealt{rosolowsky05} and references therein), although
previous studies have noted that the cloud mass distribution steepens
at high masses in the LMC (\citealt{fukui08}, \citealt{fukui10}) and
in M33 (\citealt{gratier12}). In M51, we find that the
  overall mass distribution steepens continuously with increasing
  cloud mass above our adopted sensitivity limit $3.6\times10^{5}$
  M$_{\odot}$. This is also evident for the GMC mass distributions in
the molecular ring and density wave spiral arm environments, while the
nuclear bar mass distribution exhibits a strong truncation around
$5\times10^{6}$\,M$_{\odot}$. To characterize the diverse shapes of
the GMC mass distributions and facilitate the comparison between M51
and results from other galaxies, we therefore fit the distributions with Eq.~\ref{trunclaw}
above a relatively high fiducial mass of $10^{6}$ M$_{\odot}$, where the mass distributions
show more resemblance to a truncated power-law. This limit is
significantly higher than our adopted catalog completeness limit and
roughly corresponds to the lower mass limit of the highly reliable
sample of clouds. We discuss the reasons for only fitting the mass
distributions above this relatively high mass, and the possible
effects of incompleteness on the mass distributions in
Section~\ref{sec:spectra_bias}. The fit is performed using Erik
Rosolowsky's IDL procedure \verb"MSPECFIT", which implements the
maximum likelihood method described in \citealt{rosolowsky07}. As a
goodness-of-fit test we use the KS test. The parameters of the fits to
the mass distributions are summarized in Table~\ref{gamma_values}. The
fits are overplotted on the mass distributions in
Fig.~\ref{GMC_spect2}.\\

\noindent The GMC mass spectra belonging to the different environments
of M51 show different features. The molecular ring and density-wave
spiral arm cloud distributions show similar slopes
($\gamma\approx-1.8$ to $-1.6$) and fitted maximum masses
M$_{0}>$10$^{7}$ M$_{\odot}$. The mass distributions from the
inter-arm and material arm regions, by contrast, have
$\gamma\approx-2.5$. These results indicate that the molecular gas in
the molecular ring and density-wave spiral arms is preferentially
distributed in high mass GMCs, whereas smaller clouds are the
preferred unit of molecular structure in the inter-arm and material
arm environments. The case of the nuclear bar spectrum is peculiar,
since it presents the shallowest slope ($\gamma\approx-1.3$), but also
reveals a sharp truncation for cloud masses above M$_{0}\approx
5.5\times10^{6}$ M$_{\odot}$. \\

\noindent The inter-arm and material arm spectra have N$_{0}$ close to
the unity, suggesting that a simple power-law is sufficient to
describe the mass distributions. We test this possibility finding that
upstream and material arm distributions can be well represented by
simple power-laws, as shown by the p-values of the corresponding KS
tests, which are close to 1. Even a truncated power-law, however, does
not provide a good fit for overall M51 distribution. This is not
surprising since the distribution for GMCs within the whole PAWS field
is composed of the superposition of the mass distributions from the
different M51 environments, which have different slopes and different
truncation masses.\\

\noindent The mass- and environment-dependent variations in the M51
GMC mass distributions suggest that different mechanisms regulate the
formation and destruction of GMCs in different regions of M51's inner
disk. The non-power-law shape of the mass distributions, which is most
pronounced in the central and density-wave spiral arm environments, is
suggestive of processes that promote the formation (and survival) of
intermediate and high mass clouds. The mass distributions in the
inter-arm region (especially upstream) are closer to pure power-laws,
suggesting that the mechanism(s) responsible for the curvature in the
mass distributions is not as effective in the inter-arm. The influence
of spiral structure on a GMC ensemble may therefore provide another
possible explanation for why the generic shape of the GMC mass
distribution in M51 is distinct from the simple power-law observed for
other extragalactic GMC populations, which tend to be from low-mass
dwarf galaxies (e.g. the LMC and M33, \citealt{wong11},
\citealt{gratier12}) or regions of galactic disks without strong
spiral structure (e.g. the outer Milky Way and an outer arm of M31,
\citealt{rosolowsky05}). We discuss a possible origin for the
environment-dependent changes in the shape of the mass distribution in
Section~\ref{sec:disc}.\\

\begin{table}[h]\footnotesize
\caption{Truncated power-law fits to the GMC mass spectra in different M51 environments}\label{gamma_values}
\begin{center}
\begin{tabular}{l|cccc}
\hline
\hline
\textbf{Envir.}&$\gamma$&M$_{0}$&N$_{0}$&p-value\\
&&$10^{6}$ M$_{\odot}$&&\\
\hline
\emph{All}&$ -2.29\pm 0.09$&$ 18.5\pm 3.4$&$ 17\pm 7$&$10^{-4}$\\
\hline
\emph{NB}&$ -1.33\pm 0.21$&$ 5.2\pm 0.3$&$ 90\pm 21$&1.00\\
\emph{MR}&$ -1.63\pm 0.17$&$ 15.0\pm 3.2$&$ 26\pm 20$&0.72\\
\hline
\emph{DWI}&$ -1.75\pm 0.20$&$ 12.2\pm 1.8$&$ 15\pm 12$&1.00\\
\emph{DWO}&$ -1.79\pm 0.09$&$ 11.8\pm 0.9$&$ 24\pm 9$&0.30\\
\emph{MAT}&$ -2.52\pm 0.20$&$158.6\pm 7.4$&$ 0\pm 2$&0.92\\
\hline
\emph{UPS}&$ -2.44\pm 0.40$&$ 9.3\pm 4.0$&$ 2\pm 3$&1.00\\
\emph{DNS}&$ -2.55\pm 0.23$&$ 8.3\pm 1.9$&$ 5\pm 4$&0.36\\
\hline
\end{tabular}
\end{center}

{\footnotesize Slopes $\gamma$, maximum mass $M_{0}$ and number of GMCs at the maximum mass $N_{0}$
of the truncated power-law fits to the GMC mass spectra of the different environments in
M51. The error are obtained through 50 bootstraps interaction. In the last column, we list the p-values of the KS tests as an indication of the goodness-of-fit.}
\end{table}

\newpage

\clearpage
\newpage

\subsection{Testing the Shape of the GMC Mass Distributions for Incompleteness
Effects}\label{sec:spectra_bias}

\noindent As we noted in Section~\ref{sec:masspect}, most
extragalactic GMC mass distributions that have been observed to date
are adequately represented by a simple or truncated power-law. Since
we argue that the shape of the mass spectrum yields important clues
regarding the physical mechanisms of cloud formation and destruction,
it is important to assess whether the mass distributions that we
obtain are reliable. In particular, although the mass corresponding to
the sensitivity limit of our observations ($\sim10^{5}$\,\msol)
suggests that our GMC catalog should be reasonably complete above $3.6
\times 10^{5}$\,\msol, \textsc{CPROPS} might still be unable to
distinguish clouds above this mass if they are located in a crowded
region like the spiral arms, effectively raising the completeness
limit.\\

\begin{figure}$
\begin{array}{cc}
\includegraphics[width=70mm]{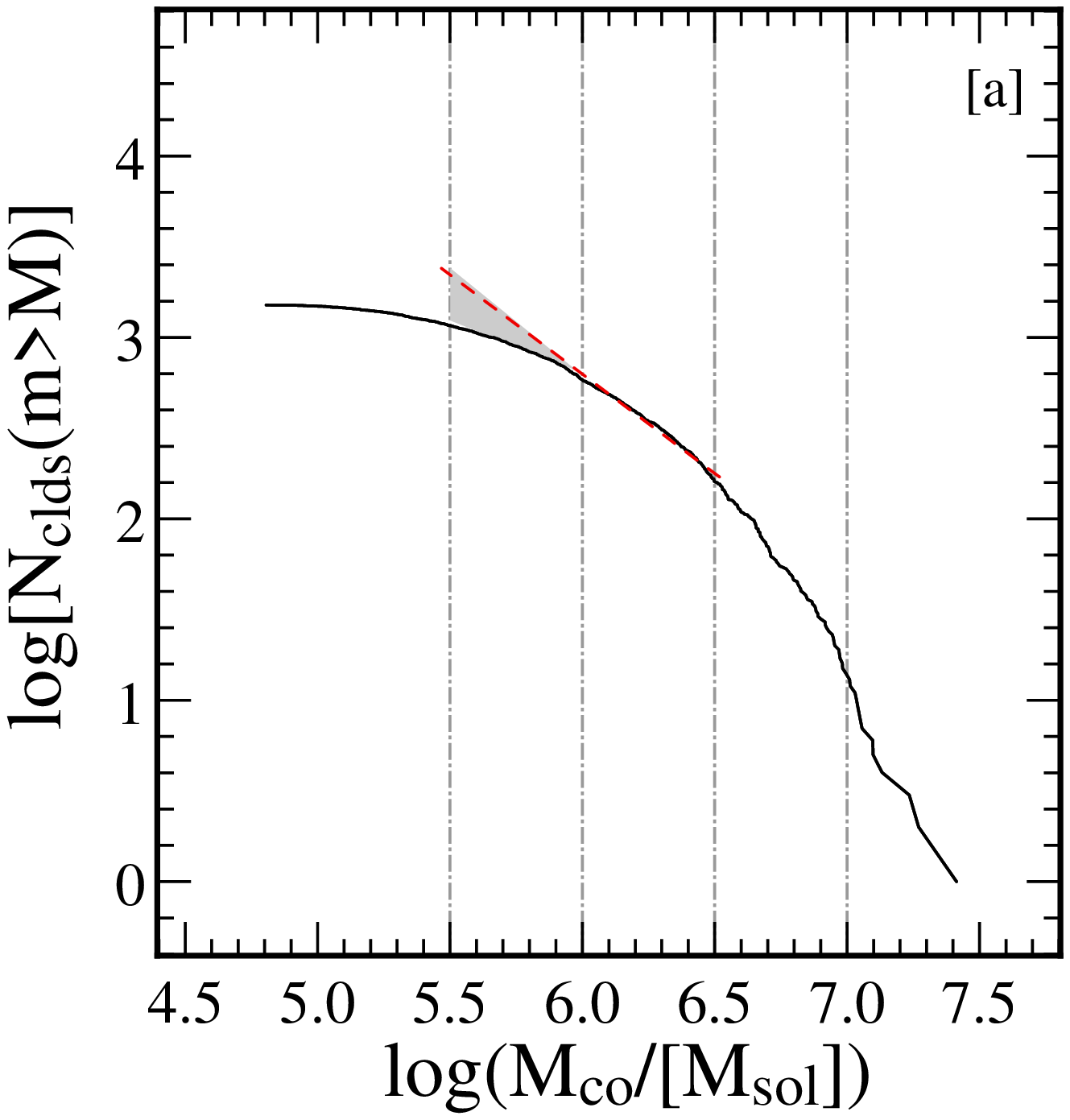} &
\includegraphics[width=70mm]{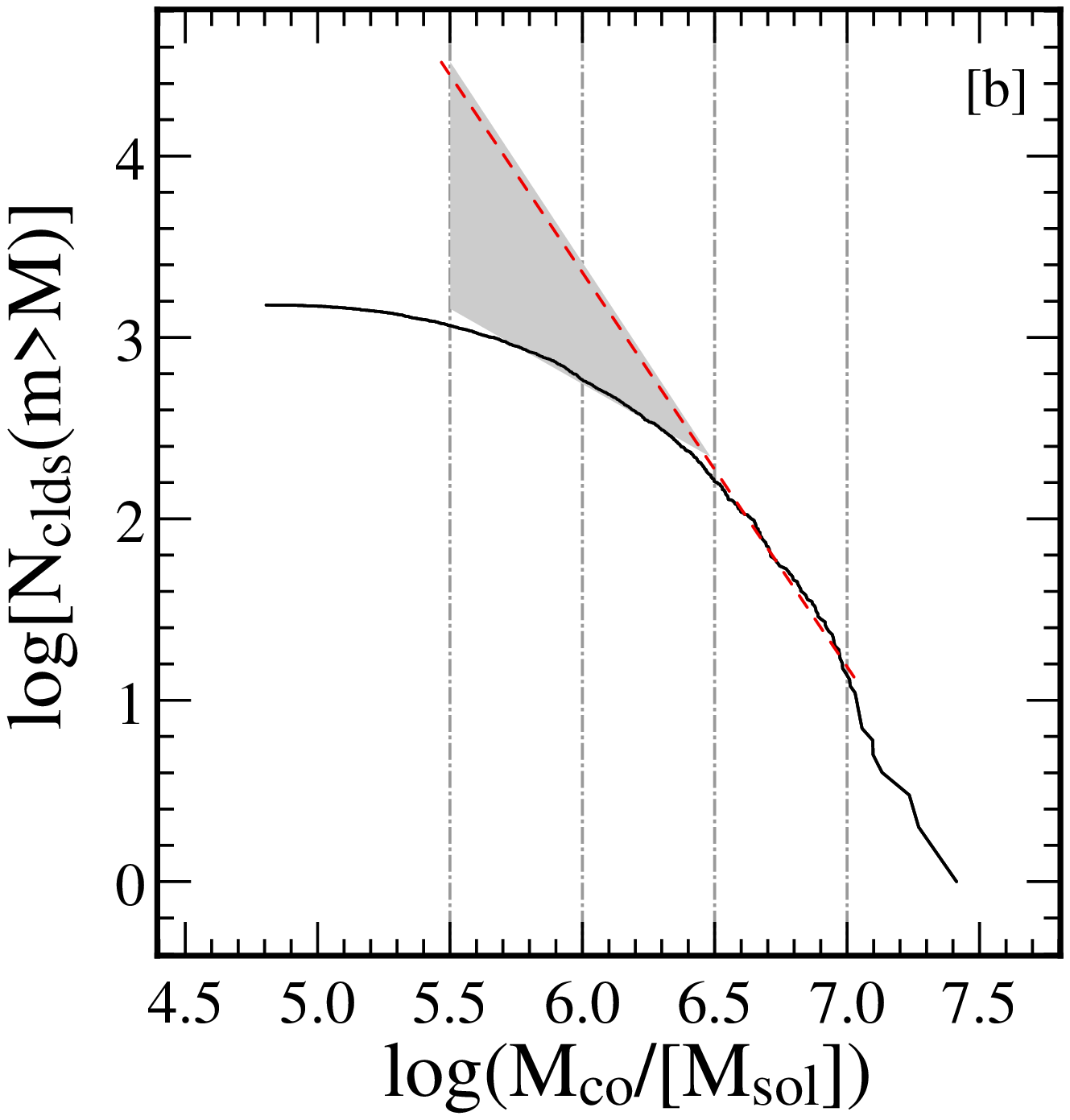} \\ \
\end{array}$
\caption{\footnotesize Schematic diagram illustrating our test for whether
  there is a genuine steepening of the GMC mass distributions in
  M51. We calculate the total number of GMCs under the assumption that
  the power-law mass distribution observed [a] across the mass range
  $\log (M) \in [6.0,6.5]$ (case A) or [b] across the mass range $\log
  (M) \in [6.5,7.0]$ continues down to $M = 10^{5.5}$\,\msol. The
  shape of the distribution at higher GMC masses is assumed to follow
  the observed distribution. The grey shaded wedge in each panel
  indicates the difference between the power-law distribution (red
  dashed line) and observed mass distribution (black solid line) in
  each case. To test whether the true GMC mass distribution could be
  consistent with the power-law mass distribution, we examine whether
  the total CO luminosity corresponding to the power-law mass
  distribution exceeds the integrated CO flux and working area flux
  within each M51 environment. }
\label{fig:missed_flux_mspec}
\end{figure}

\noindent To test whether the observed GMC mass distributions in M51
could be significantly affected by incompleteness, we estimated the
total number of GMCs with masses $M > 10^{5.5}$\,\msol\ and their
combined CO luminosity that would be expected in each M51 environment
if: (i) the true mass distribution followed a simple power-law with
the same exponent as in the intermediate mass bin down to $M =
10^{5.5}$\,\msol\ (case A); and (ii) the true mass distribution across
the mass range followed a simple power-law with the same exponent as
in the upper mass bin down to $M = 10^{5.5}$\,\msol\ (case B). A
schematic explaining the two cases is shown in
Figure~\ref{fig:missed_flux_mspec}, and the results for each M51
environment are presented in Table~\ref{tbl:missed_flux_mspec}. \\

\noindent On one hand, it is clear that there must be a genuine
steepening of the GMC mass distribution in all M51 environments. If
the mass distributions in the inner spiral arms and molecular ring
were simple power-laws with the same exponents that we observe across
the mass range $10^{6.5}$ to $10^{7}$\,\msol (i.e. case B), then the
total number of GMCs with $M > 10^{5.5}$\,\msol\ in each environment
would exceed several thousand, and the CO luminosity associated with
this mass distribution would be greater than each region's total CO
flux (measured via direct integration of the PAWS data cube) by
factors between five and ten. A similar -- though not identical --
situation applies in the material arm and inter-arm regions. The CO
luminosity corresponding to a power-law mass distribution for GMCs
with $M > 10^{5.5}$\,\msol\ with the same exponent as that in the
intermediate mass bin would not exceed (or, in the case of the
material arm, would not greatly exceed) the total CO flux of these
regions, but it would require that roughly half of the undetected GMCs
fall outside the \textsc{CPROPS} `working area', i.e. the initial mask
identifying regions of significant emission. As such, these undetected
GMCs would need to be spatially extended, low CO surface brightness
structures containing $10^{5.5}$ to $10^{6}$\,\msol\ of CO-emitting
molecular gas without an emission peak brighter than $4\sigma_{RMS} =
1.2$\,K. Since the total CO luminosity associated with this mass
distribution is comparable to the total flux of these regions,
moreover, it would also entail a strong flattening of the GMC mass
distribution for $M < 10^{5.5}$\,\msol.  A more gradual flattening of
the GMC mass distribution between $10^{5.0}$ and
$10^{6}$\,\msol\ would seem at least as plausible as the possibility
that high-mass, low-surface brightness structures are ubiquitous
throughout M51's inter-arm and material arm while clouds with $M <
10^{5.5}$\,\msol\ are intrinsically rare. \\

\noindent On the other hand, we cannot use similar arguments to rule
out that the slope of the GMC mass distributions between $10^{5.5}$ to
$10^{6}$\,\msol\ in the spiral arm and central regions could be due to
an algorithmic effect. If the mass distribution in these regions
continued with the same exponent that we observe for the intermediate
mass bin down to $10^{5.5}$\,\msol\ (or even $10^{5.0}$\,\msol), then
the constraint that the combined CO luminosity should not exceed the
observed CO flux is not violated. Indeed, the combined CO luminosity
that would be associated with GMCs with $M >
10^{5.0}$\,\msol\ assuming a simple power-law across $10^{5.0}$ to
$10^{6.5}$\,\msol\ is less than or comparable to the flux in the
working area (i.e. not only the total flux) for these environments.\\

\noindent Nevertheless, moving the completeness limit up to $10^{6}$
\Msun does not change our main conclusions about the different
physical mechanisms that regulate the formation/disruption of GMCs,
which we infer mainly from the intermediate and upper mass bins of the
mass spectra. We further note that considering only clouds with
$M_{lum}>10^{6}$ \Msun\ makes the differences in the cloud properties
described in Sections \ref{sec:props_basic}-\ref{sec:props_deriv} even
more pronounced.

\begin{deluxetable}{lcccccccc}
\tabletypesize{\footnotesize}
\rotate
\tablecaption{Results of GMC Mass Distribution Tests\label{tbl:missed_flux_mspec}}
\setlength{\tabcolsep}{0.04in}
\renewcommand{\arraystretch}{1.2}

\tablehead{
\colhead{Region}    &  \multicolumn{2}{c}{$L_{CO}$ in Environment}  & \multicolumn{2}{c}{Observed
Distribution}    & \multicolumn{2}{c}{Case A} & \multicolumn{2}{c}{Case B$^{(a)}$}\\
          &  \colhead{Total}   & \colhead{Working Area} & \colhead{$N_{GMCs}^{(b)}$} &
\colhead{$L_{CO}^{(c)}$} & \colhead{$N_{GMCs}^{(b)}$}
& \colhead{$L_{CO}^{(c)}$} & \colhead{$N_{GMCs}^{(b)}$}  & \colhead{$L_{CO}^{(c)}$}
\\
          &  \colhead{$[10^{7}\,\lcou]$} &  \colhead{$[10^{7}\,\lcou]$} & & 
\colhead{$[10^{7}\,\lcou]$} &  & \colhead{$[10^{7}\,\lcou]$} &
&  \colhead{$[10^{7}\,\lcou]$}}
\tablewidth{0pt}
\startdata
Cube      & 90.83    & 67.08  &   1160 & 47.05 & 2207 & 59.25 & 27739   & 407.9 \\
\hline
NB       &  7.48    &  6.49  &    116 &  5.07 &  270 &  6.96 &  &  \\
MR      & 17.99    & 16.35  &    160 &  9.60 &  315 & 11.44 & 5082    &  79.34 \\
\hline
DWI       & 13.13    & 11.23  &    180 &  7.58 &  280 &  8.75 & 9057    & 126.39 \\
DWO       & 18.38    & 15.73  &    260 & 11.73 &  371 & 12.76 & 8290    & 122.21 \\
MAT        &  8.06    &  5.44  &    148 &  5.64 &  537 & 10.36 &  825    &  14.19 \\
\hline
DNS      & 17.96    &  8.54  &    156 &  4.40 &  566 &  9.34 &  &  \\
UPS        &  7.79    &  3.28  &    140 &  3.03 &  478 &  7.03 &  &  \\
\enddata
\tablecomments{\footnotesize
(a) Only for environments with a maximum GMC mass greater than $10^{7}$\,\msol;
(b) number of GMCs with $M > 10^{5.5}$\,\msol\ in the distribution;
(c) combined CO luminosity of GMCs with $M > 10^{5.5}$\,\msol (see text for details).}
\end{deluxetable}

\section{Discussion}\label{sec:disc}

\subsection{An Evolutionary Scenario for the Environmental Variation of the GMC Mass Distributions in M51}\label{sec:disc_mass}

\noindent Recent studies of GMCs and their associations, i.e GMAs, in
nearby disk galaxies have provided evidence that cloud properties are
not uniform across the disk and that galactic environment (such as
bulge, disk, nuclear bars, star-forming rings, spiral arms and
inter-arm regions) might be responsible for the observed
differences. \cite{koda09}, for example, find that GMAs with masses
above $\rm 10^7\,M_{\sun}$ are exclusively located along the spiral
arms of M51. They attribute this observed spatial distribution to
large-scale dynamical processes induced by the spiral potential. In a
recent sample of five nearby galaxies from the CANON survey, a similar
trend for massive GMCs to be associated with strong spiral arms is
observed (e.g. Fig. 6 of \citealt{donovan13}). The differences in
M51's GMC properties with galactic environment that we describe in
this paper are therefore not entirely unexpected. However, our study
provides the first quantitative measure of the differences in the
cloud properties and also reveals a strong variation in the GMC mass
spectra (i.e. slope, normalization and maximum mass;
Section~\ref{sec:slope}) with galactic environment. The variations in
the mass spectrum are observational signatures of the mechanisms of
cloud formation and evolution, providing evidence for processes that
not only change the physical properties of individual clouds, but also
influence the ensemble properties of the cloud population.\\

\noindent The mass spectra of clouds in the inter-arm and density-wave
spiral arm are different. The variation in the slope $\gamma$
between the density-wave spiral arm and upstream mass spectra
(Section~\ref{sec:slope}) implies that spiral arms do not simply
gather GMCs from the upstream inter-arm environment (in this case the
slope of the mass distributions would be identical, even though the
overall normalization could change), but also modify the nature of the
constituent clouds. More precisely, the inter-arm distributions are
steep (spectral index $\gamma<-2$) and all clouds have masses lower
than 10$^{7}$\,M$_{\odot}$, characteristic of a population of clouds
that is dominated by low-mass objects. The spiral density wave mass
spectra, by contrast, are shallower ($\gamma>-2$) and have a much higher
maximum cloud mass, consistent with a cloud population mainly
constituted by high mass objects. Spiral arms, therefore, must host
processes that promote the growth of massive clouds, without providing
an effective mechanism for their destruction.\\

\begin{figure}[h!]
\begin{center}
\includegraphics[width=0.65\textwidth]{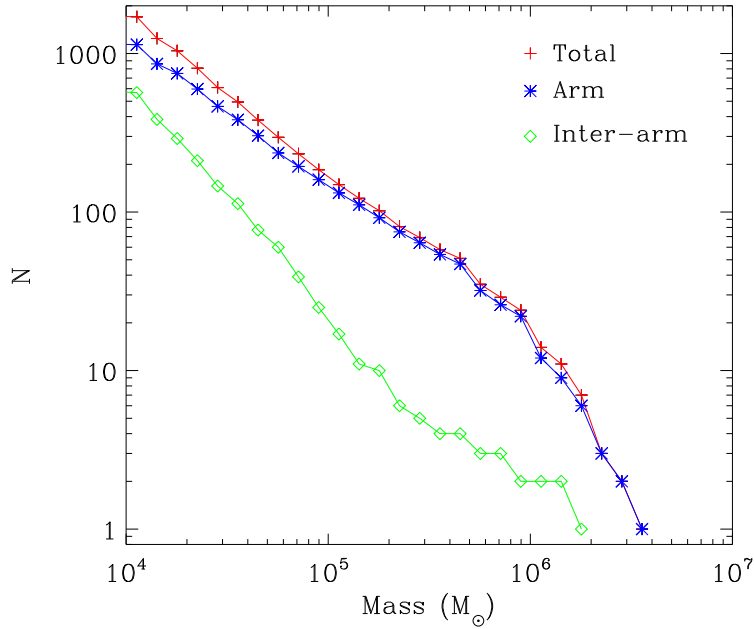} \\ \
\end{center}
\caption{\footnotesize The cumulative mass distributions for the arm
  (blue) and inter-arm (green) regions in a simulation of a two armed
  spiral galaxy. The simulation is described in Section 7 of
  \cite{dobbs12} and is presented in \cite{dobbs13}. The mass per particle
  was 312.5 M$_{\odot}$. Clouds were identified using a clump-finding
  algorithm that selects contiguous regions with $>25$ M$_{\odot}$ pc$^{-2}$, an
  approach that is not dissimilar to CPROPS. }
\label{fig:clare_sim}
\end{figure}

\noindent Within a spiral potential, Jeans instabilities are thought
to be the dominant mechanism of cloud formation
(e.g. \citealt{mckee07}). Numerical studies of gas in spiral
potentials have observed that GMCs also increase their mass through
coagulation processes (cloud collisions, accretion of small clouds,
mutual coalesence) that are aided by the converging streamlines of the
gas flow within the arms (\citealt{casoli81}, \citealt{kwan83},
\citealt{tomisaka86}, \citealt{dobbs08}, \citealt{tasker09}). Together
with those phenomena, \cite{meidt13} proposed that streaming motions
associated with the spiral potential decrease the external gas
pressure leading to increased stable masses (see also
\citealt{jog13}). Therefore, GMCs in regions of the spiral arm with
strong streaming motions can become very massive without undergoing
significant collapse. A recent numerical simulation by \cite{dobbs13}
of a two armed spiral galaxy that includes a spiral potential,
self-gravity, heating and cooling of the ISM and stellar feedback (see
Fig.~\ref{fig:clare_sim}) yields mass spectra that are similar to
those observed for the spiral arm and inter-arm region of M51. In this
simulation, cloud formation is a complex process that involves
gravitational instabilities, assembly of smaller clouds and accretion
of local interstellar gas onto the cloud. However, we note that the
number density of clouds across the entire observed mass range
increases within the spiral arm environments, i.e. low mass clouds are
also created in the arms and not just subsumed into larger
structures. This suggests that gravitational instabilities remain the
primary mechanism for GMC formation in M51's spiral arms, although
dynamical effects almost certainly play an important role in bringing
a large quantity of molecular gas to a single location, where it
subsequently fragments due to gravitational instabilities.\\

\noindent \cite{koda09} have argued that GMCs in the inter-arm regions
of M51 cannot have formed locally on an inter-arm crossing time-scale,
but are rather remnants of GMCs that were previously in the spiral
arms. The change in the GMC mass distribution between the arm and
inter-arm region suggests that GMCs undergo a disruptive process (or
processes) that preferentially affects the most massive objects when
they leave the arms. Numerical simulations of the ISM in spiral
galaxies (\citealt{dobbs06}, \citealt{dobbs13}) suggest that the
prominent ``spurs'' that emanate downstream from the spiral arms (see
\citealt{lavigne06}, \citealt{schinnerer13}) can be interpreted as
sheared GMCs or their associations due to large-scale dynamical
motions. Another possible cause of cloud destruction is feedback from
star formation. In M51, young stellar clusters and enhanced atomic gas
(HI, CII) emission (\citealt{schinnerer13}) suggest that star
formation is enhanced downstream of the outer density-wave spiral
arms. Furthermore, there is an extended, dynamically hot component of
the molecular gas in M51 (described by \citealt{pety13}) that
spatially correlates with locations of star formation, and could be
the result of galactic fountains or chimneys that have transported
some of the molecular gas away from the disk
(e.g. \citealt{putman12}). Yet star formation feedback cannot be the
primary cause for cloud disruption throughout M51's spiral arms since
the inner spiral arm segments have no evidence for high mass star
formation (\citealt{schinnerer13}).  The cloud mass distributions in
the inner and outer arms are very similar, suggesting either that star
formation feedback is not the dominant destruction mechanism in any of
the arm environments or that shear and star formation feedback yield a
similar mass distribution of cloud fragments upon the disruption of a
high mass GMC.\\

\noindent Subtle differences between the upstream and downstream GMC
mass distributions (i.e. the higher number density of low-mass
upstream clouds with respect to the downstream ones) suggest that the
disruptive events continue to act across the entire inter-arm
region. If GMCs (not the molecular gas itself) are indeed
``short-living'' entities ($\sim30$ Myr, \citealt{elmegreen00}), then
they are unable to maintain their identity throughout the whole
journey from one arm to the other (e.g. \citealt{pringle01}) causing a
transformation of the cloud population to include a higher proportion
of low mass objects. Shearing forces are strong throughout the
inter-arm region, and therefore likely to play a role in cloud
destruction. Star formation, as traced by H$\alpha$ emission, is not
entirely absent from the inter-arm region however, suggesting that
feedback also contributes to cloud destruction in this region.\\

\noindent The molecular ring is an environment that appears very
favorable for cloud formation: the mass distribution in this region is
very shallow ($\gamma\approx-1.6$) and extends to cloud masses greater
than $10^{7}$ M$_{\odot}$. The similarity between the distributions in
the molecular ring and the density-wave arm environments would seem to
suggest that cloud formation and destruction mechanisms may be
present. However, the gas dynamics in the central region are very
different from the disk. The molecular ring is coincident with a zero
torque environment caused by the overlap of resonances of the inner
bar and the spiral density wave, i.e. the combined action of outflow
driven by the nuclear bar and inflow by the spiral wave
(\citealt{meidt13}). Thus, the molecular ring zone harbors nearly
circular orbits with at most low non-circular motions (Colombo et al.,
submitted) and almost no shear (analogous to the 5\,kpc molecular ring
in the Milky Way, \citealt{dib12}). Streaming motions in the ring are
low, moreover, implying that the stable mass against cloud collapse is
determined solely by the gas density. Unlike in the inner spiral arms,
high-mass star formation is active throughout the ring and appears
coincident with regions of high gas surface density. We propose that
due to the opposing action of the bar and spiral arm torques, gas
accumulates and stalls in the molecular ring, where it develops high
densities. Gravitational instabilities then cause the gas to fragment
into clouds. In the absence of shear, star formation feedback should
be the dominant mechanism of cloud destruction in this region.\\

\noindent Finally, the formation and destruction of clouds in the
nuclear bar environment may also follow a different path than in other
parts of the PAWS field. In particular, Figure~\ref{GMC_spect} shows
the mass spectrum in the nuclear bar region has a high number density
of low and intermediate mass GMCs, but a sharp truncation at
around $10^{6.5}$ $M_{\odot}$. This implies that the bar environment
either lacks an efficient mechanism to bring small clouds together to
form larger structures, or that a very efficient mechanism for the
destruction of massive objects is active. The presence of low and
intermediate mass GMCs may be a consequence of the abundant molecular
gas reservoir collected by the nuclear bar dynamics: gas on the
leading sides of a bar loses angular momentum and is driven inwards,
as a result of negative gravitational torques
(e.g. \citealt{schwarz84}). This motion is also expected to generate
intense shear in the gas lanes (e.g \citealt{athanassoula92},
\citealt{sheth02}) that could prevent the formation of massive
objects through the inhibition of the density fluctuations that become
the seeds of massive GMCs (\citealt{hopkins12}).\\

\noindent Besides the dynamical processes discussed above, other
effects, such as variations in the interstellar radiation field
(ISRF), the molecular gas fraction and/or the $X_{CO}$ factor, could
potentially alter the properties of the M51 GMC populations. In M51,
the ISRF is decreasing from the bulge to the disk and, in particular,
the intense radiation field of the young massive stars in the
star-forming ring and/or the AGN could have a strong impact on cloud
disruption and heating of the molecular gas. Beyond the bulge region,
however, H-band observations indicate that there is no overall radial
trend in the ISRF for the range of galactocentric radii probed by PAWS
(\citealt{munoz11}). The molecular gas fraction (defined
  as $2N_{H_{2}}/(2N_{H_{2}}+N_{HI}$) over the PAWS area is very high
  ($\sim0.85$, assuming $X_{CO}=2\times10^{20}$ cm$^{-2}$ (K km
  s$^{-1}$)$^{-1}$) and does not significantly vary with radius or
  azimuth (see e.g. Fig.~2c in \citealt{koda09}). This high
molecular fraction is determined mostly by the scarcity of
HI emission within the PAWS field; halving the adopted value of
$X_{CO}$ only lowers the molecular fraction to $\sim0.7$. The
metallicity and the gas-to-dust ratio within the PAWS FoV also suggest
that variations in the $X_{CO}$ factor are unlikely to be the main
driver of the differences in the GMC properties and mass spectra that
we observe. \cite{bresolin04}, and \cite{moustakas10} both found a
metallicity close to solar with only a shallow radial gradient of $\rm
-(0.02\pm0.01)\,dex\,kpc^{-1}$, so we do not expect
metallicity-dependent variations in the conversion factor across the
PAWS FoV (see e.g. Leroy et al 2011, 2012). A recent analysis of the
Herschel FIR continuum (\citealt{cooper12}) has likewise shown that
the gas-to-dust ratio is roughly constant within the inner 13 kpc of
M51, assuming $X_{CO}=2\times10^{20}$ cm$^{-2}$ (K km
s$^{-1}$)$^{-1}$ throughout M51. If there were strong environmental
variations in the $X_{CO}$ factor within the PAWS FoV, these would
lead to corresponding spatial variations in their derived gas-to-dust
ratio map, but such variations are not observed (see Fig~16 in
\citealt{cooper12}).\\

\noindent In conclusion, the presence of spiral arms has a dramatic
effect on the GMC properties observed in the central 9\,kpc of
M51. Excluding phenomena such as a varying ISRF, molecular gas
fraction and $X_{CO}$ factor that are observed to be roughly constant
across the disk, we propose that a large amount of gas is accumulated
by the spiral arm dynamics, and subsequently fragmented by
gravitational instabilities. We further propose that the variations
in the shape of the cloud mass spectra can be interpreted as the
evolution of clouds traveling from one side of a spiral arm to the
other arm. However, further work is required to understand the
relative importance of shear and star formation feedback between and
within M51's inner spiral arms, and to characterize the effect of
these destruction processes on the shape of the GMC mass
distributions. The presence of high mass objects in the circumnuclear
ring can be best explained by gas accumulation and strong
gravitational instabilities in the absence of strong destructive
dynamical effects such as shear. It is likely that shear helps prevent
the formation of massive clouds in the nuclear bar region, but the
enhanced ISRF in M51's bulge and the AGN make it difficult to separate
their contribution from large-scale dynamical effects in this
region.\\

\clearpage
\newpage

\subsection{Larson's laws in M51}\label{sec:disc_larson}

\noindent In addition to the differences in the GMC mass
  spectra with galactic environment, the scaling relations between
  cloud properties provide further insight into the processes that
  regulate their physical properties. From our analysis in
  Section~\ref{sec:larson}, two important features of GMCs in M51
  emerge: first, both the size-velocity dispersion and CO
  luminosity-virial mass relations show a large scatter; and second,
  the GMC mass surface density varies with environment as seen by the
  radius-CO luminosity relation. Here, we argue that these results
  have a common origin, i.e. the different dynamical properties of the
  environments.

\noindent A relation between size and velocity dispersion was
identified in the early studies of Milky Way clouds (e.g. Solomon et
al. 1979; Larson 1981; Dame et al. 1986). It is often interpreted as
evidence for a cloud in virial equilibrium following the work of S87,
where the authors measured a square-root dependency between velocity
dispersion and radius of Galactic GMCs. But unlike the tight
size-velocity dispersion relation discovered by S87, the corresponding
relationship in M51 shows a large scatter. If GMCs are not strongly
bound, then they become susceptible to modification and/or disruption
by events and conditions in the surrounding interstellar medium. For
clouds where $\alpha \gg 1$, external sources of confining pressure,
such as ram pressure from inflowing material \citep[e.g.][]{heitsch09}
or the (static) weight of the surrounding gas
\citep[e.g.][]{heyeretal01} become important for their dynamical
properties and evolution.\\

\noindent The higher mass surface densities of clouds in the spiral
arms compared to the inter-arm region implies that the arm GMCs have
higher internal pressures. More precisely, we can
  estimate the internal pressure $P_{int}$ of a molecular cloud
  according to:

\begin{equation}
\frac{P_{int}}{k} = \rho_{g}\sigma_{v}^{2} = 1176\left(\frac{M}{M_{\odot}}\right)\left
(\frac{R}{{pc}}\right)^{-3}\left(\frac{\sigma_{v}}{{\rm km s}^{-1}}\right)^{2}\,\punit \, ,
\end{equation}

\noindent where $\rho_{g}$ is the H$_{2}$ volume density. For the cloud
populations in the central, inner spiral arm, material arm and
inter-arm regions of M51, we find median internal pressures of $\langle
P_{int}/k \rangle \sim 8.2\times10^{5}$, and $6.7\times10^{5}$,
$5.2\times10^{5}$,and $3.5\times10^{5}$ respectively. These
differences track the variation in the stellar mass surface density
between the different M51 environments (\citealt{meidt13}). Since the
stellar mass dominates the ambient kinetic pressure of the ISM under
the conditions that prevail in the inner disk of M51 (see
e.g. estimates for the hydrostatic midplane pressure by
\citealt{koyamaostriker09} and \citealt{elmegreen89}), the observed
variations in the GMC mass surface density may suggest that the
external ISM pressure plays a critical role in regulating the internal pressure
(and hence velocity dispersion and density) of molecular clouds in M51
(as suggested by e.g. \citealt{rosolowskyblitz05}). This interpretation
is discussed in more detail by a companion paper (Hughes et al. 2013b), 
where resolved GMC populations from a small sample of
nearby low-mass galaxies are included in the analysis.\\

\noindent In summary, our finding that the properties of GMCs in M51
vary with galactic environment argues against the view that
GMCs are long-lived, quasi-equilibrium entities, with a constant mass
surface density and isolated from their interstellar
environment. Instead, we propose that the prominent dynamical
phenomena in M51, i.e. the spiral arms and nuclear bar, are
responsible not only for efficiently transporting large quantities of
gas within the central 9\,kpc of the galactic disk, but also for
producing cloud structures that are physically different from the GMCs
observed in Local Group galaxies where such strong galactic-scale
dynamical effects are absent. Instead of isolated clouds, the GMCs
identified in high pressure, molecule-dominated environments may be
the high density peaks of a more extended molecular medium, where
large-scale dynamical effects play a larger role in controlling the
formation and evolution of GMCs than small-scale phenomena such as
star formation feedback (including stellar wind and supernova
explosions, see also \citealt{hopkins12}). In M51, star formation may
even be seen as a ``by-product'' that occurs in special places of the
galaxy where gas can accumulate and has time to virialize and
collapse, like M51's molecular ring. A corollary of our interpretation
that merits further investigation (e.g. \citealt{meidt13}) is that
only a small fraction of clouds and molecular gas may be associated
with star formation in galaxies with a strong spiral potential and
Kennicutt-Schmidt-type relations may not hold on cloud-scales in such
systems.\\

\section{Summary}\label{summary}

\noindent Using the PAWS (PdBI Arcsecond Whirlpool Survey)
observations of the $^{12}$CO(1-0) line emission in the central 9\,kpc
of M51, we cataloged a total of 1,507 GMCs using an identification
algorithm (CPROPS) that corrects for survey biases. These GMCs contain
54\% of the total flux present in the PAWS cube. Most GMCs in
M51 show a preferred orientation in the disk that roughly follows the
pattern described by the spiral arms. To investigate possible
dependencies of the GMC population on large-scale properties, the PAWS
FoV was divided in seven galactic environments. We find a distinct
dependence of GMCs properties on galactic environment that can be
summarized as follows:

\begin{enumerate}
 
\item Clouds in the density-wave spiral arms and the central region
of M51 exhibit the highest average values of peak brightness
temperature and velocity dispersion. These properties
decrease in the material arms, where clouds appear more similar to the
inter-arm ones. Inter-arm GMCs have the lowest average values of peak
brightness temperature, velocity dispersion and mass.

\item The analysis of the cloud derived properties
  suggests that there is a general decrease in H$_{2}$ masses and
  surface density of GMCs from the central to the inter-arm
  region. The densest and most massive clouds are located in the
  molecular ring and density-wave spiral arm environments.
 
\item There is no obvious size-line width relation for clouds
  in M51. The median virial parameter is $\sim1.6$, which suggests
  that the cloud population is, on average, self-gravitating.
  However, the virial mass-CO luminosity and size-velocity dispersion
  relationships show a large scatter, indicating that the GMCs are in
  diverse dynamical states, and that a significant number of clouds
  may be pressure confined and/or unbound.
 
\item The varied shapes observed for the GMC cumulative mass
  spectra can be interpreted as the result of differing mechanisms of
  GMC formation and evolution within the different M51
  environments. Cloud formation appears to be promoted in the
  molecular ring and spiral arms, where the mass spectra show a higher
  number density of GMCs and contain GMCs of especially high mass. We
  propose that the shapes of the mass spectra in M51 indicate a common
  mechanism of cloud formation (local gravitational instabilities). We
  further propose that the destruction of GMCs in M51 is mostly due to
  large-scale dynamical effects (i.e. shear), although feedback from
  high mass star formation may be more important downstream of the
  spiral arms and in the molecular ring.\\

\item The analysis of a cloud population within a complex and
  crowded environment, such as the inner region of M51, reveals
  several challenges for commonly used decomposition algorithms, like
  CPROPS, in identifying and measuring GMCs properties.

\end{enumerate}


\acknowledgements
\noindent We thank our referee (Jonathan Braine) for his thoughtful comments that greatly improved the quality of the paper. We thank the IRAM staff for their support during the observations with
the Plateau de Bure interferometer and the 30m telescope.
DC and AH acknowledge funding from the Deutsche Forschungsgemeinschaft (DFG) via grant SCHI 536/5-1
and SCHI 536/7-1 as part of the priority program SPP 1573 'ISM-SPP: Physics of the Interstellar
Medium'.
CLD acknowledges funding from the European Research Council for the FP7 ERC starting grant project
LOCALSTAR.
TAT acknowledges support from NASA grant \#NNX10AD01G.
During this work, J.~Pety was partially funded by the grant ANR-09-BLAN-0231-01 from the French {\it
Agence Nationale de la Recherche} as part of the SCHISM project (\url{http://schism.ens.fr/}).
ES, AH and DC thank NRAO for their support and hospitality during their visits in Charlottesville.
ES thanks the Aspen Center for Physics and the NSF Grant \#1066293 for hospitality during the
development and writing of this paper. DC thanks Erik Rosolowsky for help with CPROPS and Pierre
Gratier for the useful discussion. SGB acknowledges economic support from Junta de Andalucia grant P08 TIC 03531. The National Radio Astronomy Observatory is a facility of the National Science Foundation operated under cooperative agreement by Associated Universities, Inc.


\newpage


\appendix

\section{Island catalog}\label{app:island}
\noindent Islands are connected emission structures inside the working area
spanning at least one telescope beam area and one velocity
channel. Because of the high sensitivity of the PAWS cube, the island
catalog is dominated by the presence of a huge central object
that contains more than $50\%$ of the total flux present in the
data cube and more than $70\%$ of the total emission contoured by the
CPROPS island identification. It embodies almost the whole central
region and a significant portion of the spiral arms. Excluding this entity, the remaining islands are evenly distributed between the spiral arm and inter-arm regions, with only a few objects located in the central region. Approximately, 70\% of the islands are associated with a single GMC, the majority of which are located in the inter-arm region. Contrary to the single island that dominates the central and inner spiral arms, these undecomposed islands are representative of a more flocculent molecular gas environment, in which the CO emission mostly arises from discrete objects. To obtain the island catalog CPROPS was run with the following parameters:

\begin{itemize}
  \item \footnotesize \verb"THRESHOLD" = 4
  \item \footnotesize \verb"EDGE" = 1.5
  \item \footnotesize \verb"MINVCHAN" = 1
  \item \footnotesize \verb"BOOTSTRAP" = 50
  \item \footnotesize \verb"/NONUNIFORM"
  \item \footnotesize \verb"/NODECOMPOSITION"
\end{itemize}

The \verb"/NODECOMPOSITION" flag forces CPROPS to calculate the
properties of the connected regions it found without any attempt
to decompose them into substructures. A part of the full island catalog
is reported in Table~\ref{isls_catalog}. The complete version is available in electronic
format to the dedicated web-page \verb"http://www.mpia-hd.mpg.de/home/PAWS/PAWS/Data.html".

\begin{deluxetable}{ccccccccccccccc}
\tabletypesize{\scriptsize}
\rotate
\tablecaption{PAWS island catalog\label{isls_catalog}}
\tablewidth{0pt}
\setlength{\tabcolsep}{0.02in}
\renewcommand{\arraystretch}{1.2}
\tablehead{
\colhead{ID} & \colhead{RA (J2000)} & \colhead{Dec (J2000)} & \colhead{$V_{LSR}$} &
\colhead{$T_{max}$} & \colhead{$S/N$} & \colhead{$R$} & \colhead{$\sigma_{v}$} &
\colhead{$L_{CO}$} & \colhead{$M_{vir}$} &
\colhead{$\alpha$} & \colhead{PA}& \colhead{b/a}& \colhead{Reg} &\colhead{Flag}\\
& \colhead{$hh\,mm\,ss.ss$} & \colhead{$dd\,mm\,ss.ss$} & \colhead{km\,s$^{-1}$} & \colhead{K} &&
\colhead{pc} &
\colhead{km\,s$^{-1}$} & \colhead{$10^{5}$\,K\,km/s\,pc$^{2}$} & \colhead{$10^{5}$\,M$_{\odot}$} &&& \colhead{deg}&&\\
\colhead{(1)} & \colhead{(2)} & \colhead{(3)} & \colhead{(4)} & \colhead{(5)} & \colhead{(6)} &
\colhead{(7)} & 
\colhead{(8)} & \colhead{(9)} & \colhead{(10)} & \colhead{(11)} & \colhead{(12)} & \colhead{(13)} & 
\colhead{(14)} & \colhead{(15)}}
\startdata
$1$&$13^{h}29^{m}48.60^{s}$&$47^{\circ}
12'8.20"$&$-125.1$&$1.3$&$4.4$&$13\pm25$&$4.2\pm3.3$&$0.33\pm0.18$&$2.47\pm6.32$&$1.7$&$95$&$1.0$&$I
A $&$0$\\
$2$&$13^{h}29^{m}57.93^{s}$&$47^{\circ}
13'4.42"$&$-120.7$&$4.9$&$5.0$&$32\pm0$&$4.9\pm3.1$&$1.32\pm0.58$&$8.06\pm10.22$&$1.4$&$155$&$0
.6$&$IA$&$1$\\
$3$&$13^{h}29^{m}46.81^{s}$&$47^{\circ}
12'13.44"$&$-115.5$&$1.7$&$5.3$&$32\pm0$&$2.4\pm1.6$&$0.29\pm0.11$&$1.90\pm2.52$&$1.5$&$56$&$0.
8$&$SA$&$1$\\
$4$&$13^{h}29^{m}49.69^{s}$&$47^{\circ}
12'48.13"$&$-111.1$&$1.8$&$5.1$&$42\pm13$&$6.0\pm2.7$&$1.10\pm0.26$&$16.02\pm16.19$&$3.4$&$18$&
$0.4$&$IA$&$0$\\
$5$&$13^{h}29^{m}55.19^{s}$&$47^{\circ}
13'1.08"$&$-113.3$&$3.5$&$4.8$&$32\pm0$&$5.0\pm2.3$&$1.19\pm0.47$&$8.30\pm7.71$&$1.6$&$163$&$0.
5$&$IA$&$1$\\
$6$&$13^{h}29^{m}49.61^{s}$&$47^{\circ}
11'49.70"$&$-104.4$&$2.1$&$6.5$&$32\pm0$&$7.3\pm2.4$&$0.89\pm0.25$&$17.83\pm11.78$&$4.6$&$158$&
$0.5$&$CR$&$1$\\
$7$&$13^{h}29^{m}53.21^{s}$&$47^{\circ}
11'54.42"$&$-110.4$&$1.7$&$5.3$&$22\pm10$&$7.3\pm2.8$&$0.58\pm0.21$&$12.33\pm9.53$&$4.9$&$37$&$0
.5$&$CR$&$0$\\
$8$&$13^{h}29^{m}52.22^{s}$&$47^{\circ}
11'40.99"$&$0.8$&$16.5$&$41.6$&$2346\pm7$&$50.7\pm0.3$&$(6.45\pm0.24)\times10^{3}
$&$(6.26\pm0.08)\times10^{4}$ & $ 2 .
23$&$52$&$0.7$&$CR$&$0$\\
$9$&$13^{h}29^{m}54.93^{s}$&$47^{\circ}
12'11.14"$&$-112.1$&$1.4$&$5.2$&$32\pm0$&$5.4\pm3.3$&$0.34\pm0.17$&$9.78\pm11.92$&$6.7$&$15$&$0
.7$&$SA$&$1$\\
$10$&$13^{h}30^{m}0.26^{s}$&$47^{\circ}
12'54.19"$&$-110.0$&$2.6$&$5.4$&$32\pm0$&$8.0\pm3.6$&$0.98\pm0.31$&$21.70\pm19.70$&$5.1$&$132$&
$0.7$&$SA$&$1$\\
...&...&...&...&...&...&...&...&...&...&...&...&...&...&...\\
...&...&...&...&...&...&...&...&...&...&...&...&...&...&...\\
...&...&...&...&...&...&...&...&...&...&...&...&...&...&...\\
$309$&$13^{h}29^{m}50.89^{s}$&$47^{\circ}
11'45.80"$&$125.3$&$1.5$&$5.7$&$32\pm0$&$2.5\pm1.5$&$0.43\pm0.15$&$2.07\pm2.50$&$1.1$&$128$&$0.
6$&$CR$&$1$\\
\enddata
\tablecomments{\footnotesize 
(1) island identification number (\emph{ID}), 
(2) Right Ascension (\emph{RA (J2000)}), 
(3) Declination (\emph{Dec (J2000)}), 
(4) Velocity with respect to the systematic velocity of the galaxy ($V_{LSR}=472$ km/s, \citealt{shetty07}), 
(5) Peak brightness temperature ($T_{max}$),
(6) Peak signal-to-noise ratio ($S/N$),
(7) Radius (\emph{R}), 
(8) Velocity dispersion ($\sigma_{v}$), 
(9) CO luminosity ($L_{CO}$), 
(10) Mass from virial theorem ($M_{vir}$),
(11) Virial parameter ($\alpha$), 
(12) Position angle of island major axis, measured from North through West (\emph{PA}), 
(13) Ratio between minor axis and major axis ($b/a$), 
(14) Region of M51 where a given island has been identified, i.e. center (\emph{CR}), spiral arms
(\emph{SA}), inter-arm (\emph{IA}),
(15) Flag$=0$ indicates an actual measurement of the
island radius, Flag$=1$ indicates that the radius is an upper limit.}
\end{deluxetable}

\newpage

\section{GMC catalog generation}\label{app:command}
\noindent In its fundamental form the CPROPS package consists of two
sub-pipelines. The first one decomposes all
significant emission into smaller substructures. Those substructures
are used as starting seeds to derive GMC (or island) properties. The decomposition
pipeline can be tuned in a number of ways in order to accommodate the desired
analysis or the intrinsic characteristics of the emission in the data cube. The
property calculation package treats a decomposed cloud as an isolated
object completely separated from the environment in which it has been
identified. This second pipeline is practically fixed and depends
only on the cloud mask provided by the first pipeline. As a final step, CPROPS
applies a correction for the biases from instrumental resolution and
sensitivity. These processes can significantly alter the property measurements of the initial
cloud, but allow for a proper definition of the actual GMC (or island) characteristics. In
the following we summarize tests we made in order to ensure
an efficient cloud decomposition and to prove the reliability of the
catalog given the performance requirements of CPROPS.\\

\noindent To obtain the PAWS GMC catalog, CPROPS was run with the following parameters:

\begin{itemize}
  \item \footnotesize \verb"THRESHOLD" = 4
  \item \footnotesize \verb"EDGE" = 1.5
  \item \footnotesize \verb"MINVCHAN" = 1
  \item \footnotesize \verb"BOOTSTRAP" = 50
  \item \footnotesize \verb"SIGDISCONT" = 0
  \item \footnotesize \verb"/NONUNIFORM"
\end{itemize}

\noindent Due to the high resolution and large size of the PAWS data cube (935 pixel
$\times$ 601 pixel $\times$ 120 channels), CPROPS required a long computational
time to analyze the properties of the identified GMCs. To overcome
this limitation, the cube was divided in 28 sub-cubes with approximate dimensions
of 300 pixel $\times$ 300 pixel $\times$ 120 channels and every
sub-cube was analyzed individually. CPROPS decomposition was performed
in the central part (200 pixel $\times$ 200 pixel $\times$ 120
channels) of each sub-cube to avoid edge effects. The splitting scheme
was such to ensure enough overlap between sub-cubes so that objects at
the edge of the sub-cubes were not lost from the analysis. A procedure
to re-build the catalog has been used, taking into account the
astrometry of single sub-cubes. The resulting catalog contains 1606
individual GMCs, reduced to 1507 through the elimination of 99 false positives 
(see Section~\ref{sec:cat}).

\subsection{Testing CPROPS decomposition parameters}\label{app:test_sd}
\noindent In order to test the GMC identification capability of CPROPS in
different environments, three regions of the PAWS data cube have been
used: a part of the southern spiral arm (hereafter: \emph{SA1}), a
part of the northern spiral arm (hereafter: \emph{SA2}) and an
inter-arm region (hereafter: \emph{IA}). The analysis has been
performed in both the final hybrid and the PdBI-only cubes. Since the
parameters that control the box to search for a single GMC have been
already pushed to the limit (as a result of our velocity and spatial
resolution) we concentrated our test on the other decomposition
parameters \verb"SIGDISCONT" and \verb"DELTA". Our aim is to
obtain a decomposition recipe that maximizes the flux within GMCs,
without loosing objects that are identifiable by eye.\\

\noindent \verb"SIGDISCONT" is used to distinguish whether merging two kernels
significantly affects the property measurement. Numerically it is the
maximum logarithmic derivative (i.e. ``the percentage jump'')
allowed for two kernels to be said to merge seamlessly. A low value of
\verb"SIGDISCONT" means that small changes in the radius, line width,
or luminosity are registered as discontinuities and force the compared
local maxima to remain separate. \verb"DELTA" is a parameter that controls the minimum contrast (in
unit of $\sigma_{RMS}$) between a kernel and the highest shared contour level where it joins with
another kernel.

The default CPROPS decomposition in terms of GMC identification, is
performed by setting \verb"SIGDISCONT"=1 (thus only a $100\%$
variation in the moment measurement results into separating two
kernels), and to \verb"DELTA"=2 (i.e. if the uniquely
associated emission is not at least $2\sigma_{RMS}$ above the merge level with any other cloud, then the local maximum is merged with that cloud).

Several tests have been made using the default values for the
remaining parameters combined with values of \verb"DELTA" and
\verb"SIGDISCONT" (see Tables~\ref{tab:sdbc1}-\ref{tab:sdbc3}). A value of \verb"DELTA" above the
default one causes
CPROPS to merge more local maxima together in crowded regions. The final GMCs appear more extended
and the flux contained in clouds is higher. However, small and isolated objects are lost when \verb"DELTA"$>2$. Therefore for generating the catalog we maintained the
default value of \verb"DELTA"=2.

The loss of clouds is more severe when the default value of \verb"SIGDISCONT" is used. 
In this case, CPROPS rejects bright clouds, especially in SA1. However with
\verb"SIGDISCONT"$\leq0.8$ (i.e. $80\%$ of
variation in the moment measurements) these objects are recognized and
decomposed. This behavior can be understood considering the morphology
of the molecular gas within M51 and the derivative decimation, the cloud discrimination process
that involves \verb"SIGDISCONT". Through
this procedure, CPROPS analyzes the measured moment continuity of all
local maxima that are in the same island independently of their
physical distance. As can be seen from the island decomposition,
M51's spiral arms appear as a contiguous region of CO flux, thus kernels
in such a region are connected at a very low contour level (above the
threshold defined by the working area) even when they are widely separated. If one or more kernels for which a local discontinuity has already been identified exist between two contiguous local maxima, the merging of the kernels is no longer possible and the
lower of the two, in terms of peak brightness temperature, is eliminated from the
allocated maxima. Fig.~\ref{dendro} shows a dendrogram
representation of allocated maxima in a given island and the contour
relations between them. The double line represents the island, numbers
and straight vertical lines indicate the kernels: the length
represents their peak brightness temperature. Horizontal dashed lines indicate discontinuities in the measured moments registered by the
\verb"SIGDISCONT" analysis, while a continuity between two kernels
that would generate a single GMC is shown as a bold line. Kernels 1
and 7 are connected at a very low contour level, but cannot be merged
due to the presence of discontinuous maxima between them. Thus the
derivative decimation eliminates the kernel with the lower peak
temperature (number 7 in this case) even if it is a well defined
object. Setting a value of \verb"SIGDISCONT" equal to 0, kernels are
considered discontinuous by default. In this way, we force kernels to stay
separated. This allows CPROPS to
allocate kernels normally eliminated by the default decomposition and
solves the problem with discarded, but by eye-identifiable GMCs, in
the spiral arm region. The PAWS spatial and channel resolution
already furnished objects with characteristics of an average GMC by
the area and contrast decimation of kernels, therefore the
\verb"SIGDISCONT" control is unnecessary for the validation and thus the
reliability of the catalog. 

\begin{figure}[!h]
\begin{center}$
\begin{array}{ccc}
\includegraphics[width=0.3\textwidth]{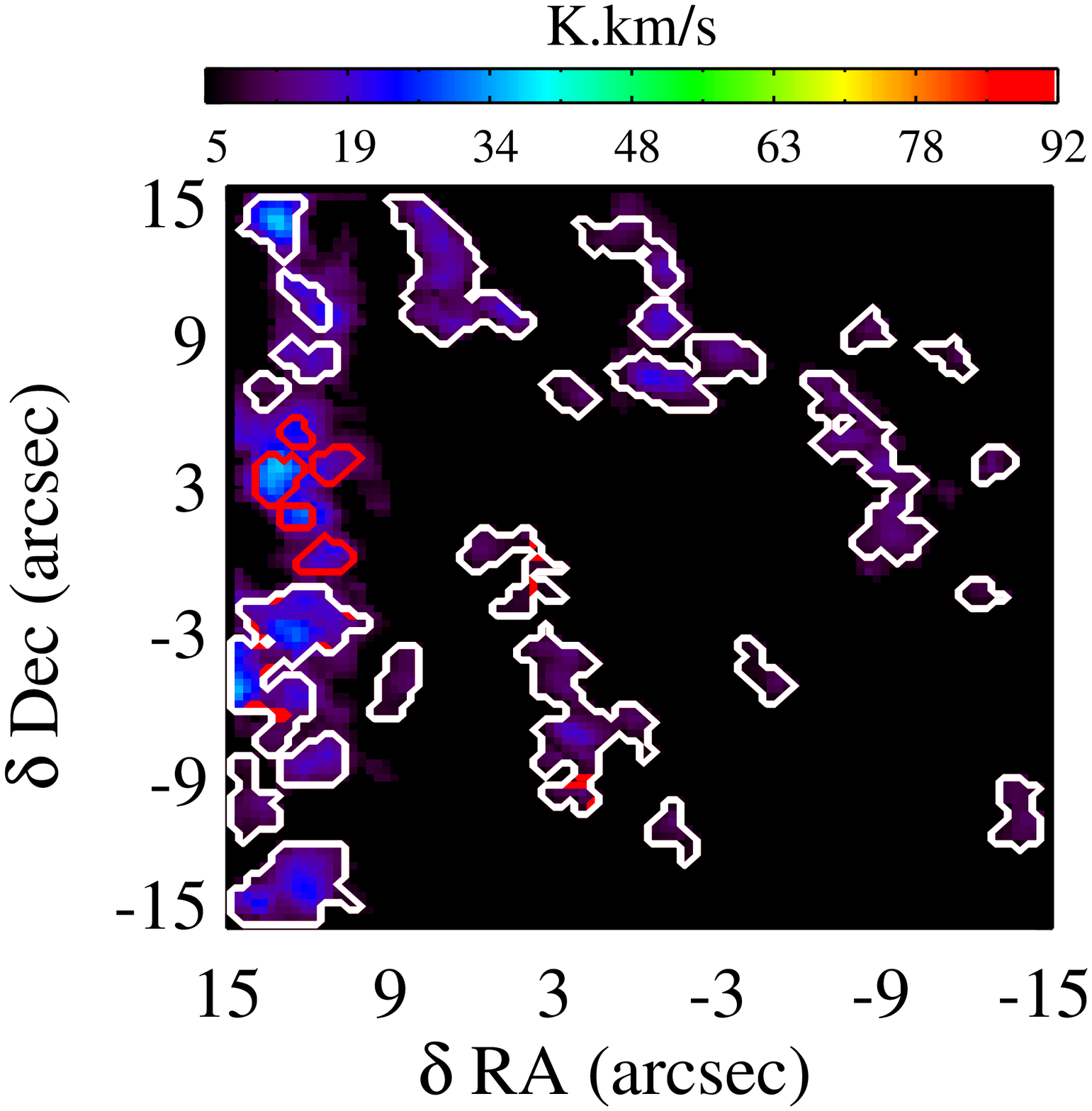} &
\includegraphics[width=0.3\textwidth]{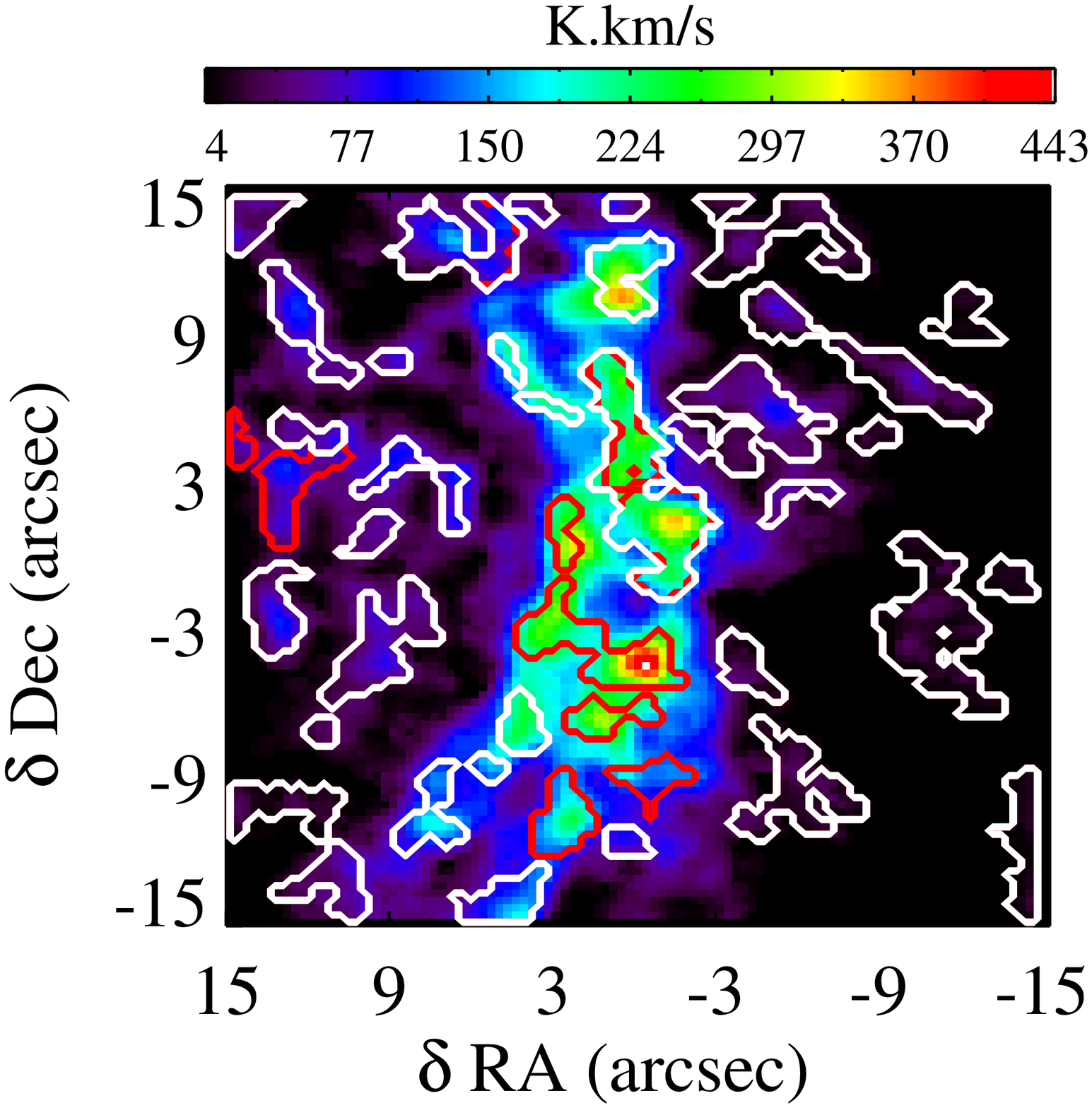} &
\includegraphics[width=0.3\textwidth]{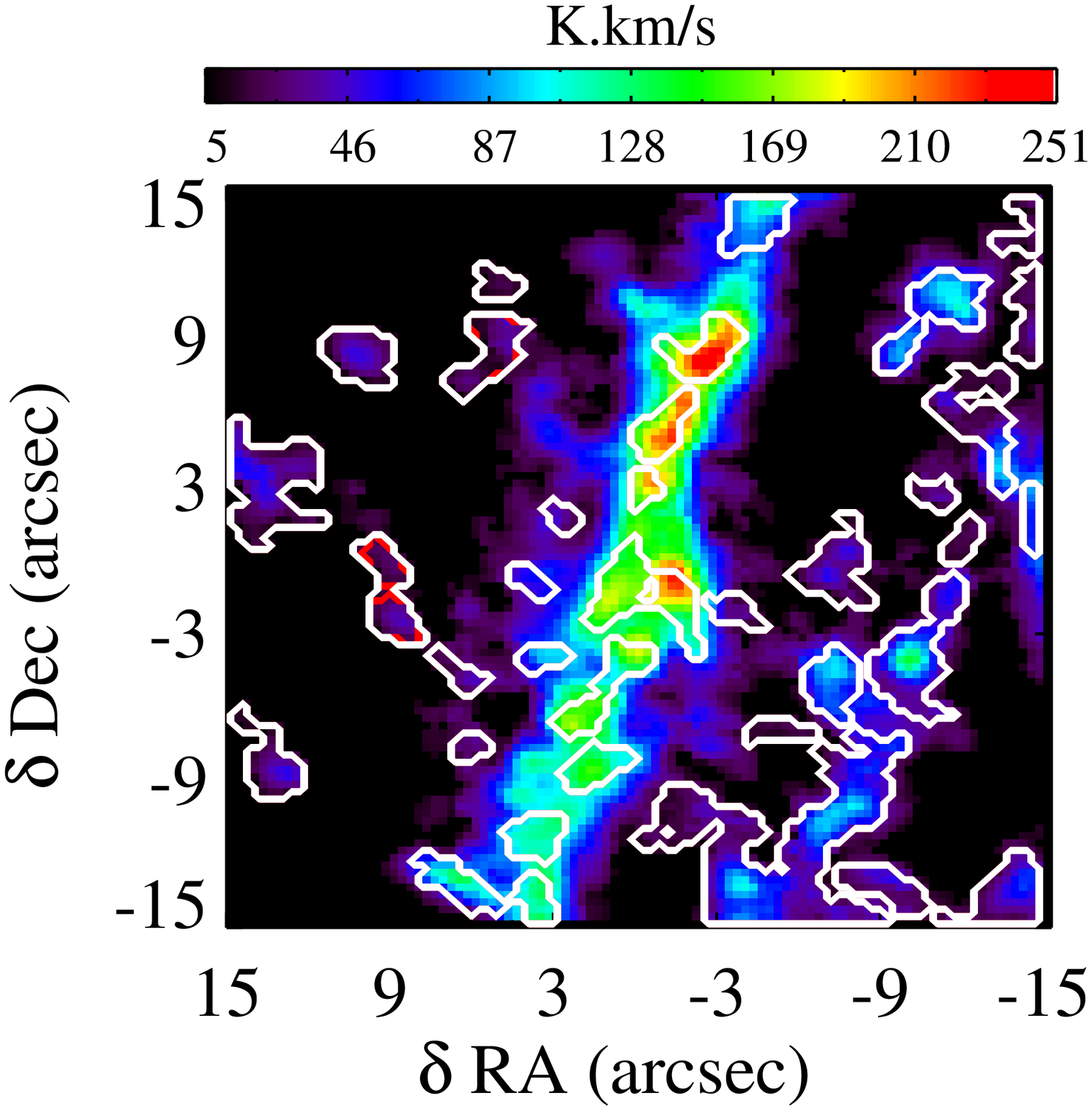}
\end{array}$
\end{center}
\caption{\footnotesize From left to right: \emph{IA}, \emph{SA1} and \emph{SA2}
subregions of the PdBI+30m cube. Red
contours show the additional objects identified using SIGDISCONT=0, white contours objects identified
using
SIGDISCONT=1 (default value). Although the decomposition for \emph{SA2} and \emph{IA} is quite
similar, many objects that can be easily identified by eye are missed in \emph{SA1} because of the
unexpected behavior of SIGDISCONT described in the text.}
\label{sigdisc}
\end{figure}

\begin{figure}[h]
\begin{center}
\includegraphics[width=0.5 \textwidth]{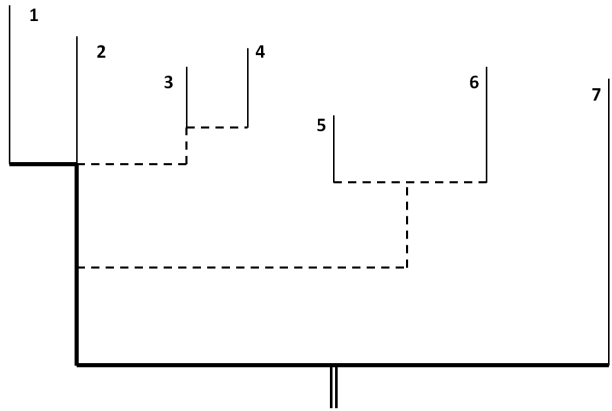}
\caption{\footnotesize Dendrogram illustration of SIGDISCONT's unexpected behavior in the
presence of
significantly extended islands. Cloud number 7 is eliminated from the catalog since it cannot merge
with cloud 1 due to discontinuous maxima between them.}
\label{dendro}
\end{center}
\end{figure}

In all environments, the flux contained in the working area is relatively high ($\sim70\%$) but the flux contained in discrete structures is much lower (20 to 30\%, depending on environment). In the spiral arm regions this percentage is always
around $20\%$. The situation for the PdBI only cube is similar, but
the flux within GMCs with respect to the total is obviously higher
(especially in the case of the inter-arm). Fig.~\ref{sigdisc} shows
the decomposition results for the default value of \verb"SIGDISCONT"
and the value used to build the catalog (\verb"SIGDISCONT"=0) for SA1,
SA2 and IA of PdBI+30m.

\clearpage
\newpage

\begin{table}[!ht]\footnotesize
\caption{CPROPS test results for the spiral arm test region SA1}
\centering
\setlength{\tabcolsep}{0.03in}
\renewcommand{\arraystretch}{0.8}
\subtable{
\begin{tabular}{|cc|ccccc|}
\hline
\multicolumn{2}{|c|}{\multirow{2}{*}{\textbf{PDBI+30m SA1}}} &
\multicolumn{5}{|c|}{\emph{SIGDISCONT}}\\
\cline{3-7}
& & 0 & 0.2 & 0.5 & 0.7 & 1 \\
\hline
\multicolumn{1}{|c|}{\multirow{6}{*}{\emph{DELTA}}} & \multicolumn{1}{|c|}{0.5}
&16&16&16&12&13\\
\multicolumn{1}{|c|}{}&\multicolumn{1}{|c|}{0.7} &16&16&16&12&13\\
\multicolumn{1}{|c|}{}&\multicolumn{1}{|c|}{1.0} &16&16&16&12&13\\
\multicolumn{1}{|c|}{}&\multicolumn{1}{|c|}{1.2} &16&16&16&12&13\\
\multicolumn{1}{|c|}{}&\multicolumn{1}{|c|}{1.5} &16&16&16&12&13\\
\multicolumn{1}{|c|}{}&\multicolumn{1}{|c|}{2.0} &16&16&16&12&13\\
\multicolumn{1}{|c|}{}&\multicolumn{1}{|c|}{2.5} &18&18&18&14&15\\
\multicolumn{1}{|c|}{}&\multicolumn{1}{|c|}{3.0} &18&18&18&14&14\\
\hline
\end{tabular}}
\subtable{
\begin{tabular}{|cc|ccccc|}
\hline
\multicolumn{2}{|c|}{\multirow{2}{*}{\textbf{PDBI only SA1}}} &
\multicolumn{5}{|c|}{\emph{SIGDISCONT}}\\
\cline{3-7}
& & 0 & 0.2 & 0.5 & 0.7 & 1 \\
\hline
\multicolumn{1}{|c|}{\multirow{6}{*}{\emph{DELTA}}} & \multicolumn{1}{|c|}{0.5}
&28&28&28&28&27\\
\multicolumn{1}{|c|}{}&\multicolumn{1}{|c|}{0.7} &28&28&28&28&27\\
\multicolumn{1}{|c|}{}&\multicolumn{1}{|c|}{1.0} &29&29&29&29&28\\
\multicolumn{1}{|c|}{}&\multicolumn{1}{|c|}{1.2} &29&29&29&29&28\\
\multicolumn{1}{|c|}{}&\multicolumn{1}{|c|}{1.5} &28&28&28&28&27\\
\multicolumn{1}{|c|}{}&\multicolumn{1}{|c|}{2.0} &29&29&29&29&28\\
\multicolumn{1}{|c|}{}&\multicolumn{1}{|c|}{2.5} &29&29&29&29&28\\
\multicolumn{1}{|c|}{}&\multicolumn{1}{|c|}{3.0} &31&31&31&31&31\\
\hline
\end{tabular}}

{\footnotesize Percentage of test cube flux contain in GMCs using different decomposition
parameter values.}
\label{tab:sdbc1}
\end{table}

\begin{table}[!h]\footnotesize
\caption{CPROPS test results for spiral arm test region SA2}
\centering
\setlength{\tabcolsep}{0.03in}
\renewcommand{\arraystretch}{0.8}
\subtable{
\begin{tabular}{|cc|ccccc|}
\hline
\multicolumn{2}{|c|}{\multirow{2}{*}{\textbf{PDBI+30m SA2}}} &
\multicolumn{5}{|c|}{\emph{SIGDISCONT}}\\
\cline{3-7}
& & 0 & 0.2 & 0.5 & 0.7 & 1 \\
\hline
\multicolumn{1}{|c|}{\multirow{6}{*}{\emph{DELTA}}} & \multicolumn{1}{|c|}{0.5}
&19&19&19&19&20\\
\multicolumn{1}{|c|}{}&\multicolumn{1}{|c|}{0.7} &19&19&19&19&20\\
\multicolumn{1}{|c|}{}&\multicolumn{1}{|c|}{1.0} &19&19&19&19&20\\
\multicolumn{1}{|c|}{}&\multicolumn{1}{|c|}{1.2} &19&19&19&19&20\\
\multicolumn{1}{|c|}{}&\multicolumn{1}{|c|}{1.5} &18&18&18&19&19\\
\multicolumn{1}{|c|}{}&\multicolumn{1}{|c|}{2.0} &19&19&19&19&19\\
\multicolumn{1}{|c|}{}&\multicolumn{1}{|c|}{2.5} &19&19&19&19&19\\
\multicolumn{1}{|c|}{}&\multicolumn{1}{|c|}{3.0} &18&18&18&19&18\\
\hline
\end{tabular}}
\subtable{
\begin{tabular}{|cc|ccccc|}
\hline
\multicolumn{2}{|c|}{\multirow{2}{*}{\textbf{PDBI only SA2}}} &
\multicolumn{5}{|c|}{\emph{SIGDISCONT}}\\
\cline{3-7}
& & 0 & 0.2 & 0.5 & 0.7 & 1 \\
\hline
\multicolumn{1}{|c|}{\multirow{6}{*}{\emph{DELTA}}} & \multicolumn{1}{|c|}{0.5}
&34&34&34&34&35\\
\multicolumn{1}{|c|}{}&\multicolumn{1}{|c|}{0.7} &34&34&34&34&35\\
\multicolumn{1}{|c|}{}&\multicolumn{1}{|c|}{1.0} &34&34&34&34&35\\
\multicolumn{1}{|c|}{}&\multicolumn{1}{|c|}{1.2} &33&33&33&33&34\\
\multicolumn{1}{|c|}{}&\multicolumn{1}{|c|}{1.5} &33&33&33&33&34\\
\multicolumn{1}{|c|}{}&\multicolumn{1}{|c|}{2.0} &32&32&32&32&33\\
\multicolumn{1}{|c|}{}&\multicolumn{1}{|c|}{2.5} &32&32&32&32&34\\
\multicolumn{1}{|c|}{}&\multicolumn{1}{|c|}{3.0} &35&35&35&35&37\\
\hline
\end{tabular}}

{\footnotesize Percentage of test cube flux contain in GMCs using different decomposition
parameter values.}
\label{tab:sdbc2}
\end{table}

\begin{table}[!hb]\footnotesize
\caption{CPROPS test results for inter-arm test region IA}
\centering
\setlength{\tabcolsep}{0.03in}
\renewcommand{\arraystretch}{0.8}
\subtable{
\begin{tabular}{|cc|ccccc|}
\hline
\multicolumn{2}{|c|}{\multirow{2}{*}{\textbf{PDBI+30m IA}}} &
\multicolumn{5}{|c|}{\emph{SIGDISCONT}}\\
\cline{3-7}
& & 0 & 0.2 & 0.5 & 0.7 & 1 \\
\hline
\multicolumn{1}{|c|}{\multirow{6}{*}{\emph{DELTA}}} & \multicolumn{1}{|c|}{0.5}
&22&22&22&22&19\\
\multicolumn{1}{|c|}{}&\multicolumn{1}{|c|}{0.7} &22&22&22&22&19\\
\multicolumn{1}{|c|}{}&\multicolumn{1}{|c|}{1.0} &22&22&22&22&19\\
\multicolumn{1}{|c|}{}&\multicolumn{1}{|c|}{1.2} &21&22&22&22&19\\
\multicolumn{1}{|c|}{}&\multicolumn{1}{|c|}{1.5} &23&23&23&23&21\\
\multicolumn{1}{|c|}{}&\multicolumn{1}{|c|}{2.0} &23&24&24&24&21\\
\multicolumn{1}{|c|}{}&\multicolumn{1}{|c|}{2.5} &24&24&24&24&21\\
\multicolumn{1}{|c|}{}&\multicolumn{1}{|c|}{3.0} &28&28&28&28&33\\
\hline
\end{tabular}}
\subtable{
\begin{tabular}{|cc|ccccc|}
\hline
\multicolumn{2}{|c|}{\multirow{2}{*}{\textbf{PDBI only IA}}} &
\multicolumn{5}{|c|}{\emph{SIGDISCONT}}\\
\cline{3-7}
& & 0 & 0.2 & 0.5 & 0.7 & 1 \\
\hline
\multicolumn{1}{|c|}{\multirow{6}{*}{\emph{DELTA}}} & \multicolumn{1}{|c|}{0.5}
&56&58&58&58&49\\
\multicolumn{1}{|c|}{}&\multicolumn{1}{|c|}{0.7} &56&58&58&58&49\\
\multicolumn{1}{|c|}{}&\multicolumn{1}{|c|}{1.0} &56&58&58&58&49\\
\multicolumn{1}{|c|}{}&\multicolumn{1}{|c|}{1.2} &56&58&58&58&49\\
\multicolumn{1}{|c|}{}&\multicolumn{1}{|c|}{1.5} &56&58&58&58&49\\
\multicolumn{1}{|c|}{}&\multicolumn{1}{|c|}{2.0} &64&66&66&66&57\\
\multicolumn{1}{|c|}{}&\multicolumn{1}{|c|}{2.5} &74&76&76&76&67\\
\multicolumn{1}{|c|}{}&\multicolumn{1}{|c|}{3.0} &73&75&75&75&67\\
\hline
\end{tabular}}

{\footnotesize Percentage of test cube flux contain in GMCs using different decomposition
parameter values.}
\label{tab:sdbc3}
\end{table}

\clearpage
\newpage

\newpage

\clearpage
\newpage

\section{Inventory of dynamically-motivated environments for M51}\label{app:env_m51}
\noindent The morphology of the grand-design spiral galaxy M51 is ideal for studying the properties
of the
molecular gas in different galactic
environments. Within the PAWS field there are three main regions where 
the molecular gas is likely subject to distinct physical
conditions (see Fig.~\ref{mask}), i.e. within the strong, nearly symmetric
\emph{spiral arms}, the \emph{inter-arm region} situated upstream and downstream of the spiral arms
and the \emph{central region},
where the gas is influenced by the 
presence of a central elliptical concentration of old stars in the form of a
nuclear bar (\citealt{rix93}). These regions can be further
divided into sub-regions, in light of the pattern of star formation
(e.g. traced by H${\alpha}$) and gas
flows (according to the profile of present-day torques; \citealt{meidt13})
within each.\\

\noindent Specifically, the central region is divided into 2 regions: 
\begin{itemize}
    \item {\it nuclear bar environment} (NB): $R_{gal}<23"$, bounded by the bar corotation
      resonance, inside of which the bar exerts negative torques and drives gas
      radially inwards
    \item {\it molecular ring environment} (MR): $23"<R_{gal}<35"$ where the influence of
      the bar and
      innermost portion of the spiral arms overlap, creating a ring-like
      accumulation of gas. The ring is sitting where the opposing forces of inner bar and the spiral density-wave cancel out.
     The high gas surface densities reached at this location result the most prominent star formation in M51. \\ 
     
\end{itemize}

\noindent Likewise, we divide the spiral arms region (SA) into three distinct environments according
to the
direction of gas flows driven in response to the underlying gravitational potential.   

\begin{itemize}
  \item {\it inner density-wave spiral arm environment} (DWI): $35"<R_{gal}<55"$ within which gas is
driven radially inward by
negative spiral-arm torquing.  This portion of the spiral arm is characterized by relatively little
star formation as traced by H${\alpha}$ and 24 $\mu m$ emission (\citealt{schinnerer13})
  \item {\it outer density-wave
    spiral arm environment} (DWO): $55"<R_{gal}<85"$ within which
gas is driven radially outward by positive spiral arm torquing.  Star formation falls on the convex
side of this portion of the spiral arms (\citealt{schinnerer13})  
  \item {\it material spiral arm environment} (MAT): $R_{gal}>85"$ beyond the boundary of positive
spiral
arm torques associated with the density wave spiral, extending to the edge of the PAWS field (within
which there is some indication that the direction of the gas flow is again reversed).\\
\end{itemize}

\noindent The width of the spiral arm environment (and each of its 3 sub-regions) is defined with
respect to
observed gas kinematics. We determine the zone of enhanced spiral streaming centered around the arm
by measuring the (rotational) auto-correlation of azimuthal streaming velocities in the PAWS field
(\citealt{meidt13}). We construct azimuthal profiles of the auto correlation signal in a
series of
radial bins and take the width of the signal at 95\% maximum as our measure of the kinematic arm
width. In testing, we find that the 95\% max-width of the CO-brightness auto correlation profile
corresponds well with the width estimated by eye from the morphology of CO brightness (\citealt{schinnerer13}).  
The average kinematic width along the two arms is centered on the spiral arm ridge
located by eye in the PAWS map of CO brightness.  Both the location of the ridge and the width are
assumed to be symmetric.\\ 

\noindent This definition of the location and width of the spiral arm ultimately yields the
definition of the
inter-arm region (IA), which we further divide in to 
\begin{itemize}
  \item {\it downstream} of the spiral arms (DNS), or the convex side where the majority of star formation related to the arms is observed in H$\alpha$ or 24 $\mu$m;
  \item {\it upstream} of the spiral arms (UPS), or the concave side basically devoid of significant star formation.
\end{itemize}

\noindent Although inside and outside corotation the gas flow direction should
change and hence the definition of up- and down-stream environments, M51 is characterized by a non-trivial dynamical structure composed by several patterns (or potential perturbations) with different pattern speeds e.g (\citealt{meidt13}, \citealt{meidt08}, \citealt{vogel93}, \citealt{elmegreen89b}, \citealt{tully74c}). 
\cite{meidt13} identified $\Omega_{b}\sim200$ km s$^{-1}$ kpc$^{-1}$ at $R_{gal}\sim20"$
corresponding to the nuclear bar corotation, $\Omega_{p,1}\sim90$ km s$^{-1}$ kpc$^{-1}$ at
$R_{gal}\sim55"$ corresponding to the inner spiral arms, $\Omega_{p,2}\sim55$ km s$^{-1}$
kpc$^{-1}$ at $R_{gal}\sim85"$ the transition between density-wave spiral arms and material arms.
This suggests that at any radius (within the PAWS FoV) a pattern is inside a corotation
resonance of another and thus the expected reversal gas flow for a single pattern is not observed.
This interpretation is supported also by the presence of the massive star formation regions along the convex side of the spiral arms only. We therefore designate the convex side of the arms as downstream, and the concave side as upstream, independent of the corotation resonances. These environments are
separated at the midpoint of the two spiral arm ridge-lines.

\newpage

\section{Kolmogorov-Smirnov test matrices}\label{app:kstest}
\noindent The tables presented here list the results of the two-sided
Kolmogorov-Smirnov (KS) tests that were carried out to evaluate differences in the cloud property distributions for different M51 environments (see Section~\ref{sec:props}). To account for measurement errors, we generated random values of a given property within the bootstrap uncertainties reported by CPROPS and we performed the test using the KSTWO procedure of the IDL
astrolibrary. The results listed in the tables are median and MAD of
p-values obtained through 100 trials. P-values lower than 0.01 show that the
cumulative distribution function of the two statistical samples are significantly
different and are indicated in bold. Values lower than 0.001 are substituted with $\mathbf<0.001$.
Differences with moderate statistical significance (up to 0.05) are indicated in italics. Results in the upper right of the matrix are for the highly reliable sample of objects ($S/N>6.5$), while results in the lower left are for the full cloud sample. CPROPS does not provide uncertainties on the peak brightness temperature measurements. We generate these using the $\sigma_{RMS}$ of the noise fluctuations along the line-of-sight where a given GMC peak temperature has been measured.

\begin{table}[h]\scriptsize
\begin{center}
\setlength{\tabcolsep}{0.04in}
\renewcommand{\arraystretch}{1.2}
\begin{tabular}{ccccccccc}
\hline
\multicolumn{1}{|c}{$\mathbf{T_{max}}$}&$\rightarrow$&\multicolumn{7}{|c|}{\textbf{Highly reliable}
($S/N>6.5$)}\\
\cline{2-9}
\multicolumn{1}{|c|}{$\downarrow$}&\multicolumn{1}{c|}{\textbf{Envir.}}&\multicolumn{1}{c}{\emph{NB}
}&\multicolumn{1}{c|}{\emph{MR}}&\multicolumn{1}{c}{\emph{DWI}}&\multicolumn{1}{c}{\emph{DWO}}
&\multicolumn{1}{c|}{\emph{MAT}}&\multicolumn{1}{c}{\emph{DNS}}&\multicolumn{1}{c|}{\emph{UPS}}\\
\cline{1-9}
\multicolumn{1}{|c|}{}&\multicolumn{1}{c|}{\emph{NB}}&\multicolumn{1}{c}{x}&\multicolumn{1}{c|}{
$\mathbf{0.007\pm0.002}$}&\multicolumn{1}{c}{$\mathbf{0.009\pm0.004}$}&\multicolumn{1}{c}{
$0.126\pm0.021$}&\multicolumn{1}{c|}{$\mathit{0.049\pm0.007}$}&\multicolumn{1}{c}{$\mathbf{<0.001}$}
&\multicolumn{1}{c|}{$\mathbf{<0.001}$}\\
\multicolumn{1}{|c|}{}&\multicolumn{1}{c|}{\emph{MR}}&\multicolumn{1}{c}{$0.066\pm0.007$}
&\multicolumn{1}{c|}{x}&\multicolumn{1}{c}{$\mathbf{<0.001}$}&\multicolumn{1}{c}{$\mathbf{<0.001}$}
&\multicolumn{1}{c|}{$\mathbf{0.002\pm0.001}$}&\multicolumn{1}{c}{$\mathbf{<0.001}$}&\multicolumn{1}
{c|}{$\mathbf{<0.001}$}\\
\cline{2-9}
\multicolumn{1}{|c|}{\textbf{Full}}&\multicolumn{1}{c|}{\emph{DWI}}&\multicolumn{1}{c}{$\mathbf{
<0.001}$}&\multicolumn{1}{c|}{$\mathbf{<0.001}$}&\multicolumn{1}{c}{x}&\multicolumn{1}{c}{
$0.093\pm0.026$}&\multicolumn{1}{c|}{$0.382\pm0.000$}&\multicolumn{1}{c}{$\mathit{0.027\pm0.014}$}
&\multicolumn{1}{c|}{$\mathbf{<0.001}$}\\
\multicolumn{1}{|c|}{}&\multicolumn{1}{c|}{\emph{DWO}}&\multicolumn{1}{c}{$\mathbf{0.002\pm0.001}$}
&\multicolumn{1}{c|}{$\mathbf{<0.001}$}&\multicolumn{1}{c}{$0.072\pm0.008$}&\multicolumn{1}{c}{x}
&\multicolumn{1}{c|}{$0.819\pm0.024$}&\multicolumn{1}{c}{$\mathbf{<0.001}$}&\multicolumn{1}{c|}{
$\mathbf{<0.001}$}\\
\multicolumn{1}{|c|}{\textbf{sample}}&\multicolumn{1}{c|}{\emph{MAT}}&\multicolumn{1}{c}{$\mathbf{
<0.001}$}&\multicolumn{1}{c|}{$\mathbf{<0.001}$}&\multicolumn{1}{c}{$0.171\pm0.107$}&\multicolumn{1}
{c}{$0.857\pm0.124$}&\multicolumn{1}{c|}{x}&\multicolumn{1}{c}{$\mathbf{0.006\pm0.002}$}
&\multicolumn{1}{c|}{$\mathbf{<0.001}$}\\
\cline{2-9}
\multicolumn{1}{|c|}{}&\multicolumn{1}{c|}{\emph{DNS}}&\multicolumn{1}{c}{$\mathbf{<0.001}$}
&\multicolumn{1}{c|}{$\mathbf{<0.001}$}&\multicolumn{1}{c}{$\mathit{0.015\pm0.007}$}&\multicolumn{1}
{c}{$\mathbf{<0.001}$}&\multicolumn{1}{c|}{$\mathbf{0.003\pm0.001}$}&\multicolumn{1}{c}{x}
&\multicolumn{1}{c|}{$\mathbf{0.008\pm0.006}$}\\
\multicolumn{1}{|c|}{}&\multicolumn{1}{c|}{\emph{UPS}}&\multicolumn{1}{c}{$\mathbf{<0.001}$}
&\multicolumn{1}{c|}{$\mathbf{<0.001}$}&\multicolumn{1}{c}{$\mathbf{0.001\pm0.001}$}&\multicolumn{1}
{c}{$\mathbf{<0.001}$}&\multicolumn{1}{c|}{$\mathbf{<0.001}$}&\multicolumn{1}{c}{$\mathbf{<0.001}$}
&\multicolumn{1}{c|}{x}\\
\hline
\end{tabular}
\label{KS_TMAX}
\end{center}
\end{table}

\begin{table}[h]\scriptsize
\begin{center}
\setlength{\tabcolsep}{0.04in}
\renewcommand{\arraystretch}{1.2}
\begin{tabular}{ccccccccc}
\hline
\multicolumn{1}{|c}{$\mathbf{R}$}&$\rightarrow$&\multicolumn{7}{|c|}{\textbf{Highly reliable}
($S/N>6.5$)}\\
\cline{2-9}
\multicolumn{1}{|c|}{$\downarrow$}&\multicolumn{1}{c|}{\textbf{Envir.}}&\multicolumn{1}{c}{\emph{NB}
}&\multicolumn{1}{c|}{\emph{MR}}&\multicolumn{1}{c}{\emph{DWI}}&\multicolumn{1}{c}{\emph{DWO}}
&\multicolumn{1}{c|}{\emph{MAT}}&\multicolumn{1}{c}{\emph{DNS}}&\multicolumn{1}{c|}{\emph{UPS}}\\
\cline{1-9}
\multicolumn{1}{|c|}{}&\multicolumn{1}{c|}{\emph{NB}}&\multicolumn{1}{c}{x}&\multicolumn{1}{c|}{
$0.600\pm0.098$}&\multicolumn{1}{c}{$0.296\pm0.149$}&\multicolumn{1}{c}{$0.586\pm0.558$}
&\multicolumn{1}{c|}{$0.455\pm0.427$}&\multicolumn{1}{c}{$\mathit{0.038\pm0.056}$}&\multicolumn{1}{
c|}{$0.581\pm0.459$}\\
\multicolumn{1}{|c|}{}&\multicolumn{1}{c|}{\emph{MR}}&\multicolumn{1}{c}{$0.797\pm0.187$}
&\multicolumn{1}{c|}{x}&\multicolumn{1}{c}{$0.603\pm0.313$}&\multicolumn{1}{c}{$0.333\pm0.291$}
&\multicolumn{1}{c|}{$0.287\pm0.226$}&\multicolumn{1}{c}{$\mathbf{0.005\pm0.007}$}&\multicolumn{1}{
c|}{$0.457\pm0.315$}\\
\cline{2-9}
\multicolumn{1}{|c|}{\textbf{Full}}&\multicolumn{1}{c|}{\emph{DWI}}&\multicolumn{1}{c}{
$0.672\pm0.322$}&\multicolumn{1}{c|}{$0.357\pm0.433$}&\multicolumn{1}{c}{x}&\multicolumn{1}{c}{
$0.295\pm0.308$}&\multicolumn{1}{c|}{$0.384\pm0.469$}&\multicolumn{1}{c}{$\mathbf{0.002\pm0.003}$}
&\multicolumn{1}{c|}{$0.357\pm0.384$}\\
\multicolumn{1}{|c|}{}&\multicolumn{1}{c|}{\emph{DWO}}&\multicolumn{1}{c}{$0.202\pm0.197$}
&\multicolumn{1}{c|}{$0.215\pm0.184$}&\multicolumn{1}{c}{$0.272\pm0.156$}&\multicolumn{1}{c}{x}
&\multicolumn{1}{c|}{$0.885\pm0.114$}&\multicolumn{1}{c}{$0.125\pm0.181$}&\multicolumn{1}{c|}{
$0.945\pm0.054$}\\
\multicolumn{1}{|c|}{\textbf{sample}}&\multicolumn{1}{c|}{\emph{MAT}}&\multicolumn{1}{c}{
$0.279\pm0.309$}&\multicolumn{1}{c|}{$0.296\pm0.167$}&\multicolumn{1}{c}{$0.283\pm0.198$}
&\multicolumn{1}{c}{$0.983\pm0.019$}&\multicolumn{1}{c|}{x}&\multicolumn{1}{c}{$0.535\pm0.479$}
&\multicolumn{1}{c|}{$0.934\pm0.057$}\\
\cline{2-9}
\multicolumn{1}{|c|}{}&\multicolumn{1}{c|}{\emph{DNS}}&\multicolumn{1}{c}{$\mathbf{0.001\pm0.002}$}
&\multicolumn{1}{c|}{$\mathbf{0.002\pm0.002}$}&\multicolumn{1}{c}{$\mathbf{0.006\pm0.009}$}
&\multicolumn{1}{c}{$0.082\pm0.119$}&\multicolumn{1}{c|}{$0.071\pm0.093$}&\multicolumn{1}{c}{x}
&\multicolumn{1}{c|}{$0.274\pm0.202$}\\
\multicolumn{1}{|c|}{}&\multicolumn{1}{c|}{\emph{UPS}}&\multicolumn{1}{c}{$0.089\pm0.130$}
&\multicolumn{1}{c|}{$0.083\pm0.120$}&\multicolumn{1}{c}{$0.106\pm0.130$}&\multicolumn{1}{c}{
$0.402\pm0.348$}&\multicolumn{1}{c|}{$0.432\pm0.228$}&\multicolumn{1}{c}{$0.643\pm0.225$}
&\multicolumn{1}{c|}{x}\\
\hline
\end{tabular}
\label{KS_R}
\caption{\footnotesize Kolmogorov-Smirnov test for \emph{Peak temperature} (top) and \emph{Radius} (bottom).}
\end{center}
\end{table}

\clearpage
\newpage

\begin{table}[h]\scriptsize
\begin{center}
\setlength{\tabcolsep}{0.04in}
\renewcommand{\arraystretch}{1.2}
\begin{tabular}{ccccccccc}
\hline
\multicolumn{1}{|c}{$\mathbf{\sigma_{v}}$}&$\rightarrow$&\multicolumn{7}{|c|}{\textbf{Highly reliable}
($S/N>6.5$)}\\
\cline{2-9}
\multicolumn{1}{|c|}{$\downarrow$}&\multicolumn{1}{c|}{\textbf{Envir.}}&\multicolumn{1}{c}{\emph{NB}
}&\multicolumn{1}{c|}{\emph{MR}}&\multicolumn{1}{c}{\emph{DWI}}&\multicolumn{1}{c}{\emph{DWO}}
&\multicolumn{1}{c|}{\emph{MAT}}&\multicolumn{1}{c}{\emph{DNS}}&\multicolumn{1}{c|}{\emph{UPS}}\\
\cline{1-9}
\multicolumn{1}{|c|}{}&\multicolumn{1}{c|}{\emph{NB}}&\multicolumn{1}{c}{x}&\multicolumn{1}{c|}{
$0.071\pm0.056$}&\multicolumn{1}{c}{$0.173\pm0.065$}&\multicolumn{1}{c}{$0.734\pm0.279$}
&\multicolumn{1}{c|}{$0.128\pm0.109$}&\multicolumn{1}{c}{$\mathbf{<0.001}$}&\multicolumn{1}{c|}{
$0.060\pm0.084$}\\
\multicolumn{1}{|c|}{}&\multicolumn{1}{c|}{\emph{MR}}&\multicolumn{1}{c}{$0.231\pm0.126$}
&\multicolumn{1}{c|}{x}&\multicolumn{1}{c}{$0.545\pm0.202$}&\multicolumn{1}{c}{$0.107\pm0.102$}
&\multicolumn{1}{c|}{$\mathbf{0.007\pm0.009}$}&\multicolumn{1}{c}{$\mathbf{<0.001}$}&\multicolumn{1}
{c|}{$\mathbf{<0.001}$}\\
\cline{2-9}
\multicolumn{1}{|c|}{\textbf{Full}}&\multicolumn{1}{c|}{\emph{DWI}}&\multicolumn{1}{c}{
$0.486\pm0.547$}&\multicolumn{1}{c|}{$0.188\pm0.102$}&\multicolumn{1}{c}{x}&\multicolumn{1}{c}{
$0.395\pm0.195$}&\multicolumn{1}{c|}{$0.050\pm0.045$}&\multicolumn{1}{c}{$\mathbf{<0.001}$}
&\multicolumn{1}{c|}{$\mathbf{0.002\pm0.002}$}\\
\multicolumn{1}{|c|}{}&\multicolumn{1}{c|}{\emph{DWO}}&\multicolumn{1}{c}{$0.595\pm0.196$}
&\multicolumn{1}{c|}{$0.115\pm0.114$}&\multicolumn{1}{c}{$0.700\pm0.135$}&\multicolumn{1}{c}{x}
&\multicolumn{1}{c|}{$0.052\pm0.053$}&\multicolumn{1}{c}{$\mathbf{<0.001}$}&\multicolumn{1}{c|}{
$\mathbf{0.008\pm0.011}$}\\
\multicolumn{1}{|c|}{\textbf{sample}}&\multicolumn{1}{c|}{\emph{MAT}}&\multicolumn{1}{c}{$\mathbf{
0.007\pm0.010}$}&\multicolumn{1}{c|}{$\mathbf{<0.001}$}&\multicolumn{1}{c}{$\mathbf{0.002\pm0.003}$}
&\multicolumn{1}{c}{$\mathbf{<0.001}$}&\multicolumn{1}{c|}{x}&\multicolumn{1}{c}{$\mathit{
0.018\pm0.024}$}&\multicolumn{1}{c|}{$0.154\pm0.161$}\\
\cline{2-9}
\multicolumn{1}{|c|}{}&\multicolumn{1}{c|}{\emph{DNS}}&\multicolumn{1}{c}{$\mathbf{<0.001}$}
&\multicolumn{1}{c|}{$\mathbf{<0.001}$}&\multicolumn{1}{c}{$\mathbf{<0.001}$}&\multicolumn{1}{c}{
$\mathbf{<0.001}$}&\multicolumn{1}{c|}{$0.069\pm0.100$}&\multicolumn{1}{c}{x}&\multicolumn{1}{c|}{
$0.175\pm0.232$}\\
\multicolumn{1}{|c|}{}&\multicolumn{1}{c|}{\emph{UPS}}&\multicolumn{1}{c}{$\mathit{0.017\pm0.026}$}
&\multicolumn{1}{c|}{$\mathbf{<0.001}$}&\multicolumn{1}{c}{$\mathit{0.024\pm0.025}$}&\multicolumn{1}
{c}{$\mathbf{0.004\pm0.005}$}&\multicolumn{1}{c|}{$0.247\pm0.255$}&\multicolumn{1}{c}{$\mathit{
0.022\pm0.018}$}&\multicolumn{1}{c|}{x}\\
\hline
\end{tabular}
\label{KS_SV}
\end{center}
\end{table}

\begin{table}[h]\scriptsize
\begin{center}
\setlength{\tabcolsep}{0.04in}
\renewcommand{\arraystretch}{1.2}
\begin{tabular}{ccccccccc}
\hline
\multicolumn{1}{|c}{$\mathbf{b/a}$}&$\rightarrow$&\multicolumn{7}{|c|}{\textbf{Highly reliable}
($S/N>6.5$)}\\
\cline{2-9}
\multicolumn{1}{|c|}{$\downarrow$}&\multicolumn{1}{c|}{\textbf{Envir.}}&\multicolumn{1}{c}{\emph{NB}
}&\multicolumn{1}{c|}{\emph{MR}}&\multicolumn{1}{c}{\emph{DWI}}&\multicolumn{1}{c}{\emph{DWO}}
&\multicolumn{1}{c|}{\emph{MAT}}&\multicolumn{1}{c}{\emph{DNS}}&\multicolumn{1}{c|}{\emph{UPS}}\\
\cline{1-9}
\multicolumn{1}{|c|}{}&\multicolumn{1}{c|}{\emph{NB}}&\multicolumn{1}{c}{x}&\multicolumn{1}{c|}{
$0.503\pm0.133$}&\multicolumn{1}{c}{$0.981\pm0.008$}&\multicolumn{1}{c}{$0.965\pm0.026$}
&\multicolumn{1}{c|}{$0.129\pm0.029$}&\multicolumn{1}{c}{$0.227\pm0.075$}&\multicolumn{1}{c|}{
$0.634\pm0.069$}\\
\multicolumn{1}{|c|}{}&\multicolumn{1}{c|}{\emph{MR}}&\multicolumn{1}{c}{$0.606\pm0.306$}
&\multicolumn{1}{c|}{x}&\multicolumn{1}{c}{$0.382\pm0.031$}&\multicolumn{1}{c}{$0.127\pm0.044$}
&\multicolumn{1}{c|}{$0.300\pm0.092$}&\multicolumn{1}{c}{$0.503\pm0.249$}&\multicolumn{1}{c|}{
$0.988\pm0.011$}\\
\cline{2-9}
\multicolumn{1}{|c|}{\textbf{Full}}&\multicolumn{1}{c|}{\emph{DWI}}&\multicolumn{1}{c}{
$0.808\pm0.103$}&\multicolumn{1}{c|}{$0.596\pm0.096$}&\multicolumn{1}{c}{x}&\multicolumn{1}{c}{
$0.847\pm0.129$}&\multicolumn{1}{c|}{$0.106\pm0.073$}&\multicolumn{1}{c}{$0.191\pm0.027$}
&\multicolumn{1}{c|}{$0.537\pm0.003$}\\
\multicolumn{1}{|c|}{}&\multicolumn{1}{c|}{\emph{DWO}}&\multicolumn{1}{c}{$0.889\pm0.066$}
&\multicolumn{1}{c|}{$0.576\pm0.100$}&\multicolumn{1}{c}{$0.903\pm0.088$}&\multicolumn{1}{c}{x}
&\multicolumn{1}{c|}{$\mathit{0.037\pm0.018}$}&\multicolumn{1}{c}{$0.071\pm0.023$}&\multicolumn{1}{
c|}{$0.263\pm0.045$}\\
\multicolumn{1}{|c|}{\textbf{sample}}&\multicolumn{1}{c|}{\emph{MAT}}&\multicolumn{1}{c}{$\mathbf{
0.009\pm0.004}$}&\multicolumn{1}{c|}{$\mathit{0.012\pm0.013}$}&\multicolumn{1}{c}{$\mathbf{
0.005\pm0.002}$}&\multicolumn{1}{c}{$\mathbf{<0.001}$}&\multicolumn{1}{c|}{x}&\multicolumn{1}{c}{
$0.841\pm0.103$}&\multicolumn{1}{c|}{$0.764\pm0.160$}\\
\cline{2-9}
\multicolumn{1}{|c|}{}&\multicolumn{1}{c|}{\emph{DNS}}&\multicolumn{1}{c}{$\mathit{0.025\pm0.017}$}
&\multicolumn{1}{c|}{$0.108\pm0.038$}&\multicolumn{1}{c}{$\mathit{0.037\pm0.014}$}&\multicolumn{1}{c
}{$\mathit{0.013\pm0.004}$}&\multicolumn{1}{c|}{$0.409\pm0.119$}&\multicolumn{1}{c}{x}&\multicolumn{
1}{c|}{$0.973\pm0.028$}\\
\multicolumn{1}{|c|}{}&\multicolumn{1}{c|}{\emph{UPS}}&\multicolumn{1}{c}{$0.130\pm0.100$}
&\multicolumn{1}{c|}{$0.307\pm0.245$}&\multicolumn{1}{c}{$0.158\pm0.153$}&\multicolumn{1}{c}{
$\mathit{0.039\pm0.045}$}&\multicolumn{1}{c|}{$0.495\pm0.078$}&\multicolumn{1}{c}{$0.971\pm0.032$}
&\multicolumn{1}{c|}{x}\\
\hline
\end{tabular}
\label{KS_RAT}
\end{center}
\end{table}

\begin{table}[h]\scriptsize
\begin{center}
\setlength{\tabcolsep}{0.04in}
\renewcommand{\arraystretch}{1.2}
\begin{tabular}{ccccccccc}
\hline
\multicolumn{1}{|c}{$\mathbf{\phi}$}&$\rightarrow$&\multicolumn{7}{|c|}{\textbf{Highly reliable}
($S/N>6.5$)}\\
\cline{2-9}
\multicolumn{1}{|c|}{$\downarrow$}&\multicolumn{1}{c|}{\textbf{Envir.}}&\multicolumn{1}{c}{\emph{NB}
}&\multicolumn{1}{c|}{\emph{MR}}&\multicolumn{1}{c}{\emph{DWI}}&\multicolumn{1}{c}{\emph{DWO}}
&\multicolumn{1}{c|}{\emph{MAT}}&\multicolumn{1}{c}{\emph{DNS}}&\multicolumn{1}{c|}{\emph{UPS}}\\
\cline{1-9}
\multicolumn{1}{|c|}{}&\multicolumn{1}{c|}{\emph{NB}}&\multicolumn{1}{c}{x}&\multicolumn{1}{c|}{
$\mathbf{<0.001}$}&\multicolumn{1}{c}{$\mathbf{<0.001}$}&\multicolumn{1}{c}{$\mathbf{<0.001}$}
&\multicolumn{1}{c|}{$\mathbf{<0.001}$}&\multicolumn{1}{c}{$\mathbf{<0.001}$}&\multicolumn{1}{c|}{
$\mathbf{<0.001}$}\\
\multicolumn{1}{|c|}{}&\multicolumn{1}{c|}{\emph{MR}}&\multicolumn{1}{c}{$\mathbf{<0.001}$}
&\multicolumn{1}{c|}{x}&\multicolumn{1}{c}{$\mathbf{<0.001}$}&\multicolumn{1}{c}{$\mathbf{<0.001}$}
&\multicolumn{1}{c|}{$\mathbf{<0.001}$}&\multicolumn{1}{c}{$\mathbf{<0.001}$}&\multicolumn{1}{c|}{
$\mathbf{<0.001}$}\\
\cline{2-9}
\multicolumn{1}{|c|}{\textbf{Full}}&\multicolumn{1}{c|}{\emph{DWI}}&\multicolumn{1}{c}{$\mathbf{
<0.001}$}&\multicolumn{1}{c|}{$\mathbf{0.001\pm0.000}$}&\multicolumn{1}{c}{x}&\multicolumn{1}{c}{
$\mathbf{<0.001}$}&\multicolumn{1}{c|}{$\mathbf{<0.001}$}&\multicolumn{1}{c}{$\mathbf{<0.001}$}
&\multicolumn{1}{c|}{$\mathbf{<0.001}$}\\
\multicolumn{1}{|c|}{}&\multicolumn{1}{c|}{\emph{DWO}}&\multicolumn{1}{c}{$\mathbf{<0.001}$}
&\multicolumn{1}{c|}{$\mathbf{<0.001}$}&\multicolumn{1}{c}{$\mathbf{<0.001}$}&\multicolumn{1}{c}{x}
&\multicolumn{1}{c|}{$\mathbf{0.008\pm0.000}$}&\multicolumn{1}{c}{$\mathbf{0.003\pm0.000}$}
&\multicolumn{1}{c|}{$\mathbf{0.006\pm0.000}$}\\
\multicolumn{1}{|c|}{\textbf{sample}}&\multicolumn{1}{c|}{\emph{MAT}}&\multicolumn{1}{c}{$\mathbf{
<0.001}$}&\multicolumn{1}{c|}{$\mathbf{<0.001}$}&\multicolumn{1}{c}{$\mathbf{<0.001}$}&\multicolumn{
1}{c}{$\mathbf{<0.001}$}&\multicolumn{1}{c|}{x}&\multicolumn{1}{c}{$\mathbf{0.001\pm0.000}$}
&\multicolumn{1}{c|}{$0.060\pm0.000$}\\
\cline{2-9}
\multicolumn{1}{|c|}{}&\multicolumn{1}{c|}{\emph{DNS}}&\multicolumn{1}{c}{$\mathbf{<0.001}$}
&\multicolumn{1}{c|}{$\mathbf{<0.001}$}&\multicolumn{1}{c}{$\mathbf{<0.001}$}&\multicolumn{1}{c}{
$\mathbf{<0.001}$}&\multicolumn{1}{c|}{$\mathbf{<0.001}$}&\multicolumn{1}{c}{x}&\multicolumn{1}{c|}{
$0.863\pm0.000$}\\
\multicolumn{1}{|c|}{}&\multicolumn{1}{c|}{\emph{UPS}}&\multicolumn{1}{c}{$\mathbf{<0.001}$}
&\multicolumn{1}{c|}{$\mathbf{<0.001}$}&\multicolumn{1}{c}{$\mathbf{<0.001}$}&\multicolumn{1}{c}{
$\mathbf{<0.001}$}&\multicolumn{1}{c|}{$\mathbf{0.010\pm0.000}$}&\multicolumn{1}{c}{$0.547\pm0.000$}
&\multicolumn{1}{c|}{x}\\
\hline
\end{tabular}
\label{KS_PA}
\caption{\footnotesize Kolmogorov-Smirnov test for \emph{Velocity dispersion} (top), \emph{Axis ratio} (middle) and \emph{Orientation} (bottom).}
\end{center}
\end{table}

\clearpage
\newpage

\begin{table}[h]\scriptsize
\begin{center}
\setlength{\tabcolsep}{0.04in}
\renewcommand{\arraystretch}{1.2}
\begin{tabular}{ccccccccc}
\hline
\multicolumn{1}{|c}{$\mathbf{M_{lum}}$}&$\rightarrow$&\multicolumn{7}{|c|}{\textbf{Highly reliable}
($S/N>6.5$)}\\
\cline{2-9}
\multicolumn{1}{|c|}{$\downarrow$}&\multicolumn{1}{c|}{\textbf{Envir.}}&\multicolumn{1}{c}{\emph{NB}
}&\multicolumn{1}{c|}{\emph{MR}}&\multicolumn{1}{c}{\emph{DWI}}&\multicolumn{1}{c}{\emph{DWO}}
&\multicolumn{1}{c|}{\emph{MAT}}&\multicolumn{1}{c}{\emph{DNS}}&\multicolumn{1}{c|}{\emph{UPS}}\\
\cline{1-9}
\multicolumn{1}{|c|}{}&\multicolumn{1}{c|}{\emph{NB}}&\multicolumn{1}{c}{x}&\multicolumn{1}{c|}{
$\mathit{0.040\pm0.042}$}&\multicolumn{1}{c}{$0.286\pm0.154$}&\multicolumn{1}{c}{$0.055\pm0.033$}
&\multicolumn{1}{c|}{$0.484\pm0.387$}&\multicolumn{1}{c}{$\mathbf{<0.001}$}&\multicolumn{1}{c|}{
$\mathbf{0.004\pm0.006}$}\\
\multicolumn{1}{|c|}{}&\multicolumn{1}{c|}{\emph{MR}}&\multicolumn{1}{c}{$0.196\pm0.126$}
&\multicolumn{1}{c|}{x}&\multicolumn{1}{c}{$0.097\pm0.106$}&\multicolumn{1}{c}{$0.564\pm0.211$}
&\multicolumn{1}{c|}{$\mathbf{0.002\pm0.002}$}&\multicolumn{1}{c}{$\mathbf{<0.001}$}&\multicolumn{1}
{c|}{$\mathbf{<0.001}$}\\
\cline{2-9}
\multicolumn{1}{|c|}{\textbf{Full}}&\multicolumn{1}{c|}{\emph{DWI}}&\multicolumn{1}{c}{$\mathit{
0.024\pm0.032}$}&\multicolumn{1}{c|}{$\mathit{0.017\pm0.018}$}&\multicolumn{1}{c}{x}&\multicolumn{1}
{c}{$0.237\pm0.136$}&\multicolumn{1}{c|}{$0.238\pm0.261$}&\multicolumn{1}{c}{$\mathbf{0.001\pm0.002}
$}&\multicolumn{1}{c|}{$\mathit{0.017\pm0.021}$}\\
\multicolumn{1}{|c|}{}&\multicolumn{1}{c|}{\emph{DWO}}&\multicolumn{1}{c}{$\mathit{0.044\pm0.057}$}
&\multicolumn{1}{c|}{$0.233\pm0.089$}&\multicolumn{1}{c}{$0.257\pm0.139$}&\multicolumn{1}{c}{x}
&\multicolumn{1}{c|}{$\mathit{0.029\pm0.031}$}&\multicolumn{1}{c}{$\mathbf{<0.001}$}&\multicolumn{1}
{c|}{$\mathbf{<0.001}$}\\
\multicolumn{1}{|c|}{\textbf{sample}}&\multicolumn{1}{c|}{\emph{MAT}}&\multicolumn{1}{c}{$\mathbf{
0.002\pm0.002}$}&\multicolumn{1}{c|}{$\mathbf{<0.001}$}&\multicolumn{1}{c}{$0.256\pm0.153$}
&\multicolumn{1}{c}{$\mathit{0.037\pm0.043}$}&\multicolumn{1}{c|}{x}&\multicolumn{1}{c}{$\mathit{
0.011\pm0.015}$}&\multicolumn{1}{c|}{$\mathit{0.031\pm0.040}$}\\
\cline{2-9}
\multicolumn{1}{|c|}{}&\multicolumn{1}{c|}{\emph{DNS}}&\multicolumn{1}{c}{$\mathbf{<0.001}$}
&\multicolumn{1}{c|}{$\mathbf{<0.001}$}&\multicolumn{1}{c}{$\mathbf{<0.001}$}&\multicolumn{1}{c}{
$\mathbf{<0.001}$}&\multicolumn{1}{c|}{$\mathbf{0.008\pm0.011}$}&\multicolumn{1}{c}{x}&\multicolumn{
1}{c|}{$0.364\pm0.181$}\\
\multicolumn{1}{|c|}{}&\multicolumn{1}{c|}{\emph{UPS}}&\multicolumn{1}{c}{$\mathbf{<0.001}$}
&\multicolumn{1}{c|}{$\mathbf{<0.001}$}&\multicolumn{1}{c}{$\mathbf{<0.001}$}&\multicolumn{1}{c}{
$\mathbf{<0.001}$}&\multicolumn{1}{c|}{$\mathit{0.011\pm0.015}$}&\multicolumn{1}{c}{$0.547\pm0.230$}
&\multicolumn{1}{c|}{x}\\
\hline
\end{tabular}
\label{KS_MLUM}
\end{center}
\end{table}


\begin{table}[h]\scriptsize
\begin{center}
\setlength{\tabcolsep}{0.04in}
\renewcommand{\arraystretch}{1.2}
\begin{tabular}{ccccccccc}
\hline
\multicolumn{1}{|c}{$\mathbf{M_{vir}}$}&$\rightarrow$&\multicolumn{7}{|c|}{\textbf{Highly reliable}
($S/N>6.5$)}\\
\cline{2-9}
\multicolumn{1}{|c|}{$\downarrow$}&\multicolumn{1}{c|}{\textbf{Envir.}}&\multicolumn{1}{c}{\emph{NB}
}&\multicolumn{1}{c|}{\emph{MR}}&\multicolumn{1}{c}{\emph{DWI}}&\multicolumn{1}{c}{\emph{DWO}}
&\multicolumn{1}{c|}{\emph{MAT}}&\multicolumn{1}{c}{\emph{DNS}}&\multicolumn{1}{c|}{\emph{UPS}}\\
\cline{1-9}
\multicolumn{1}{|c|}{}&\multicolumn{1}{c|}{\emph{NB}}&\multicolumn{1}{c}{x}&\multicolumn{1}{c|}{
$0.165\pm0.125$}&\multicolumn{1}{c}{$\mathit{0.026\pm0.033}$}&\multicolumn{1}{c}{$0.548\pm0.518$}
&\multicolumn{1}{c|}{$0.072\pm0.105$}&\multicolumn{1}{c}{$\mathbf{<0.001}$}&\multicolumn{1}{c|}{
$\mathit{0.036\pm0.054}$}\\
\multicolumn{1}{|c|}{}&\multicolumn{1}{c|}{\emph{MR}}&\multicolumn{1}{c}{$0.397\pm0.140$}
&\multicolumn{1}{c|}{x}&\multicolumn{1}{c}{$0.418\pm0.337$}&\multicolumn{1}{c}{$0.086\pm0.087$}
&\multicolumn{1}{c|}{$\mathbf{<0.001}$}&\multicolumn{1}{c}{$\mathbf{<0.001}$}&\multicolumn{1}{c|}{
$\mathbf{<0.001}$}\\
\cline{2-9}
\multicolumn{1}{|c|}{\textbf{Full}}&\multicolumn{1}{c|}{\emph{DWI}}&\multicolumn{1}{c}{
$0.142\pm0.076$}&\multicolumn{1}{c|}{$0.442\pm0.206$}&\multicolumn{1}{c}{x}&\multicolumn{1}{c}{
$\mathit{0.019\pm0.022}$}&\multicolumn{1}{c|}{$\mathbf{<0.001}$}&\multicolumn{1}{c}{$\mathbf{<0.001}
$}&\multicolumn{1}{c|}{$\mathbf{<0.001}$}\\
\multicolumn{1}{|c|}{}&\multicolumn{1}{c|}{\emph{DWO}}&\multicolumn{1}{c}{$0.415\pm0.397$}
&\multicolumn{1}{c|}{$\mathit{0.021\pm0.030}$}&\multicolumn{1}{c}{$0.089\pm0.120$}&\multicolumn{1}{c
}{x}&\multicolumn{1}{c|}{$0.069\pm0.053$}&\multicolumn{1}{c}{$\mathbf{<0.001}$}&\multicolumn{1}{c|}{
$\mathit{0.025\pm0.014}$}\\
\multicolumn{1}{|c|}{\textbf{sample}}&\multicolumn{1}{c|}{\emph{MAT}}&\multicolumn{1}{c}{$\mathbf{
0.004\pm0.006}$}&\multicolumn{1}{c|}{$\mathbf{<0.001}$}&\multicolumn{1}{c}{$\mathbf{<0.001}$}
&\multicolumn{1}{c}{$\mathit{0.013\pm0.014}$}&\multicolumn{1}{c|}{x}&\multicolumn{1}{c}{
$0.109\pm0.064$}&\multicolumn{1}{c|}{$0.343\pm0.136$}\\
\cline{2-9}
\multicolumn{1}{|c|}{}&\multicolumn{1}{c|}{\emph{DNS}}&\multicolumn{1}{c}{$\mathbf{<0.001}$}
&\multicolumn{1}{c|}{$\mathbf{<0.001}$}&\multicolumn{1}{c}{$\mathbf{<0.001}$}&\multicolumn{1}{c}{
$\mathbf{<0.001}$}&\multicolumn{1}{c|}{$0.150\pm0.140$}&\multicolumn{1}{c}{x}&\multicolumn{1}{c|}{
$0.281\pm0.196$}\\
\multicolumn{1}{|c|}{}&\multicolumn{1}{c|}{\emph{UPS}}&\multicolumn{1}{c}{$0.052\pm0.076$}
&\multicolumn{1}{c|}{$\mathbf{0.004\pm0.006}$}&\multicolumn{1}{c}{$\mathbf{0.003\pm0.004}$}
&\multicolumn{1}{c}{$0.095\pm0.080$}&\multicolumn{1}{c|}{$0.548\pm0.140$}&\multicolumn{1}{c}{
$\mathit{0.031\pm0.032}$}&\multicolumn{1}{c|}{x}\\
\hline
\end{tabular}
\label{KS_MVIR}
\end{center}
\end{table}

\begin{table}[h]\scriptsize
\begin{center}
\setlength{\tabcolsep}{0.04in}
\renewcommand{\arraystretch}{1.2}
\begin{tabular}{ccccccccc}
\hline
\multicolumn{1}{|c}{$\mathbf{\Sigma_{H2}}$}&$\rightarrow$&\multicolumn{7}{|c|}{\textbf{High
reliable} ($S/N>6.5$)}\\
\cline{2-9}
\multicolumn{1}{|c|}{$\downarrow$}&\multicolumn{1}{c|}{\textbf{Envir.}}&\multicolumn{1}{c}{\emph{NB}
}&\multicolumn{1}{c|}{\emph{MR}}&\multicolumn{1}{c}{\emph{DWI}}&\multicolumn{1}{c}{\emph{DWO}}
&\multicolumn{1}{c|}{\emph{MAT}}&\multicolumn{1}{c}{\emph{DNS}}&\multicolumn{1}{c|}{\emph{UPS}}\\
\cline{1-9}
\multicolumn{1}{|c|}{}&\multicolumn{1}{c|}{\emph{NB}}&\multicolumn{1}{c}{x}&\multicolumn{1}{c|}{
$\mathit{0.050\pm0.073}$}&\multicolumn{1}{c}{$0.746\pm0.288$}&\multicolumn{1}{c}{$0.063\pm0.081$}
&\multicolumn{1}{c|}{$0.623\pm0.277$}&\multicolumn{1}{c}{$0.125\pm0.146$}&\multicolumn{1}{c|}{
$0.075\pm0.099$}\\
\multicolumn{1}{|c|}{}&\multicolumn{1}{c|}{\emph{MR}}&\multicolumn{1}{c}{$0.122\pm0.166$}
&\multicolumn{1}{c|}{x}&\multicolumn{1}{c}{$\mathbf{0.003\pm0.004}$}&\multicolumn{1}{c}{
$0.115\pm0.098$}&\multicolumn{1}{c|}{$\mathit{0.049\pm0.034}$}&\multicolumn{1}{c}{$\mathbf{<0.001}$}
&\multicolumn{1}{c|}{$\mathbf{<0.001}$}\\
\cline{2-9}
\multicolumn{1}{|c|}{\textbf{Full}}&\multicolumn{1}{c|}{\emph{DWI}}&\multicolumn{1}{c}{
$0.296\pm0.318$}&\multicolumn{1}{c|}{$\mathbf{<0.001}$}&\multicolumn{1}{c}{x}&\multicolumn{1}{c}{
$\mathit{0.018\pm0.026}$}&\multicolumn{1}{c|}{$0.485\pm0.225$}&\multicolumn{1}{c}{$0.168\pm0.115$}
&\multicolumn{1}{c|}{$0.073\pm0.033$}\\
\multicolumn{1}{|c|}{}&\multicolumn{1}{c|}{\emph{DWO}}&\multicolumn{1}{c}{$0.355\pm0.252$}
&\multicolumn{1}{c|}{$0.161\pm0.089$}&\multicolumn{1}{c}{$\mathbf{0.005\pm0.007}$}&\multicolumn{1}{c
}{x}&\multicolumn{1}{c|}{$0.192\pm0.248$}&\multicolumn{1}{c}{$\mathbf{<0.001}$}&\multicolumn{1}{c|}{
$\mathbf{<0.001}$}\\
\multicolumn{1}{|c|}{\textbf{sample}}&\multicolumn{1}{c|}{\emph{MAT}}&\multicolumn{1}{c}{
$0.523\pm0.306$}&\multicolumn{1}{c|}{$\mathit{0.026\pm0.019}$}&\multicolumn{1}{c}{$0.149\pm0.095$}
&\multicolumn{1}{c}{$0.109\pm0.139$}&\multicolumn{1}{c|}{x}&\multicolumn{1}{c}{$\mathit{
0.025\pm0.018}$}&\multicolumn{1}{c|}{$\mathit{0.042\pm0.029}$}\\
\cline{2-9}
\multicolumn{1}{|c|}{}&\multicolumn{1}{c|}{\emph{DNS}}&\multicolumn{1}{c}{$\mathit{0.041\pm0.061}$}
&\multicolumn{1}{c|}{$\mathbf{<0.001}$}&\multicolumn{1}{c}{$0.363\pm0.323$}&\multicolumn{1}{c}{
$\mathbf{<0.001}$}&\multicolumn{1}{c|}{$0.074\pm0.071$}&\multicolumn{1}{c}{x}&\multicolumn{1}{c|}{
$0.387\pm0.100$}\\
\multicolumn{1}{|c|}{}&\multicolumn{1}{c|}{\emph{UPS}}&\multicolumn{1}{c}{$0.086\pm0.127$}
&\multicolumn{1}{c|}{$\mathbf{<0.001}$}&\multicolumn{1}{c}{$0.086\pm0.116$}&\multicolumn{1}{c}{
$\mathbf{<0.001}$}&\multicolumn{1}{c|}{$0.055\pm0.064$}&\multicolumn{1}{c}{$0.501\pm0.285$}
&\multicolumn{1}{c|}{x}\\
\hline
\end{tabular}
\label{KS_SD}
\caption{\footnotesize Kolmogorov-Smirnov test for \emph{Luminosity mass} (top), \emph{Virial mass} (middle) and \emph{Surface density} (bottom).}
\end{center}
\end{table}

\clearpage
\newpage

\begin{table}[h]\scriptsize
\begin{center}
\setlength{\tabcolsep}{0.04in}
\renewcommand{\arraystretch}{1.2}
\begin{tabular}{ccccccccc}
\hline
\multicolumn{1}{|c}{$\mathbf{c}$}&$\rightarrow$&\multicolumn{7}{|c|}{\textbf{Highly reliable}
($S/N>6.5$)}\\
\cline{2-9}
\multicolumn{1}{|c|}{$\downarrow$}&\multicolumn{1}{c|}{\textbf{Envir.}}&\multicolumn{1}{c}{\emph{NB}
}&\multicolumn{1}{c|}{\emph{MR}}&\multicolumn{1}{c}{\emph{DWI}}&\multicolumn{1}{c}{\emph{DWO}}
&\multicolumn{1}{c|}{\emph{MAT}}&\multicolumn{1}{c}{\emph{DNS}}&\multicolumn{1}{c|}{\emph{UPS}}\\
\cline{1-9}
\multicolumn{1}{|c|}{}&\multicolumn{1}{c|}{\emph{NB}}&\multicolumn{1}{c}{x}&\multicolumn{1}{c|}{
$0.172\pm0.114$}&\multicolumn{1}{c}{$0.265\pm0.205$}&\multicolumn{1}{c}{$0.360\pm0.120$}
&\multicolumn{1}{c|}{$0.495\pm0.471$}&\multicolumn{1}{c}{$\mathit{0.013\pm0.019}$}&\multicolumn{1}{
c|}{$0.107\pm0.133$}\\
\multicolumn{1}{|c|}{}&\multicolumn{1}{c|}{\emph{MR}}&\multicolumn{1}{c}{$0.140\pm0.136$}
&\multicolumn{1}{c|}{x}&\multicolumn{1}{c}{$0.684\pm0.215$}&\multicolumn{1}{c}{$0.331\pm0.185$}
&\multicolumn{1}{c|}{$\mathbf{0.006\pm0.008}$}&\multicolumn{1}{c}{$\mathbf{<0.001}$}&\multicolumn{1}
{c|}{$\mathbf{<0.001}$}\\
\cline{2-9}
\multicolumn{1}{|c|}{\textbf{Full}}&\multicolumn{1}{c|}{\emph{DWI}}&\multicolumn{1}{c}{
$0.265\pm0.121$}&\multicolumn{1}{c|}{$0.365\pm0.235$}&\multicolumn{1}{c}{x}&\multicolumn{1}{c}{
$0.511\pm0.174$}&\multicolumn{1}{c|}{$\mathit{0.023\pm0.030}$}&\multicolumn{1}{c}{$\mathbf{<0.001}$}
&\multicolumn{1}{c|}{$\mathbf{<0.001}$}\\
\multicolumn{1}{|c|}{}&\multicolumn{1}{c|}{\emph{DWO}}&\multicolumn{1}{c}{$0.304\pm0.152$}
&\multicolumn{1}{c|}{$0.411\pm0.232$}&\multicolumn{1}{c}{$0.909\pm0.105$}&\multicolumn{1}{c}{x}
&\multicolumn{1}{c|}{$\mathit{0.029\pm0.033}$}&\multicolumn{1}{c}{$\mathbf{<0.001}$}&\multicolumn{1}
{c|}{$\mathbf{0.009\pm0.014}$}\\
\multicolumn{1}{|c|}{\textbf{sample}}&\multicolumn{1}{c|}{\emph{MAT}}&\multicolumn{1}{c}{
$0.424\pm0.520$}&\multicolumn{1}{c|}{$\mathbf{0.004\pm0.005}$}&\multicolumn{1}{c}{$\mathbf{
0.009\pm0.013}$}&\multicolumn{1}{c}{$\mathit{0.014\pm0.020}$}&\multicolumn{1}{c|}{x}&\multicolumn{1}
{c}{$0.202\pm0.115$}&\multicolumn{1}{c|}{$0.311\pm0.119$}\\
\cline{2-9}
\multicolumn{1}{|c|}{}&\multicolumn{1}{c|}{\emph{DNS}}&\multicolumn{1}{c}{$\mathit{0.023\pm0.033}$}
&\multicolumn{1}{c|}{$\mathbf{<0.001}$}&\multicolumn{1}{c}{$\mathbf{<0.001}$}&\multicolumn{1}{c}{
$\mathbf{<0.001}$}&\multicolumn{1}{c|}{$0.158\pm0.118$}&\multicolumn{1}{c}{x}&\multicolumn{1}{c|}{
$0.497\pm0.510$}\\
\multicolumn{1}{|c|}{}&\multicolumn{1}{c|}{\emph{UPS}}&\multicolumn{1}{c}{$0.447\pm0.182$}
&\multicolumn{1}{c|}{$\mathit{0.015\pm0.015}$}&\multicolumn{1}{c}{$\mathit{0.029\pm0.032}$}
&\multicolumn{1}{c}{$0.080\pm0.112$}&\multicolumn{1}{c|}{$0.676\pm0.149$}&\multicolumn{1}{c}{
$0.103\pm0.088$}&\multicolumn{1}{c|}{x}\\
\hline
\end{tabular}
\label{KS_C}
\end{center}
\end{table}

\begin{table}[h]\scriptsize
\begin{center}
\setlength{\tabcolsep}{0.04in}
\renewcommand{\arraystretch}{1.2}
\begin{tabular}{ccccccccc}
\hline
\multicolumn{1}{|c}{$\mathbf{\alpha}$}&$\rightarrow$&\multicolumn{7}{|c|}{\textbf{Highly reliable}
($S/N>6.5$)}\\
\cline{2-9}
\multicolumn{1}{|c|}{$\downarrow$}&\multicolumn{1}{c|}{\textbf{Envir.}}&\multicolumn{1}{c}{\emph{NB}
}&\multicolumn{1}{c|}{\emph{MR}}&\multicolumn{1}{c}{\emph{DWI}}&\multicolumn{1}{c}{\emph{DWO}}
&\multicolumn{1}{c|}{\emph{MAT}}&\multicolumn{1}{c}{\emph{DNS}}&\multicolumn{1}{c|}{\emph{UPS}}\\
\cline{1-9}
\multicolumn{1}{|c|}{}&\multicolumn{1}{c|}{\emph{NB}}&\multicolumn{1}{c}{x}&\multicolumn{1}{c|}{
$0.411\pm0.122$}&\multicolumn{1}{c}{$0.362\pm0.249$}&\multicolumn{1}{c}{$0.079\pm0.079$}
&\multicolumn{1}{c|}{$0.137\pm0.077$}&\multicolumn{1}{c}{$0.070\pm0.023$}&\multicolumn{1}{c|}{
$0.430\pm0.128$}\\
\multicolumn{1}{|c|}{}&\multicolumn{1}{c|}{\emph{MR}}&\multicolumn{1}{c}{$0.791\pm0.186$}
&\multicolumn{1}{c|}{x}&\multicolumn{1}{c}{$\mathit{0.049\pm0.039}$}&\multicolumn{1}{c}{
$0.284\pm0.205$}&\multicolumn{1}{c|}{$0.245\pm0.266$}&\multicolumn{1}{c}{$0.304\pm0.248$}
&\multicolumn{1}{c|}{$0.648\pm0.426$}\\
\cline{2-9}
\multicolumn{1}{|c|}{\textbf{Full}}&\multicolumn{1}{c|}{\emph{DWI}}&\multicolumn{1}{c}{
$0.196\pm0.166$}&\multicolumn{1}{c|}{$\mathit{0.025\pm0.022}$}&\multicolumn{1}{c}{x}&\multicolumn{1}
{c}{$\mathit{0.023\pm0.026}$}&\multicolumn{1}{c|}{$\mathit{0.018\pm0.022}$}&\multicolumn{1}{c}{
$\mathit{0.018\pm0.017}$}&\multicolumn{1}{c|}{$0.080\pm0.088$}\\
\multicolumn{1}{|c|}{}&\multicolumn{1}{c|}{\emph{DWO}}&\multicolumn{1}{c}{$0.578\pm0.138$}
&\multicolumn{1}{c|}{$0.360\pm0.133$}&\multicolumn{1}{c}{$\mathit{0.044\pm0.031}$}&\multicolumn{1}{c
}{x}&\multicolumn{1}{c|}{$0.063\pm0.084$}&\multicolumn{1}{c}{$0.170\pm0.196$}&\multicolumn{1}{c|}{
$0.258\pm0.224$}\\
\multicolumn{1}{|c|}{\textbf{sample}}&\multicolumn{1}{c|}{\emph{MAT}}&\multicolumn{1}{c}{
$0.372\pm0.132$}&\multicolumn{1}{c|}{$0.396\pm0.309$}&\multicolumn{1}{c}{$\mathit{0.013\pm0.013}$}
&\multicolumn{1}{c}{$0.127\pm0.128$}&\multicolumn{1}{c|}{x}&\multicolumn{1}{c}{$0.648\pm0.252$}
&\multicolumn{1}{c|}{$0.752\pm0.291$}\\
\cline{2-9}
\multicolumn{1}{|c|}{}&\multicolumn{1}{c|}{\emph{DNS}}&\multicolumn{1}{c}{$0.178\pm0.228$}
&\multicolumn{1}{c|}{$0.217\pm0.215$}&\multicolumn{1}{c}{$\mathbf{0.005\pm0.005}$}&\multicolumn{1}{c
}{$0.339\pm0.427$}&\multicolumn{1}{c|}{$0.677\pm0.237$}&\multicolumn{1}{c}{x}&\multicolumn{1}{c|}{
$0.777\pm0.212$}\\
\multicolumn{1}{|c|}{}&\multicolumn{1}{c|}{\emph{UPS}}&\multicolumn{1}{c}{$0.456\pm0.262$}
&\multicolumn{1}{c|}{$0.172\pm0.089$}&\multicolumn{1}{c}{$0.425\pm0.286$}&\multicolumn{1}{c}{
$\mathit{0.026\pm0.025}$}&\multicolumn{1}{c|}{$\mathit{0.039\pm0.036}$}&\multicolumn{1}{c}{$\mathbf{
0.009\pm0.006}$}&\multicolumn{1}{c|}{x}\\
\hline
\end{tabular}
\label{KS_ALPHA}
\caption{\footnotesize Kolmogorov-Smirnov test for \emph{Scaling parameter} (top) and \emph{Virial parameter} (bottom).}
\end{center}
\end{table}

\clearpage
 
\end{document}